\documentclass[10pt,twocolumn]{emulateapj}

\usepackage{amsmath}
\usepackage{amssymb}
\usepackage{anysize}
\usepackage{color}
\usepackage{enumitem}
\usepackage{graphicx}
\usepackage{natbib}

\DeclareMathOperator\arcsinh{arcsinh}

\setlist[itemize]{leftmargin=*}
\setlist[enumerate]{leftmargin=*}

\begin{document}
\renewcommand{\vec}[1]{\boldsymbol{#1}}
\newcommand{\hvec}[1]{\hat{\boldsymbol{#1}}}
\newcommand{\tens}[1]{\mathsf{#1}}
\newcommand{\sgn}{\textrm{sgn}}
\newcommand{\pd}[2]{\frac{\partial #1}{\partial #2} }
\newcommand{\HALF}{\frac{1}{2}}
\newcommand{\THREEHALF}{\frac{3}{2}}
\newcommand{\DS}{\displaystyle}
\newcommand{\Fluid}{\color{blue}Fluid\color{black}}
\newcommand{\Particles}{\color{red}Particles\color{black}}
\newcommand{\C}{{\mathbb C}}
\newcommand{\E}{E}
\newcommand{\F}{{\cal F}}
\newcommand{\T}{{\cal T}}
\newcommand{\Pt}{{\cal P}}
\newcommand{\V}{V}
\newcommand{\U}{U}
\newcommand{\av}[1]{\left<#1\right>}
\newcommand{\indx}{{\vec{\mathrm i}}}
\newcommand{\CR}{{\textsc{\tiny CR}}}
\newcommand{\quotes}[1]{``#1''}
\newcommand*\blue{\color{blue}}

\title{A Particle Module for the PLUTO Code: I - an implementation of the MHD-PIC equations}

\author{A. Mignone}
\affil{Physics Department, University of Turin,
       via Pietro Giuria 1 (10125) Torino, Italy}

\author{G. Bodo}
\affil{INAF, Osservatorio Astrofisico di Torino, Strada Osservatorio 20,
       Pino Torinese 10025, Italy}

\author{B. Vaidya}
\affil{Centre of Astronomy, Indian Institute of Technology Indore, Khandwa Road, Simrol , Indore 453552, India}

\and

\author{G. Mattia}
\affil{Physics Department, University of Turin,
       via Pietro Giuria 1 (10125) Torino, Italy}

\altaffiltext{1}{AAS Journals Data Scientist}

\begin{abstract}

We describe an implementation of a particle physics module available for the PLUTO code, appropriate for the dynamical evolution of a plasma consisting of a thermal fluid and a non-thermal component represented by relativistic charged particles, or cosmic rays (CR).
While the fluid is approached using standard numerical schemes for magnetohydrodynamics, CR particles are treated kinetically using conventional Particle-In-Cell (PIC) techniques.

The module can be used to describe either test particles motion in the fluid electromagnetic field or to solve the fully coupled MHD-PIC system of equations with particle backreaction on the fluid as originally introduced by \cite{Bai_etal.2015}.
Particle backreaction on the fluid is included in the form of momentum-energy feedback and by introducing the CR-induced Hall term in Ohm's law.
The hybrid MHD-PIC module can be employed to study CR kinetic effects on scales larger than the (ion) skin depth provided the Larmor gyration scale is properly resolved.
When applicable, this formulation avoids to resolve microscopic scales offering a substantial computational saving with respect to PIC simulations.

We present a fully-conservative formulation which is second-order accurate in time and space and extends to either Runge-Kutta (RK) or corner-transport-upwind (CTU) time-stepping schemes (for the fluid) while a standard Boris integrator is employed for the particles.
For highly-energetic relativistic CRs and in order to overcome the time step restriction a novel sub-cycling strategy that retains second-order accuracy in time is presented.
Numerical benchmarks and applications including Bell instability, diffusive shock acceleration and test particle acceleration in reconnecting layers are discussed.

\end{abstract}


\section{Introduction}
\label{sec:Introduction}
%
%
%
%

High-energy astrophysical phenomena are connected with environments where matter exists under extreme conditions leading to powerful releases of electromagnetic radiation from radio, to optical, $X$-rays and $\gamma$-rays wavebands.
Typical examples are found in blazar jets \citep{Bottcher.2007,Giannios.2013}, gamma-ray bursts \citep[GRB, see e.g.][]{Giannios.2008, McKUzd.2012, BenPir.2014, BenGia.2017}, pulsar wind nebulae \citep[PWN, see e.g.][and reference therein]{Bucciantini_etal.2011, Kargaltsev_etal.2015, Olmi_etal.2016} and supernovae remnants \citep[SNR, see e.g.][and reference therein]{Amato_Blasi.2009,Morlino_etal.2013,Caprioli_Spitkovsky.2014c} among others.
The observed radiation presents typical signatures of non-thermal emission processes such as syhncrotron and inverse Compton, typically arising from charged particles accelerated by electromagnetic fields.

A comprehensive modeling of such systems is a challenging task because physical mechanisms operate over an enormous range of spatial and temporal scales stretching from the microphysical scale - where energy dissipation occurs and emission originates - to the macroscopic scale - where dynamics trigger dissipation.
Owing to the complexity of the interactions, state of the art modeling and key achievements in this field have been obtained mostly through time-dependent numerical computations.
For these reasons, our current understanding of astrophysical systems is limited by the range of scales beyond which one or more model assumptions breaks down or when computational resources become prohibitive.

On the one hand, fluid models such as magnetohydrodynamics (MHD) have been extensively applied to investigate the large-scale dynamics of high-energy astrophysical environments in jets \citep[e.g.][]{Rossi_etal.2008, Mignone_etal.2010,Mignone_etal.2013, MLNH.2012,Porth.2013,EHK.2016, DTG.2017}, PWN \citep[e.g.][]{delZanna_etal.2006} and also supernovae remnants \citep{Orlando_Drake_Laming.2009,Miceli_etal.2016}.
By its own nature, however, the fluid approach is applicable on scales much larger than the Larmor radius and it cannot capture important kinetic effects relevant to the micro-scale.
On the other hand, Particle-in-Cell \citep[PIC, see the book by][]{Birsdall_and_Langdon.2004} codes provide the most self-consistent approach to model plasma dynamics at small scales \citep[e.g.][]{Chang_etal.2008,Sironi_etal.2013, Sironi_Spitkovsky.2014}.
However, PIC codes must resolve the electron skin depth which, in most cases, is several orders of magnitude smaller than the overall size of a typical astrophysical system.
Even with the most powerful supercomputers, PIC simulations become prohibitively expensive to describe astrophysical systems at larger scales.
Alternatively, hybrid codes that treat ions as particles and electrons as fluid \citep{Gargate_etal.2007,Kunz_etal.2014} are commonly used in space physics and laboratory plasma.
Hybrid methods cannot capture kinetic effects at the electron scale and the temporal and spatial scales are limited in resolution by the ion inertial length.

Recently \cite{Bai_etal.2015} have proposed yet another approach, called the MHD-PIC method, for describing the interaction between a collisionless thermal plasma and a population of non-thermal cosmic rays particles (CR, typically ions).
The same approach has also been recently employed in the work by \cite{vanMarle_etal.2018} to study magnetic field amplification and particle acceleration near non-relativistic astrophysical shocks.
In that study, the authors generalize the MHD-PIC approach to any type of suprathermal particles (electrons and ions).
The MHD-PIC model, which can be formally derived by considering a three-component plasma in which thermal electrons are massless, does not capture the electron physics and it can be used to describe non-thermal ions kinetic effects on scales that are not tied to the inertial skin depth but only to the gyration radius.

In the present work we describe the numerical implementation of the MHD-PIC particle module in the PLUTO code for astrophysical fluid-dynamics \citep{PLUTO.2007, PLUTO.2012} while providing, at the same time, some new implementation strategies allowing our hybrid framework to be employed with more general second-order time-stepping schemes and to improve in terms of accuracy.
In addition, the presented module can also be used to model the dynamics of charged test particles in a time-dependent magnetized fluid, with straightforward generalization to the relativistic case.
Our numerical framework is part of a more general fluid-particle module and a companion work \citep[][Paper II]{Vaidya_etal.2018} presents a different hybrid scheme for the modeling of non-thermal spectral signatures from highly energetic electrons, embedded in a thermal MHD plasma using Lagrangian particles with a time-dependent energy distribution.
A brief description together with applications to astrophysical jets have been presented in \cite{Vaidya_etal.2016}.

The paper is organized as follows.
The MHD-PIC equations describing the evolution of the composite fluid+CR system are discusses in Section \ref{sec:MHD-PIC} while the numerical implementation is described in Section \ref{sec:NumImplementation}.
Numerical benchmarks and applications, for both the full MHD-PIC composite system (including feedback) and the test-particle implementation (without feedback), are presented in Section \ref{sec:tests}.
A summary is given in Section \ref{sec:summary}.

\section{The MHD-PIC Equations}
\label{sec:MHD-PIC}
%

The MHD-PIC approach was recently developed by \cite{Bai_etal.2015} to describe the dynamical interaction between a thermal plasma and a non-thermal population of collisionless cosmic rays (CR henceforth).
While the thermal component, that comprises ions and massless electrons, is described through a fluid approach making use of shock-capturing MHD methods, CR particles (representing energetic ions or electrons) are treated kinetically using a conventional PIC techniques.
This formalism aims at capturing the kinetic effects of CR particles without the need to resolve the plasma skin depth, as it is typically required by PIC codes.
In the MHD-PIC formalism, instead, only the Larmor (gyration) scale must be properly resolved.
This extends the range of applicability to much larger spatial (and temporal) scales when compared to the standard PIC approach, inasmuch the particle gyroradius largely exceeds the plasma skin depth $c/\omega_{pi} \approx 2.27\times 10^7/\sqrt{n_i}\, {\rm cm}$.

The thermal plasma is described by the single fluid model which is obtained by averaging the two fluid equations for ions and massless electrons.
The derivation, that may also be found on many plasma physics textbooks, is given in the Appendix (\ref{app:MHD-PIC}) for lengthy reasons.
The relevant equations, given by (\ref{eq:app:sF_continuity}), (\ref{eq:app:sF_momentum}) and (\ref{eq:app:sF_energy}), include the effect of CR's through the condition of charge neutrality for the composite system (fluid+CR) and the definition of the total current density,
\begin{align}
  &\DS q_g + q_\CR = q_e + q_i + q_\CR = 0
  \label{eq:qtot} \\ \noalign{\medskip}
  &\DS\vec{J}_g + \vec{J}_\CR = \vec{J} = \frac{c}{4\pi}\nabla\times\vec{B} \,,
  \label{eq:Jtot}
\end{align}
where $q_g=q_i + q_e$ is the charge density of the thermal plasma (Equation \ref{eq:app:sF_cc}), $q_\CR$ is CR charge density while $\vec{J}_\CR\equiv q_\CR \vec{v}_\CR$ is CR current density.

Neglecting contributions from the heat flux vector and the viscous stress tensor and taking advantage of (\ref{eq:qtot}) and (\ref{eq:Jtot}), straightforward manipulation of Equations (\ref{eq:app:sF_continuity})-(\ref{eq:app:sF_energy}), properly augmented with Faraday's law of induction, leads to the quasi-conservative form of the MHD-PIC equations:
\begin{align}
  \pd{\rho}{t} &+ \nabla\cdot(\rho\vec{v}_g) = 0
  \label{eq:MHD-PIC:rho} \\ \noalign{\medskip}
  \pd{(\rho\vec{v}_g)}{t} &+ \nabla\cdot\tens{T}_m = -\vec{F}_{\CR}
   \label{eq:MHD-PIC:rhov} \\ \noalign{\medskip}
  \pd{\vec{B}}{t} & + \nabla\times c\vec{\E} = \vec{0}
  \label{eq:MHD-PIC:B} \\ \noalign{\medskip}
%
  \pd{E_g}{t}  &+ \nabla\cdot\left(\rho H\vec{v}_g + \vec{S}\right)
        = -\vec{v}_g\cdot\vec{F}_{\CR} \,.
  \label{eq:MHD-PIC:E}
\end{align}
Here $\rho$ and $\vec{v}_g$ represent, respectively, the single fluid density and velocity (Equations \ref{eq:app:sF_density} and \ref{eq:app:sF_speed}) which, in the limit of massless electrons, can be trivially identified with those of the ions, i.e., $\rho \to \rho^{(i)}$ and $\vec{v}_g \to \vec{v}^{(i)}$.
The total energy density $E_g$ is expressed as the sum of kinetic, thermal and magnetic contributions,
\begin{equation}\label{eq:Eg}
 E_g = \frac{1}{2}\rho\vec{v}_g^2 + \frac{3}{2} p + \frac{\vec{B}^2}{8\pi}\,,
\end{equation}
while $\rho H$ is the gas enthalpy:
\begin{equation}\label{eq:rhoH}
  \rho H = \left(\frac{1}{2}\rho\vec{v}_g^2 + \frac{5}{2} p\right)\vec{v}_g\,,
\end{equation}
The gas pressure $p$, as shown in the Appendix, can be expressed by the sum of the ions and pressure terms of the original two fluid equations.
Finally, $\tens{T}_m$ defines the momentum flux tensor,
\begin{equation}\label{eq:Tm}
  \tens{T}_m = \rho\vec{v}_g\vec{v}_g
  -\frac{\vec{B}\vec{B}}{4\pi} + \tens{I}\left(p + \frac{\vec{B}^2}{8\pi}\right)
  \,,
\end{equation}
where $\tens{I}$ is the unit tensor, $\vec{B}$ is the magnetic field, $\vec{S} = c\vec{E}\times\vec{B}/4\pi$ is the Poynting vector and $\vec{E}$ is the electric field.

The force experienced by the fluid from the CR appears on the right hand side of Equation (\ref{eq:MHD-PIC:rhov}) and it is the opposite of the Lorentz force experienced by the particles:
\begin{equation} \label{eq:Fcr}
  \vec{F}_\CR = q_\CR\vec{E}  + \frac{1}{c} \vec{J}_\CR \times \vec{B}.
\end{equation}
Thus the last term on the right hand side of Equation (\ref{eq:MHD-PIC:E}) is interpreted as the opposite of the energy gained by the CRs due to the work done by the Lorentz force \citep[see Eq. 17 in][]{Bai_etal.2015}.

The electric field $\vec{\E}$ can be directly obtained from Ohm's law which is expressed by the electron equation of motion (\ref{eq:app:2F_momentum}) in the limit $\rho^{(e)}\to 0$.
Using the definition of the total current (\ref{eq:Jtot}), the second in Equation (\ref{eq:app:sF_cc}) and the definition of $\vec{J}_\CR$ together with the fact that $\vec{v}^{(i)} \to \vec{v}_g$,  yields
\begin{equation} \label{eq:Ohm_general}
  \begin{array}{lcl}
  \DS c \vec{\E} & = & \DS
    -\vec{v}_g \times \vec{B}
    -\frac{1}{q_e} \vec{J} \times \vec{B}  \\ \noalign{\medskip}
   &   &\DS
    - \frac{q_\CR}{|q_e|} (\vec{v}_\CR - \vec{v}_g)\times\vec{B} 
    + \frac{c}{q_e} \nabla \cdot\Pt^{(e)} \,.
 \end{array}
\end{equation}
In Equation (\ref{eq:Ohm_general}) the first term  on the right hand side is the standard convective term, the second is the Hall term, the third  describes the relative drift between CR and fluid and will be referred to as the CR-Hall term, and the last term is the electron pressure term.
As noted by \cite{Bai_etal.2015}, at scales much larger than the ion skin depth,  both the standard Hall term and the electron pressure terms can be safely neglected and the final form of Ohm's law is then
\begin{equation}\label{eq:Ohm}
  c \vec{\E} = - \vec{v}_g \times \vec{B} - R (\vec{v}_\CR - \vec{v}_g) \times \vec{B}
  \,, 
\end{equation}
where
\begin{equation}\label{eq:R}
  R = \frac{q_\CR}{|q_e|} = \frac{q_\CR}{q_i + q_\CR}
\end{equation}
is the charge density ratio between CR and electrons and the regime in which the described formalism is valid demands $R \ll 1$.
The second equality can be recovered with the aid of the charge neutrality condition, Equation (\ref{eq:qtot}).

Using the expression for the electric field (\ref{eq:Ohm}), the CR force (Equation \ref{eq:Fcr}) can be rewritten as
\begin{equation}\label{eq:Fcr2}
  \vec{F}_\CR = (1-R)\left(q_{\CR}\vec{\E}_0
                            + \frac{1}{c}\vec{J}_{\CR}\times\vec{B}\right)\,,
\end{equation}
where $\vec{\E}_0 = -\vec{v}_g\times\vec{B}/c$ is the convective electric field.
Expression (\ref{eq:Fcr2}) is more convenient for computational purposes.
Likewise, combining Equations (\ref{eq:Fcr}) and (\ref{eq:Fcr2}) yields the following expression for the total electric field
\begin{equation}\label{eq:CR_emf}
  \vec{\E} = \vec{\E}_0 - \frac{\vec{F}_\CR}{q_i} \,.
\end{equation}

The charge density of the thermal ions $q_i$ is expressed in terms of the charge to mass ratio for the ions,
\begin{equation}\label{eq:alpha_i}
  \alpha_i \equiv \left(\frac{e}{mc}\right)_i\,,
\end{equation}
so that $q_i/c = \alpha_i \rho$ where $\rho$ is the gas density.
A similar expression holds for the CR charge density, see Section \ref{sec:particles}.

\subsection{Particle Equations of Motion}
\label{sec:particles}
%
%
%
%

According to the Particle-In-Cell (PIC) formalism \citep[for a review see, e.g.,][]{Lapenta.2012}, computational particles (CR) represents clouds of physical particles that are close to each other in phase space.
CR particles are defined in terms of their spatial coordinates $\vec{x}_p$ and velocity $\vec{v}_p$ which are governed by the equation of motion
\begin{equation}\label{eq:CR}
  \left\{\begin{array}{lcl}
   \DS \frac{d\vec{x}_p}{dt}         &=& \vec{v}_p
   \\ \noalign{\medskip}
    \DS \frac{d(\gamma\vec{v})_p}{dt} &=& \DS
       \alpha_p\left(c\vec{\E} + \vec{v}_p\times\vec{B}\right)
 \end{array}\right. 
\end{equation}
where $\gamma = 1/\sqrt{1-\vec{v}^2_p/\C^2}$ is the Lorentz factor whereas
\begin{equation}\label{eq:alpha_p}
  \alpha_p \equiv \left(\frac{e}{mc}\right)_p
\end{equation}
is the CR charge to mass ratio.
Here and in what follows, the suffix $p$ will be used to label a single particle.
Using Equation (\ref{eq:alpha_p}), the charge density of an individual particle $q_p$ can be written as $q_p/c = \alpha_p\varrho_p$ where  $\varrho_{p}$ is the actual mass density contribution of a single CR particle.

Since the actual speed of light does not explicitly appears in the MHD equations, we use $\C$ to specify an artificial value for the speed of light which, for consistency reasons, must be greater than any characteristic signal velocity.
The electric and magnetic fields $\vec{\E}$ and $\vec{B}$ are computed from the magnetized fluid and must be properly interpolated at the particle position.
This is described in Section \ref{sec:mapping}.

\section{Numerical Implementation}
\label{sec:NumImplementation}
%
%
%

We now provide a detailed description of the numerical method employed for the solution of the MHD-PIC equations, Equations. (\ref{eq:MHD-PIC:rho})--(\ref{eq:MHD-PIC:E}), in the PLUTO code.
The solution methods features a MHD solver already present in the code (modified by the presence of additional terms describing the particle backreaction onto the gas) coupled to a particle integrator.

Fluid quantities such as density, magnetic field, and so forth, are discretized on a computational grid with cell indices $\vec{i}\equiv(i,j,k)$ and stored as three-dimensional arrays.
On the contrary, particles (being meshless quantities) are held in memory using a doubly linked list consisting of sequentially linked node structures.
Each node contains the particle itself and pointers to the previous and to the next node in the sequence.
In a linked list, elements can be inserted or removed in a straightforward way and shuffling operations can be easily performed by changing pointers. 
Besides, different types of particle data structures can be employed. 
These features make the linked list approach very flexible and we have adopted as a general implementation strategy shared by all particles modules in the PLUTO code, including the Lagrangian particle module described in paper II.


\subsection{MHD Integrators}
\label{sec:MHD_integrators}
%
%
%

The numerical solution of the MHD-PIC equations has been implemented by modifying two of the available second-order time-stepping algorithms available with the code.
The first one features the corner transport upwind (CTU) scheme \citep{Colella.1990, GarSto.2005, PLUTO.2012} and also present an extension of the scheme to the standard second-order total variation diminishing (TVD) Runge-Kutta (RK2).

Both implementations are second-order accurate in time and space and conserve momentum and energy to machine accuracy for the composite gas+particle system.

The magnetic field is evolved using constrained-transport (CT) although our formulation can be extended to other divergence-cleaning methods \cite[such as][]{MigTze.2010, MigTzeBod.2010} in a straightforward manner.

\subsubsection{CTU Time Stepping.}
%

We now provide a schematic description of the the CTU method while refer the reader to Appendix \ref{app:CTU-CT} for a more detailed description. 
The scheme consists of a first predictor step where time-centered states are constructed according to
\begin{equation}\label{eq:CTU_pred}
  U^{n+\HALF}_\indx =   U^n_\indx
                   + \frac{\Delta t^n}{2}{\cal L}_\indx(U^*, \vec{F}^n_\CR)
                   + \frac{\Delta t^n}{2}S_{\CR,\indx}^n  \,,
\end{equation}
where $U=(\rho,\, \rho\vec{v}_g,\, \vec{B},\, E_g)$ denotes the array of conserved quantities, $U^*$ is the normal predictor state, ${\cal L}$ is a conservative flux-difference operator:
\begin{equation}\label{eq:Lop}
  {\cal L}_\indx(U^*, \vec{F}^n_\CR) =  -\sum_d\frac{1}{\Delta x_d}
              \Big(\F_{\vec{i}+\HALF\hvec{e}_d} - \F_{\vec{i}+\HALF\hvec{e}_d}\Big)
\end{equation}
with $d=x,y,z$ labeling the direction while
\begin{equation}
  S^n_{\CR,\indx} = \left(0,
                          -\vec{F}_\CR,
                           \vec{0},
                          -\vec{F}_\CR\cdot\vec{v}_g\right)^n_\indx
\end{equation}
accounts for the source terms in the momentum and energy equations, (\ref{eq:MHD-PIC:rhov}) and (\ref{eq:MHD-PIC:E}), respectively.
The fluxes $\F_{\vec{i}\pm\HALF\hvec{e}_d}$ in Equation (\ref{eq:Lop}) are computed by solving a Riemann problem at cell interfaces and by adding the CR contribution terms in the induction and energy equations.

The CR force term is computed using Equation (\ref{eq:Fcr2}) by depositing individual particle charges and currents on the grid:
\begin{equation}\label{eq:q+J_deposit}
  \begin{array}{lcl}
    \DS\left(\frac{q_\CR}{c}\right)_\indx & = &\DS
    \sum_p W(\vec{x}_\indx-\vec{x}_p) \alpha_p\varrho_p
  \\ \noalign{\medskip}
    \DS\left(\frac{\vec{J}_\CR}{c}\right)_\indx & = &\DS
    \sum_p W(\vec{x}_\indx-\vec{x}_p) \alpha_p\varrho_p\vec{v}_p
             \end{array}
\end{equation}
where $\alpha_p$ is defined in Equation (\ref{eq:alpha_p}) whereas $W()$ are weight functions (see Section \ref{sec:mapping}). 

Particles are then evolved for a full step (see Section \ref{sec:particle_mover}) using the electromagnetic fields at the mid-point time level:
\begin{equation}\label{eq:CTU_particle_advance}
  \left(\begin{array}{l}
  \vec{x}_p \\ \noalign{\medskip}
  \vec{u}_p
  \end{array}\right)^{n+1}
   =
  \left(\begin{array}{l}
  \vec{x}_p \\ \noalign{\medskip}
  \vec{u}_p
  \end{array}\right)^n
   +
  \Delta t^n
  \left(\begin{array}{c}
  \vec{v}_p \\ \noalign{\medskip}
  \vec{a}_p
  \end{array}\right)^{n+\HALF}\,,
\end{equation}
where $\vec{x}_p$ and $\vec{u}_p = \gamma_p\vec{v}_p$ are, respectively, the spatial coordinate and four-velocity of the $p-$th particle.
Here $\vec{a}_p\equiv\vec{a}_p(\vec{x}_p,\vec{u}_p,U,\vec{F}_\CR)$ is a compact expression for the Lorentz acceleration, given by the second Equation in (\ref{eq:CR}), showing its dependence on both particles and fluid quantities.
After CR have been evolved for a full time step, the total momentum and energy change of a single particle can be computed as 
\begin{equation}\label{eq:delta_mE}
  \begin{array}{lcl}
  \Delta\vec{m}_p &=& \varrho_p (\vec{u}^{n+1}_p - \vec{u}^n_p)    \\ \noalign{\medskip}
  \Delta E_{k,p}    &=& \varrho_p (E^{n+1}_{k,p} - E^n_{k,p})  \,,
  \end{array}
\end{equation}
where $E_{k,p} = (\gamma_p - 1)\C^2$ is the (specific) kinetic energy of a single particle.
We then deposit the opposite of these quantities on the grid, allowing momentum and energy feedback to be computed from the particles location at the half-step:
\begin{equation}\label{eq:CTU_feedback}
  S^{n+\HALF}_{\CR,\indx} = 
  -\sum_p \frac{W(\vec{x}_\indx - \vec{x}_p^{n+\HALF})}{\Delta t^n}
  \left(\begin{array}{c}
   0              \\ \noalign{\medskip}
  \Delta\vec{m}_p \\ \noalign{\medskip}
   \vec{0}        \\ \noalign{\medskip}
  \Delta E_{k,p} 
  \end{array}\right) \,.
\end{equation}
As pointed out by \cite{Bai_etal.2015}, this ensures exact conservation of total momentum and energy of the composite gas+CR system.

In the corrector step fluid quantities are finally evolved for a full step, 
\begin{equation}\label{eq:CTU_corr}
  \begin{array}{lcl}
  U^{n+1}_\indx = U^n_\indx
                & + & \Delta t^n{\cal L}_\indx(U^{n+\HALF},\vec{F}^{n+\HALF}_\CR)
  \\ \noalign{\medskip}
                & + & \Delta t^n S_{\CR,\indx}^{n+\HALF}  \,,
  \end{array}
\end{equation}
where $\vec{F}^{n+\HALF}_\CR$ is given by the opposite of the momentum component of the source term (\ref{eq:CTU_feedback}).
This completes our derivation of the CTU scheme (more detailed can be found in the Appendix \ref{app:CTU-CT}).

\subsubsection{Runge-Kutta Time Stepping.}
%

Runge-Kutta (RK) time stepping methods are based on the method of lines in which the spatial discretization is considered separately from the temporal evolution that is left continuous in time.
Equations (\ref{eq:MHD-PIC:rho})-(\ref{eq:MHD-PIC:E}) are then discretized as regular ordinary differential equations based on predictor-corrector steps.

We consider the second-order RK method (RK2) which consists of a first predictor step, in which the fluid is advanced by a \emph{full} step:
\begin{equation}\label{eq:RK2_pred}
  U^{*}_\indx = U^n_\indx + \Delta t^n{\cal L}_\indx(U^n, \vec{F}^n_\CR)
                          + \Delta t^nS_{\CR,\indx}^n \,.
\end{equation}

Particles are then evolved using Equation (\ref{eq:CTU_particle_advance}), where the half-time level fluid variables are computed from the arithmetic average of conservative variables at level $n$ and the predicted ones,
\begin{equation}
  U^{n+\HALF}_\indx = \frac{U^n_\indx + U^*_\indx}{2}\,.
\end{equation}

The final corrector step employs a trapezoidal rule for the flux terms and a midpoint rule for the sources:
\begin{equation}\label{eq:RK2-v1}
  \begin{array}{lcl}
  U^{n+1}_\indx = U^n_\indx &+&\DS
                   \Delta t^n\frac{  {\cal L}_\indx(U^n,\vec{F}^n_\CR)
                                   + {\cal L}_\indx(U^*,\vec{F}^*_\CR)}{2}
  \\ \noalign{\medskip}     &+&\DS
                           \Delta t^nS^{n+\HALF}_{\CR,\indx} \,.
  \end{array}
\end{equation}
In the previous equation, $\vec{F}^*_\CR = 2\vec{F}^{n+\HALF}_\CR - \vec{F}^n_\CR$ is obtained by simple extrapolation while $\vec{F}^{n+\HALF}_\CR$ is computed using the opposite of the momentum component in Equation (\ref{eq:CTU_feedback}).
Momentum and energy feedback at the half-time level are accounted for by $S^{n+\HALF}_{\CR,\indx}$ and computed as for the CTU scheme using Equation (\ref{eq:CTU_feedback}).

For implementation purposes it is more convenient to rewrite Equation (\ref{eq:RK2-v1}) using (\ref{eq:RK2_pred}) as
\begin{equation}\label{eq:RK2-v2}
  U^{n+1}_\indx = \frac{U^n_\indx + U^*_\indx}{2}
                + \Delta t\frac{{\cal L}_\indx(U^*,\vec{F}^*_\CR)
                + S^*_{\CR,\indx}}{2}
\end{equation}
where $S^*_{\CR,\indx} = 2S^{n+\HALF}_{\CR,\indx} - S^n_{\CR,\indx}$.

\subsection{Particle Mover}
\label{sec:particle_mover}
%
%

Particle's position and velocity are assumed to be known at the same time level $n$ rather than being staggered in time.
This allows the code to employ variable time step as it is typically the case in fluid simulations.
Equation (\ref{eq:CR}) is solved by means of a standard Boris pusher which essentially an implicit position-Verlet algorithm, cast as
\begin{equation}\label{eq:Boris}
  \begin{array}{lcll}
    \DS \vec{x}_p^{n+\HALF} &=&\DS\vec{x}_p^n + \frac{\Delta t^n}{2}\vec{v}_p^n
     & \;\mathrm{(drift)}
    \\ \noalign{\medskip}
    \DS \vec{u}^{-}_p       &=&\DS\vec{u}^n_p + \frac{h}{2}c\vec{\E}^{n+\HALF}
     & \;\mathrm{(kick)}
    \\ \noalign{\medskip}
    \DS \vec{u}^{+}_p       &=&\DS\vec{u}^-_p
        + 2\frac{\vec{u}_p^- + \vec{u}_p^-\times\vec{b}}{1+\vec{b}^2}\times\vec{b}
     & \;\mathrm{(rotate)}
    \\ \noalign{\medskip}
    \DS \vec{u}^{n+1} &=& \DS\vec{u}_p^+ + \frac{h}{2}c\vec{\E}^{n+\HALF}  
     & \;\mathrm{(kick)}
    \\ \noalign{\medskip}
    \DS \vec{x}_p^{n+1} &=& \DS   \vec{x}_p^{n+\HALF}
                                + \frac{\Delta t^n}{2}\vec{v}_p^{n+1}  
     & \;\mathrm{(drift)}
  \end{array}  
\end{equation}
where $\vec{u}_p = \gamma_p\vec{v}_p$ is the particle four-velocity, $h = \alpha_p\Delta t^n$ while $\vec{b} = (h/2)\vec{B}^{n+\HALF}/\gamma^{n+\HALF}$.
Electromagnetic fields are interpolated at the particle half-step position $\vec{x}^{n+\HALF}$ using Equation (\ref{eq:interpolation}).
Since interpolation at the particle position does not necessarily preserve the orthogonality between $\vec{\E}$ and $\vec{B}$ (when the electric field is obtained from Equation \ref{eq:Ohm} or \ref{eq:CR_emf}), a cleaning step is required to remove non-orthogonal components from the electric field:
\begin{equation}\label{eq:CR_cleaning}
 \vec{\E}\, \leftarrow\, \vec{\E} - (\vec{\E}\cdot\vec{B})\frac{\vec{B}}{B^2}  
\end{equation}
Note that the rotation does not change the particle energy and therefore $\gamma^{n+\HALF}$ is obtained directly from $\vec{u}^-_p$.

Time step restriction is computed by requiring that no particle travels more than $N_{\rm max}$ zones and that the Larmor scale is resolved with more than 1 cycle:
\begin{equation}\label{eq:particles_dt}
  \Delta t^{-1}_{p} = \max_p\left[
    \max_d\left(
        \frac{|\hvec{e}_d\cdot\vec{v}_{p}^{n+\HALF}|}{N_{\rm max}\Delta x_d}
        \right), \,
     \frac{\Omega_{\perp,p}}{\epsilon_L} \right]
\end{equation}
where the first maximum extends to all particles, $\vec{v}_p^{n+\HALF}$ is the half-time level averaged velocity, $\Omega_{\perp,p} = \alpha_pB_\perp/\gamma_p$ is the Larmor frequency while
\begin{equation}
   B_\perp = \sqrt{\vec{B}^2 - \frac{(\vec{v}_p\cdot\vec{B})^2}
                                    {\vec{v}_p\cdot\vec{v}_p}}
\end{equation}
is the transverse component of magnetic field.
In Equation (\ref{eq:particles_dt}), we choose $N_{\rm max} = 1.8$ and $\epsilon_L= 0.3$ as safety factors.

\subsubsection{CR Predictor Step}
\label{sec:CR_predictor}
%
%
We note that the particle mover given by Equation (\ref{eq:Boris}) requires knowledge of the electric field at the half time step.
In the case of test particles, this is not a problem, since the electric field depends solely on the fluid and can be easily be determined.

However, in the MHD-PIC system, the electric field (see Equation \ref{eq:Ohm}), is comprised of the convective and CR Hall terms, but only the former is known at the half time level $t^n+\Delta t^n/2$ while the latter can only be computed at the base time level $t^n$.
Formally, therefore, we expect the integration scheme to be only first-order accurate in time.
We point out that, for the conditions under which the MHD-PIC formalism is valid ($R\ll 1$), the CR Hall term is generally unimportant but it may become comparable to the convective term for large CR streaming velocities.

In order to achieve full second order accuracy, we propose a predictor step where particles are evolved for a half time increment using a first order explicit-implicit scheme,
\begin{equation}\label{eq:CR_predictor_eq}
  \vec{u}_p^{*,n} = \vec{u}_p^n + \frac{h}{2}\left(
        c\vec{\E}^{n}  
      +  \frac{\vec{u}_p^{*,n}}{\gamma_p^{*,n}}\times\vec{B}^{*,n}\right)
\end{equation}
where the notation $\vec{u}_p^{*,n}$ stands for the half-time level predicted value while $h=\alpha_p \Delta t$.
Since the $\vec{v}\times\vec{B}$ term does not alter the velocity magnitude, we compute the Lorentz factor by first applying a kick to the particle velocity, $\vec{u}_p^{*,-} =\vec{u}_p^n + (h/2)c\vec{\E}^{n}$ and then compute
\begin{equation}
  \gamma^{*,n} = \sqrt{1 + \left(\frac{\vec{u}^{*,-}}{\C}\right)^2} \,.
\end{equation} 
Equation (\ref{eq:CR_predictor_eq}) can then be solved to obtain
\begin{equation}\label{eq:CR_predictor}
  \vec{u}_p^{*,n} = \left(\tens{I} - \tens{M}^{*,n}\right)^{-1}\vec{u}_p^{*,-}
\end{equation}
where 
\begin{equation}
  \left(\tens{I} - \tens{M}^{*,n}\right)^{-1}_{ij}
    =
  \frac{\delta_{ij} + b_ib_j - \tens{M}^{*,n}_{ij}}{1 + \vec{b}^2}  \,.
\end{equation}
with $\tens{M} = \tens{I}\times\vec{b}$ and $\vec{b} = (h/2)\vec{B}^{*,n}/\gamma^{*,n}$.
The value of the magnetic field is interpolated at the particle position using the half-time step magnetic field $\vec{B}^{n+\HALF}$ already available from the MHD integrator.

After the predictor step, the full electric field can be evaluated using Equation (\ref{eq:CR_emf}) and particle positions and velocities can be restored to their initial values $\vec{x}_p^n$, $\vec{v}_p^n$.
Note that, since only an approximate value of the solution is needed, we do not apply a cleaning step to make $\vec{\E}$ and $\vec{B}$ orthogonal during the predictor step.
We point out that the predictor step is only used to predict the half-time level approximation to the electromotive force but not for the actual evolution of the CR particles, which are advanced according to the Boris pusher, Equation (\ref{eq:Boris}).

\subsubsection{Particle Sub-Cycling}
\label{sec:subcycling}
%
%

\begin{figure}[!ht]
  \centering
  \includegraphics[width=0.5\textwidth]{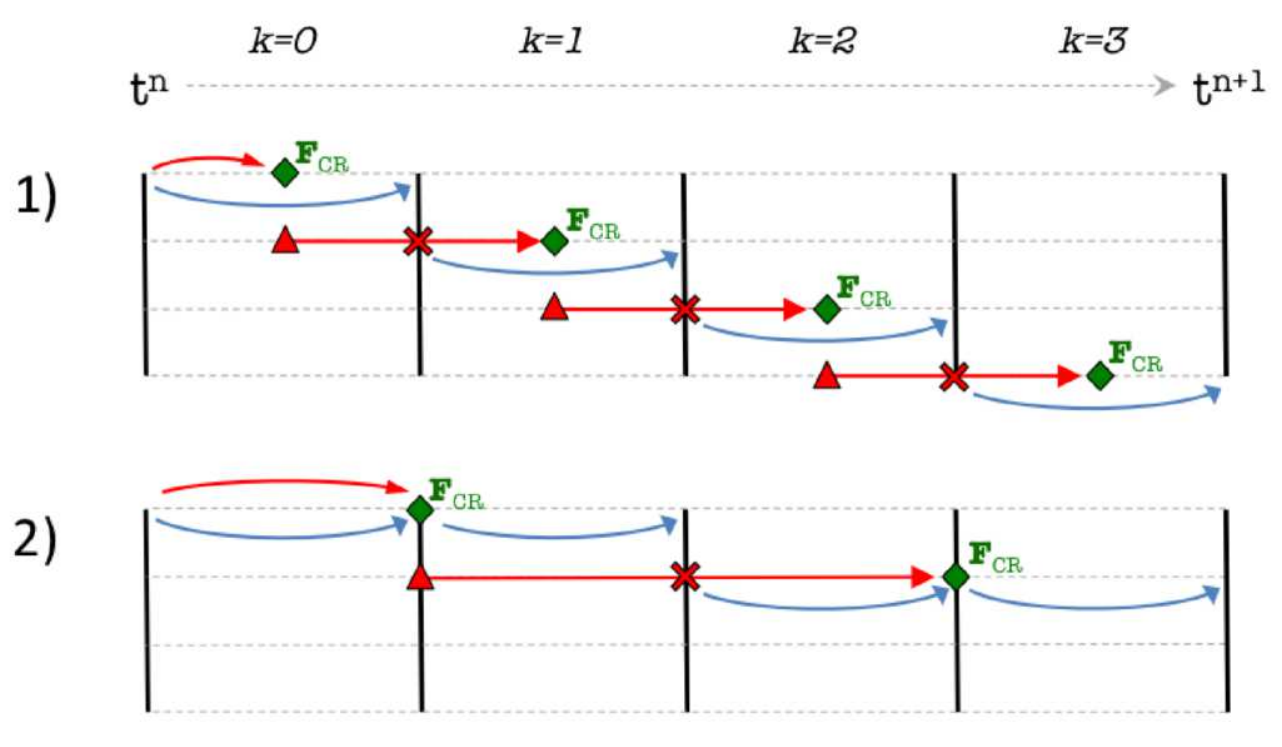}
  \caption{\footnotesize Subcycling methods 1 (top) and 2 (bottom) for
  $N_{\rm sub} = 4$.
  Blue arrows represent application of the Boris pusher; red triangles
  (momentum variation) and crosses (direct computation of the CR force)
  joint by a red line are used to extrapolate $\vec{F}_\CR$ at the next half
  sub-step (green diamonds, method 1) or full sub-step (green diamonds,
  method 2).
  The curved red line at the beginning of the cycle represents
  application of the predictor step.
  }
  \label{fig:subcycling}
\end{figure}

At large energies, the particle evolution time-scale can become considerably shorter than the fluid dynamical time, slowing down the total computational time.
To overcome this issue, we allow multiple particle time steps to be taken during a single fluid update.
Our approach improves over that of \cite{Bai_etal.2015} in several aspects.

Let $N_{\rm sub}$ be the number of steps involved during the sub-cycling.
The particle pusher Equation (\ref{eq:Boris}) is now applied $N_{\rm sub}$ times over equally spaced time intervals $[t^{n+\theta k},\,  t^{n+\theta(k+1)}]$ of length $\theta\Delta t$ where $\theta = 1/N_{\rm sub}$ and $k=0,...,N_{\rm sub}-1$.
During subcycling, electric and magnetic fields are kept constant to the predicted half-time level with the exception of the CR-Hall term (the second term in Equation \ref{eq:CR_emf}) which is recomputed in order to maintain second order accuracy.

We elaborated two forms of subcycling taking advantage of momentum and energy deposition (needed for feedback) which are accumulated at each fluid step.

\begin{enumerate}
  \item The first strategy recomputes the force at each sub-step and can
  be used with an even or odd number of steps. 
  After solution values have evolved to the intermediate level
  $(\vec{x}_p,\vec{v}_p)^{n+\theta k}$, we recompute the CR force using Equation
  (\ref{eq:Fcr2}), correct the electric field using Equation (\ref{eq:CR_emf})
  and, for $k>0$, predict the midpoint force value for the next sub-step
  using time extrapolation:
  \begin{equation}
     \vec{F}_\CR^{n+\theta(k+\HALF)} = 2\vec{F}_\CR^{n+\theta k}
    - \left(\frac{\Delta\vec{m}}{\theta\Delta t}\right)^{n+\theta(k-\HALF)}
  \end{equation}
  where $\Delta\vec{m}^{n+\theta(k-\HALF)}$ is momentum difference over the previous
  sub-cycle.
  At the beginning of the cycle ($k=0$), we employ the predictor step
  given by Equation (\ref{eq:CR_predictor}) with $\theta\Delta t/2$.
  This method is represented in the top panel of Figure \ref{fig:subcycling}.

  \item The second strategy recomputes the electric field every other
  sub-step thus leading to a more efficient scheme that can be used
  with an even number of sub-steps.
  The electric field is extrapolated in time (when $k=2,4,6,...$) by a
  full sub-step by taking advantage of the total momentum variation
  accumulated until then:
  \begin{equation}
     \vec{F}_\CR^{n+\theta(k+1)}
   =   \frac{k + 2}{k}\vec{F}_\CR^{n+\theta k}
     - \frac{2}{k}\left(\frac{1}{k\theta\Delta t}\sum_{j=1}^{k}
       \Delta\vec{m}^{n+\theta(j-\HALF)}\right)
  \end{equation}
  where the summation represents the total momentum change accumulated over
  $k$ substeps.
  Note that, when $k$ is odd, we do not recompute $\vec{F}_\CR$.

  At the beginning of the cycle ($k=0$), we employ the predictor step
  given by Equation (\ref{eq:CR_predictor}) with time step $\theta\Delta t$.
  This method is represented in the bottom panel of Figure \ref{fig:subcycling}
  in the case $N_{\rm sub} = 4$.

\end{enumerate}

\subsection{Connection between Grid and Particle Quantities}
\label{sec:mapping}
%
%
%

An important step of the algorithm requires depositing particle quantities to the grid and interpolating fluid quantities at the particle locations.

Let $q_p$ be a quantity associated with a particle (e.g. charge or velocity), then  deposition in cell $(i,j,k)$ is achieved by a weighted sum 
\begin{equation}\label{eq:deposition}
  Q_{ijk} = \sum_{p=1}^{N_p} W(\vec{x}_{\indx}-\vec{x}_p)q_p
\end{equation}
where $W(\vec{x}_{\indx}-\vec{x}_p) = W(x_i - x_p)W(y_j - y_p)W(z_k - z_p)$ is the product of three one-dimensional weight functions.
Within PLUTO, we have implemented traditional shape functions such as `Nearest Neighbour Point' (NGP), `Cloud-In-Cell' (CIC) and `Triangular Shape Cloud' (TSC).
Explicit formula for the weight can be found, e.g., in \cite{Haugbolle_etal.2013}.
In practice, since the weight functions have a finite stencil that extends over 3 zones, each particle can give a non-zero contribution only to the computational zone hosting the particle, its left and right neighbours.
If $\delta = (x_p-x_i)/\Delta x$ is the distance between the particle and the $i-$th zone such that $\delta \in[-1/2,1/2]$, the corresponding weights $W_i$, $W_{i-1}$ and $W_{i+1}$ are computed as
\begin{itemize}
  \item Nearest grid point ({\bf NGP}):
  \begin{equation}
    W_{i\pm1} = 0;\quad
    W_{i} = 1;\quad
  \end{equation}

  \item Cloud in cell ({\bf CIC}):
  \begin{equation}
    W_{i\pm1} = \frac{|\delta| \pm \delta}{2};\quad
    W_{i} = 1-|\delta|;
  \end{equation}

  \item Triangular Shape Cloud ({\bf TSC}):
  \begin{equation}
    W_{i\pm1} = \frac{1}{2}\left(\frac{1}{2}\pm\delta\right)^2 \,;\quad
    W_{i} = \frac{3}{4} - \delta^2       \,;\quad
  \end{equation}
\end{itemize}
Note that $W_{i-1}+W_{i}+W_{i+1} = 1$ when $\delta \in [-1/2,1/2]$.

Particle interpolation (also referred to as \emph{field weighting}) is the opposite process of interpolating grid (fluid) quantities at a given particle position:
\begin{equation}\label{eq:interpolation}
  q_p = \sum_{ijk}W(\vec{x}_{ijk}-\vec{x}_p)Q_{ijk}
\end{equation}
where only neighbour cells give a non-zero contribution to the particle.
For consistency the same weighting scheme must be used for particles and field \citep[see][]{Birsdall_and_Langdon.2004}.

\section{Numerical Benchmarks and Code Performance}
\label{sec:tests}
%
%
%
%

In this section we present selected numerical benchmarks in order to verify the correctness and accuracy of our MHD-PIC and test-particle model implementations.

Before proceeding we point out that while the ideal MHD equations are notoriously scale-invariant, the presence of a non-zero source terms on the right hand side of the momentum and energy Equations (\ref{eq:MHD-PIC:rhov}) and (\ref{eq:MHD-PIC:E}) breaks down this property.
If we denote with $L_0$, $\rho_0$ and $V_0$ our physical reference units for lengths, density and velocity (respectively), a straightforward analysis shows that the source terms (and similarly the particle equation of motion \ref{eq:CR}) are rescaled by a factor $L_0\omega_{pi}/c$ where $\omega_{pi}$ is the ion plasma frequency.
This naturally suggests the ion skin depth $c/\omega_{pi}$ as the natural reference length.
In addition, if the Alfv{\'e}n velocity $v_A$ is used as the reference speed, time will be conveniently expressed in units of the inverse Larmor frequency $\Omega^{-1}_L = c/(\omega_{pi}v_A)$.

\subsection{Particle Gyration}
\label{sec:gyration}
%

\begin{figure*}
  \centering
    \includegraphics[width=0.9\textwidth]{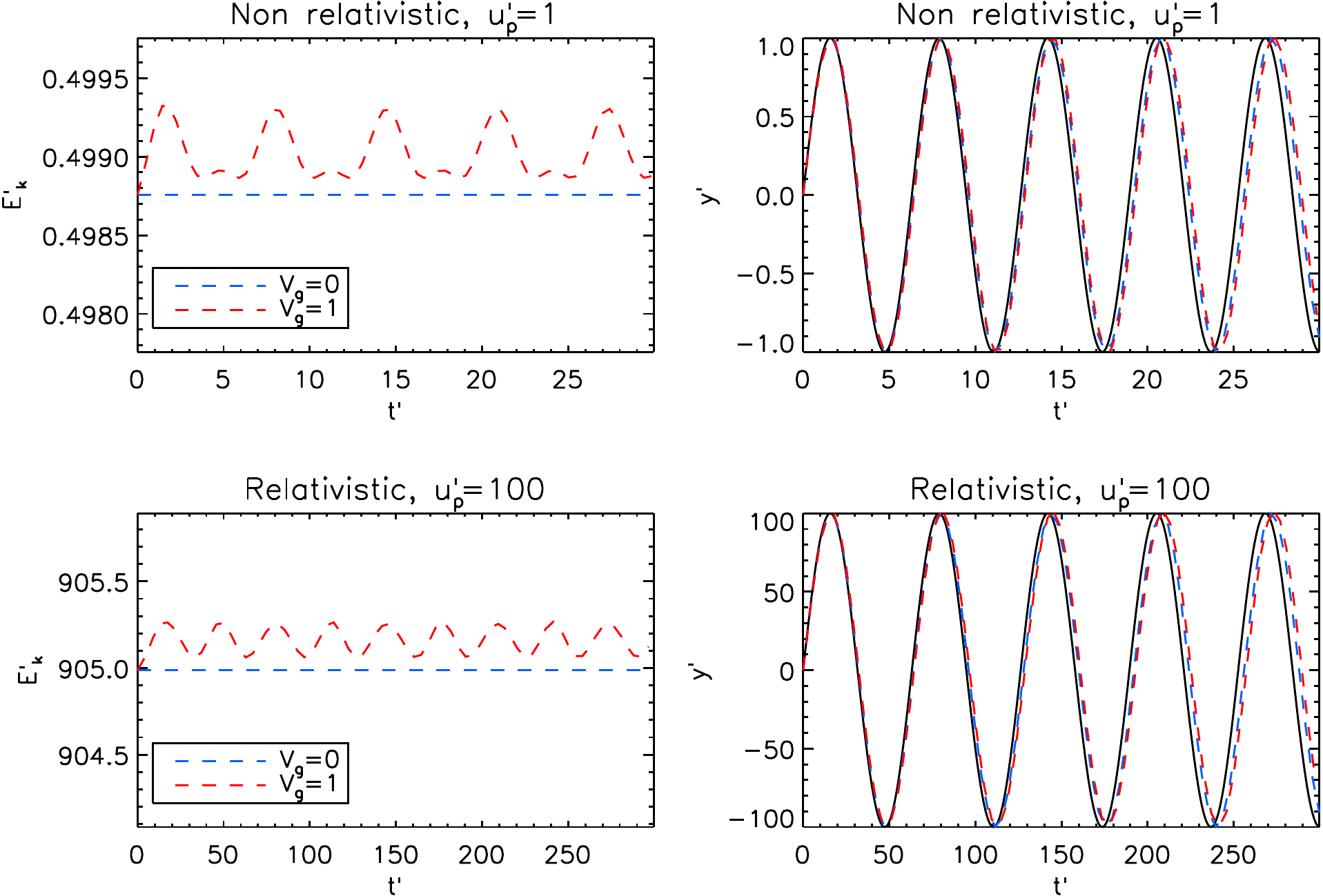}
  \caption{\footnotesize Particle kinetic energy (left panels) and position
          (right panels) as a function of time for the gyration test problem.
           Quantities are plotted in the fluid rest-frame.
           The top and bottom panels show, respectively, the results for a
           non-relativistic ($u'_p=1$) and relativistic ($u'_p=100$) test
           particle.
           Blue and red dashed lines corresponds, respectively, to zero
           background flow velocity ($V_g=0$) and mildly relativistic flow
           ($V_g = 1$).
           \label{fig:gyration}}
\end{figure*}

We begin by considering the gyration of a single test particle in a constant magnetic field directed along the vertical axis, $\vec{B} = (0,0,B_0)$.
Fluid backreaction is not included as in \cite{Bai_etal.2015}.
We solve the MHD-PIC equations in a reference frame $\Sigma$ where the background fluid has constant density and pressure as it is uniformly advected in the $x$ direction with velocity $\vec{v}_g=(V_g,0,0)$.

The motion of the particle is more conveniently described in the fluid comoving frame $\Sigma'$, where the inductive electric field vanishes and the particle equation of motion reduces to
\begin{equation}\label{eq:gyration}
  \frac{d(\gamma'_p \vec{v}'_p)}{dt'} = \alpha_p \vec{v}'_p\times\vec{B}' \,
\end{equation}
where primed quantities are now in the fluid rest frame.
The general solution of Equation (\ref{eq:gyration}), for a point charge located at the origin of $\Sigma'$, is a simple gyration:
\begin{equation}
  \left\{\begin{array}{lcl}
    x'_p(t) & = & \DS \frac{  v'^{\,0}_{p,x}\sin(\Omega' t')
                            + v'^{\,0}_{p,y}[1 - \cos(\Omega' t')]}{\Omega'}
    \\ \noalign{\medskip}
    y'_p(t) & = & \DS \frac{  v'^{\,0}_{p,x}[\cos(\Omega' t')-1]
                            + v'^{\,0}_{p,y}\sin(\Omega' t')}{\Omega'}
    \\ \noalign{\medskip}
    v'_{x,p}(t) & = & \DS   v'^{\,0}_{p,x}\cos(\Omega' t')
                          + v'^{\,0}_{p,y}\sin(\Omega' t')
    \\ \noalign{\medskip}
    v'_{y,p}(t) & = & \DS - v'^{\,0}_{p,x}\sin(\Omega' t')
                          + v'^{\,0}_{p,y}\cos(\Omega' t')
  \end{array}\right.
\end{equation}
where $v'^{\,0}_{x,p}$ and $v'^{\,0}_{y,p}$ are the Cartesian components of the particle initial velocity $\vec{v}'_p$ while 
\begin{equation}
  \Omega'_L = \frac{\alpha_pB'_0}{\gamma'_p}\,,\qquad
   B'_0     = B_0\sqrt{1 - \left(\frac{V_g}{\C}\right)^2}
\end{equation}
are the Larmor gyrofrequency and magnetic field in the $\Sigma'$ frame.
The gyration radius is $R'_L = v'_p/\Omega'$ and the particle kinetic energy must be conserved in this frame, i.e.
\begin{equation}\label{eq:E'_p}
  E'_{k,p} = (\gamma'_p - 1)\C^2 = const\,.
\end{equation}

For the present test, we prescribe $\vec{v}'_p = (0,\,u'_p/\gamma'_p,\,0)$ where $u'_p$ is the particle four velocity, $\gamma'_p=\sqrt{1+(u'_p/\C)^2}$ its Lorentz factor and we set $\C = 10$, $\alpha_p = B_0 = 1$.
Velocity components in the lab frame are easily found through a Lorentz transformation.
Following \cite{Bai_etal.2015} we consider both non-relativistic ($u'_p=1$) and relativistic ($u'_p=100$) test particles, with or without drift velocity.

In order to mimic the variable time step generally expected in fluid simulations, we set the time step to be $\Delta t = \Delta t_0(1 + 0.2\cos\varphi)$, where $\varphi$ is a random number in the range $[0,2\pi]$ and $\Delta t_0 = 0.5$ (non-relativistic particle) or $\Delta t_0 = 5$ (relativistic particle).
With this choice $\Omega'_L\Delta t \approx 0.5$ in both cases.
Particle sub-cycling is not employed.
As pointed out in \cite{Bai_etal.2015}, a relatively large time step has been chosen to amplify the error.

\paragraph{Non Relativistic Particle.}
In the top panels of Figure \ref{fig:gyration} we plot, in the co-moving frame, the energy (left) and $y'$ coordinate (right) as a function of time for a non-relativistic particle with $u'_p = 1$ and $V_g = 0$ (no drift, blue dashed line) or $V_g=1$ (drift, red dashed line).
The particle initial energy in the comoving frame is therefore $E'_{k,p} \approx 0.4988$ while its gyration radius is $R'_L=1$.
Energy is conserved exactly in absence of drift, while it shows small-amplitude oscillations corresponding to a relative error $\approx 0.1 \%$ when $V_g=1$.
Phase errors are also within an acceptable level and results are in good agreement with \cite{Bai_etal.2015}.

\paragraph{Relativistic Particle.}
In the bottom panels of Figure \ref{fig:gyration} we show the evolution of a test particle with initial velocity $u'_p = 100$ ($\gamma'_p\approx 10$).
In this case, $\Omega'\approx 0.1$ and we set $\Delta t_0 = 5$ so that $\Delta t_0\Omega'_L = 0.5$ as in the previous case.
From Equation (\ref{eq:E'_p}) we have $E'_{k,p}\approx 904.99$ while the gyration radius is $R'_L=100$.
Exact energy conservation is achieved during the evolution in absence of drift, while small-amplitude oscillations are present when $V_g=1$.
The relative errors are $\lesssim 0.05\%$, slightly less than before.
Phase errors are comparable to the non-relativistic case.

Our results well agree with those of \cite{Bai_etal.2015}.
We conclude that, since in most astrophysical applications $\Omega_L\Delta t < 0.5$, we can safely depend on our implementation of the Boris algorithm.

\subsection{Particle Motion in non-Orthogonal Electric and Magnetic Fields}
\label{sec:parallel_EB}
%

\begin{figure*}
  \centering
  \includegraphics[width=0.9\textwidth]{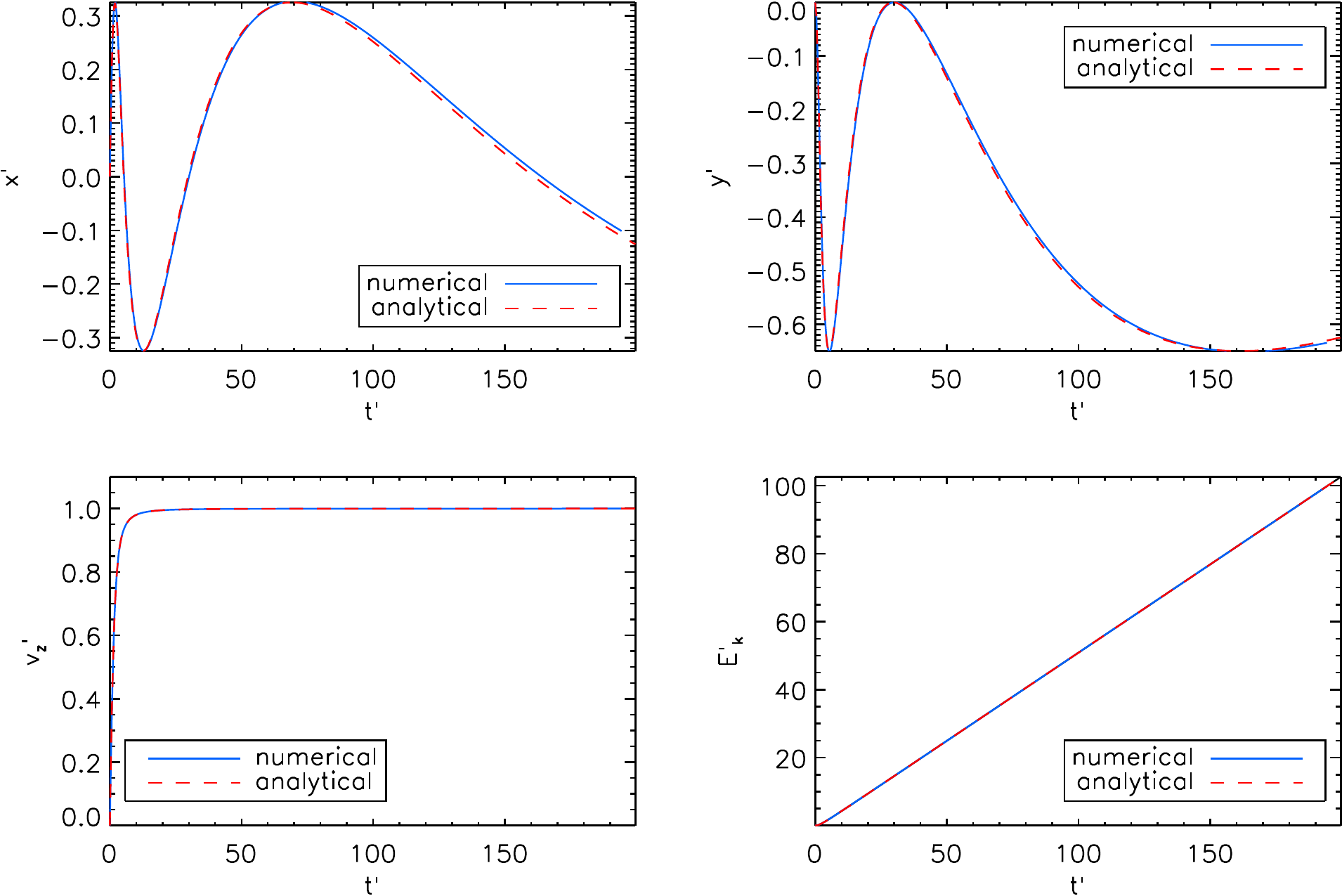}\hfill
  \caption{\footnotesize Particle position (top panels), velocity
   (bottom left panel) and energy density (bottom right panel) as a function
   of time for a charged particle in generic electric and magnetic fields.
   Quantities are plotted in the frame where both electric and magnetic field are
   along $z$-axis. \label{fig:parallel_EB}}
\end{figure*}
Next, we consider the motion of a relativistic charged particle in non-orthogonal electric and magnetic fields.

The initial condition consists of a spatially uniform plasma with constant density and pressure with $\vec{\E} = (0,\, \E_{y},\, \E_{z})$, $\vec{B} = (0,\,0,\, B_{z})$.
The fluid is assumed to be at rest.
A single test particle is initially placed at the origin, with velocity along the $x$ axis, i.e., $\vec{v}_p(0) = v_0\hvec{e}_x$.

This problem has an exact solution which is best expressed in a frame of reference where the electric and magnetic field are parallel \citep{LL.1975}, through a Lorentz boost with velocity
\begin{equation}
 \frac{\vec{V}}{c} = \frac{   \vec{E}^{2} + \vec{B}^{2}
                            - \DS\sqrt{\left(\vec{E}^{2} - \vec{B}^{2}\right)^{2}
                           + 4(\vec{\E}\cdot\vec{B})^{2}}}
                          {2|\vec{\E}\times\vec{B}|^{2}}(\vec{\E}\times\vec{B}).
\end{equation}
In this frame, the electric and magnetic fields become parallel with components $\tilde{\vec{\E}} = (0,\, \tilde{\E}_y,\, \tilde{\E}_z)$ and $\tilde{\vec{B}} = (0,\, \tilde{B}_y,\, \tilde{B}_z)$.
A rotation is then performed in order to place the electric (and thus the magnetic) field along $z$-axis:
\begin{equation}
  \left(\begin{array}{l}
   \E_{y}' \\ \noalign{\medskip}
   \E_{z}'
  \end{array}\right)
   = 
  \left(\begin{array}{ll}
    \cos\theta & \sin\theta \\ \noalign{\medskip}
   -\sin\theta & \cos\theta
  \end{array}\right)
  \left(\begin{array}{l}
  \tilde{\E}_{y} \\ \noalign{\medskip}
  \tilde{\E}_{z}
 \end{array}\right)
\end{equation}
where $\theta = \arctan\left(-\tilde{\E}_{y}/\tilde{\E}_{z}\right)$.
In the $\Sigma'$ frame, it is straightforward to show that the particle initial velocity is still directed along the $x'$ direction.
The solution of Equation (\ref{eq:CR}) for a charged particle located at the origin of $\Sigma'$ \citep{LL.1975} can be written in terms of perpendicular components
\begin{equation}\label{eq:parEB_perp}
  \begin{array}{lcl}
  \vec{x}'_{p,\perp}(t') &=& \DS \frac{v'_0}{\Omega'_L}
              \Big[\sin\phi(t'),\, \cos\phi(t') - 1,\, 0\Big]
  \\ \noalign{\medskip}
  \vec{v}'_{p,\perp}(t') &=& \DS \frac{v'_0}{\Omega'_L}
              \Big[\cos\phi(t'),\, -\sin\phi(t'),\, 0\Big]\frac{d\phi}{dt'}
  \end{array}
\end{equation}
and parallel components
\begin{equation}\label{eq:parEB_par}
  \left\{\begin{array}{lcl}
   z'_p(t') &=& \DS c\tau_\E\cosh\left(\frac{\E'}{B'}\phi(t')\right)
   \\ \noalign{\medskip}
   v'_{z,p}(t') &=& \DS  \frac{c}{\Omega'_L}
                         \sinh\left(\DS\frac{\E'}{B'}\phi(t')\right)\,
                   \frac{d\phi}{dt'}
  \end{array}\right. \,.
\end{equation}
In the above expressions, $\E' \equiv \E'_z$ while $\Omega'_L$ and $\tau_\E$ are the gyrofrequency and the acceleration time-scale in the $\Sigma'$ frame defined, respectively, as
\begin{equation}
  \Omega'_L = \frac{\alpha_pB'}{\gamma_0'},\qquad
  \tau_\E  = \frac{1}{\alpha_p\E'} \,.
\end{equation}

Finally, $\phi(t')$ is given by 
\begin{equation}\label{eq:parallel_EB_phi}
  \phi(t') = \frac{B'}{\E'}\arcsinh\left(\frac{c^2}{E'_{k0}} \frac{t'}{\tau_\E}\right)\,.
\end{equation}
where $E'_{k0} = (\gamma'_0 - 1)c^2$ represents the initial particle kinetic energy (per unit mass).
Notice that our solution has been derived under the assumption that the particle initial velocity lies in the $x$ direction only.

Equations (\ref{eq:parEB_perp}) and (\ref{eq:parEB_par}) describe a stretched helical trajectory with exponentially increasing pitch.
Note that $t'$ is the time coordinate in the $\Sigma'$ frame.

For the present test we prescribe $\vec{v}_p(0) = (0.5,0,0)$, $\vec{\E} = (0,0.3,0.5)$, $\vec{B} = (0,0,1)$ and set the charge to mass ratio as well as the artificial speed of light $\C$ equal to 1.
We integrate the particle equation of motion until $t_s = 200$ and, as in the previous test, we set the time step to be $\Delta t = \Delta t_{0}(1 + 0.2 \cos\varphi)$, where $\varphi$ is a random number between $0$ and $2\pi$, and $\Delta t_{0} = 0.5$.
Particle sub-cycling is not employed.

In the top panels of Figure \ref{fig:parallel_EB} we plot, in the $\Sigma'$ frame, the $x$-coordinate (left panel) and the $y$-coordinate (right panel) as a function of time for a relativistic particle in a generic electromagnetic field.
Likewise we plot, in the bottom panels, the $z$-component of the velocity (left panel) and the particle energy (right panel).
The relative error is computed by transforming the energy in the $\Sigma'$ frame and then taking the maximum value over time:
\begin{equation}
  \Delta_{L1} = \max\left[\frac{\left|E'_{k,p}(t'^n) - E'^{\rm ex}_{k,p}(t'^n)\right|}
                               {E'^{\rm ex}_{k,p}(t'^n)}\right]
\end{equation}
where $E'_{k,p} = (\gamma' - 1)\C^2$ is the particle (specific) kinetic energy while
\begin{equation}
  E'^{\rm ex}_{k,p}(t') = \C^2\sqrt{(\gamma'_0)^2 + \left(\frac{t'}{\tau_\E}\right)^2} - \C^2
\end{equation}
is the exact expression for the particle energy as a function of time.
We obtain $\Delta_{L1} \lesssim 0.1\%$, showing a good agreement between the analytic and the numerical solution.

\subsection{Fluid-Particle Relative Drift}
\label{sec:relative_drift}
%

We now assess the temporal accuracy of our integration schemes by considering the evolution of the full gas-CR system starting from a spatially uniform distribution of gas and particles.
The computational domain is a doubly periodic $2$-D box defined by $x,y \in [-1,1]$ with constant magnetic field and orthogonal to the plane of computation, $\vec{B} = (0,\,0,\,B_0)$.
We choose a frame of reference where the total (gas+CR) momentum is zero so that, at $t=0$, gas and particles stream in opposite directions, 
\begin{equation}
  \vec{v}_g(0) = -\frac{\varrho_p}{\rho}v_0\hvec{e}_x,\quad
  \vec{v}_p(0) = v_0\hvec{e}_x
\end{equation}
Density and pressure of the fluid are set to unity.
The evolution of the composite (gas + particles) system, which now includes CR feedback, is governed by the MHD-PIC equations (\ref{eq:MHD-PIC:rho})--(\ref{eq:MHD-PIC:E}) which, in absence of spatial gradients, reduce to 
\begin{equation}\label{eq:rd}
  \left\{\begin{array}{lcl}
    \DS \frac{d\vec{v}_g}{dt} &=& \DS 
          \alpha_i R\left(\vec{v}_g - \vec{v}_p\right)\times\vec{B}
   \\ \noalign{\medskip}
    \DS \frac{d\vec{v}_p}{dt} &=& \DS
          \alpha_p (1-R)\left[ (\vec{v}_p-\vec{v}_g)\times\vec{B}\right]
  \end{array}\right.
\end{equation}
where $\alpha_i$ and $\alpha_p$ denotes, as usual, the charge to mass density ratios of the ions and the CR particles, respectively.
Note that, in writing Equation (\ref{eq:rd}), we have tacitly assumed that $\vec{v}_p \equiv \vec{v}_\CR$ in absence of spatial dependence.
Also, the expression for the CR force appearing on the right hand side of Equation (\ref{eq:MHD-PIC:rhov}) has been rewritten by combining Equation (\ref{eq:CR_emf}) with $\vec{\E}$ given by Equation (\ref{eq:Ohm}), yielding
\begin{equation}\label{eq:Fcr3}
  \vec{F}_\CR = \frac{q_i}{c}R\left(\vec{v}_\CR - \vec{v}_g\right) \times\vec{B} \,.
\end{equation}

The system of ordinary differential equations (\ref{eq:rd}) with the initial conditions previously specified has an exact analytic solution given by:
\begin{equation}\label{eq:rd-exact}
  \left\{\begin{array}{lcl}
    \DS \vec{v}_g^{\rm ex} (t) &=& \DS
         -\frac{\Omega_g}{\Omega_p}\vec{v}^{\rm ex}_p(t)
   \\ \noalign{\medskip}
    \DS \vec{v}^{\rm ex}_p (t) &=& \DS
         v_0 \Big[ \cos(\Omega t)\hvec{e}_x
                  -\sin(\Omega t)\hvec{e}_y\Big]
  \end{array}\right.
\end{equation}
where $\Omega = \Omega_g + \Omega_p$, $\Omega_g = \alpha_iRB_0$, $\Omega_p = \alpha_p(1-R)B_0$.
Note also that, from the definition of $R = \alpha_p\varrho_p/(\alpha_i\rho_i + \alpha_p\varrho_p)$ we have that $\rho\Omega_g = \varrho_p\Omega_p$.
Equation (\ref{eq:rd-exact}) shows that both particles and gas trace clock-wise circular orbits with the same period but different radii.
It can also be easily verified that the total (gas+particle) momentum remain constant in time, as expected.

\begin{figure*}
  \centering
  \includegraphics[width=0.9\textwidth]{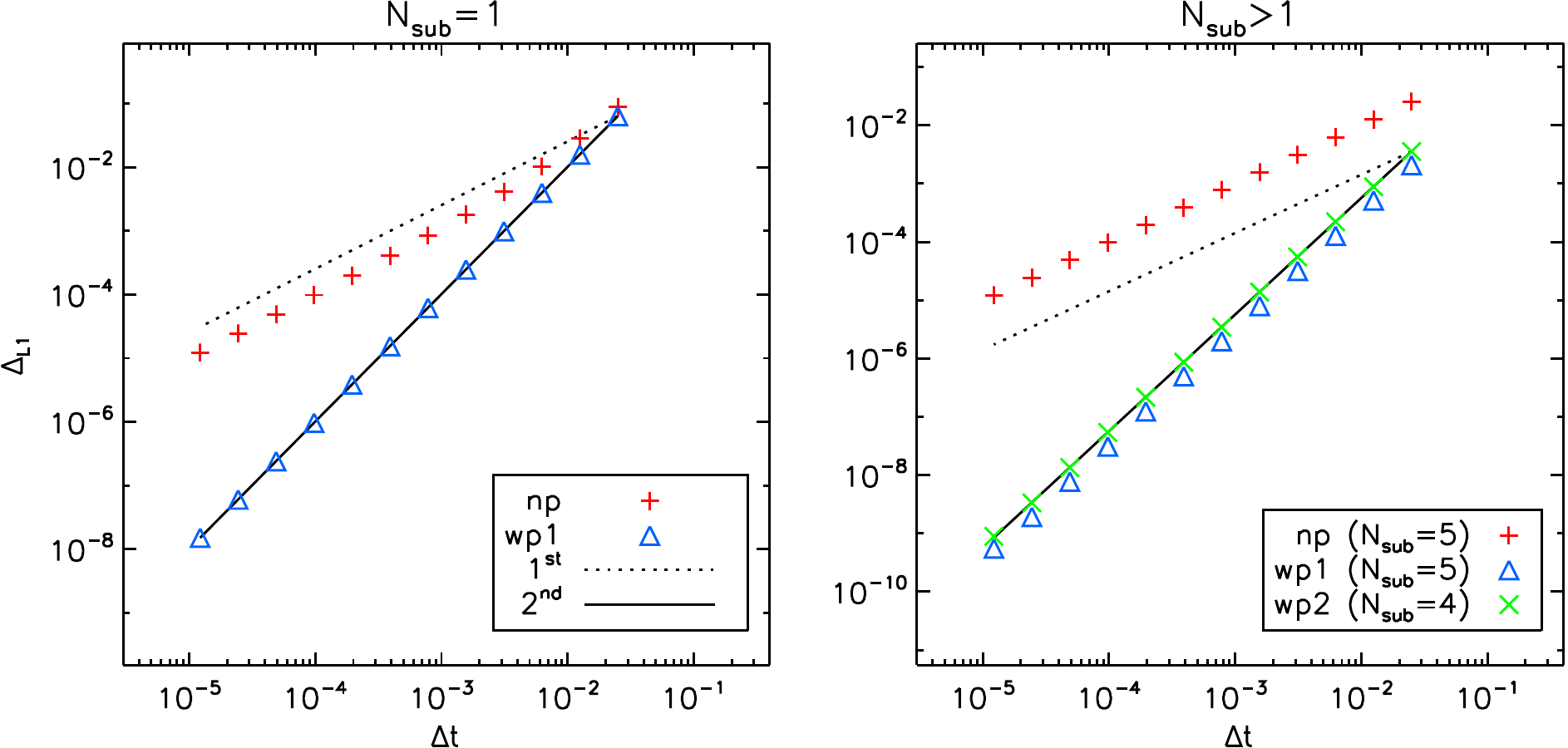}\hfill
  \caption{\footnotesize $L_1$ norm errors for the fluid-particle relative drift
           problem using the standard particle update \emph{without} sub-cycling
           (left panel) and with sub-cycling (right panel).
           Results after one period obtained without the predictor step
          (\texttt{np}) are shown using red plus signs while blue triangles
           and green crosses correspond
           to computations obtained by including the predictor step
           with sub-cycling methods 1 (\texttt{wp1}) or 2 (\texttt{wp2}),
           see \S\ref{sec:subcycling}.
           Computations using sub-cycling employ $N_{\rm sub}=5$ except for
           sub-cycling method 2 for which we set $N_{\rm sub}= 4$.
           The black dashed (solid) line gives the expected convergence
           rate for a $1^{\rm st}$ ($2^{\rm nd}$) temporally-accurate scheme.
           \label{fig:relative_drift}}
\end{figure*}

We choose $\alpha_p = \alpha_i = 1$, $B_0= 2\pi$ so that our units are such that $\Omega = 2\pi$.
For the test considered here, we set $v_0=5$ and $\varrho_p = 10^{-2}\rho$.
We use $8\times 8$ grid zones and 1 particle per cell.
The system is evolved for exactly one period $T = 1$ using constant time steps $\Delta t = 1/N_t$, where $N_t$ is the number of (fluid) time-steps: $N_t = 40,\, 80,\, 160,\, ...,\, 81920$.
For the sake of comparison, we have repeated computations with and without the predictor step (see sect. \ref{sec:CR_predictor}) and also by varying the number of sub-steps used during particle sub-cycling.
Both sub-cycling methods, illustrated in Section \ref{sec:subcycling}, have been compared with $N_{\rm sub} = 1$ and $N_{\rm sub} = 5$ ($N_{\rm sub} = 4$ for method 2).  
The error is computed at the end of each computation using the $L_1$ norm
\begin{equation}\label{eq:rd-error}
  \Delta_{L1} =    \Big|\vec{v}_g(T) - \vec{v}_g^{\rm ex}(T)\Big|
                 + \Big|\vec{v}_p(T) - \vec{v}^{\rm ex}_p(T)\Big|
\end{equation}
and it is plotted in Figure \ref{fig:relative_drift} without sub-cycling (left panel, $N_{\rm sub} = 1$) and with sub-cycling (right, $N_{\rm sub} > 1$).

Results obtained without the predictor step (labeled with \lq\texttt{np}\rq, red plus signs) show essentially first-order accuracy regardless of sub-cycling ($N_{\rm sub}=1$ or $N_{\rm sub}= 5$ in the left and right panel, respectively).
On the contrary, including the predictor step noticeably improves the overall scheme's convergence yielding genuine second-order temporal accuracy.
This holds when $N_{\rm sub} = 1$ (left panel) and also when $N_{\rm sub} > 1$ (right panel).
Results obtained with sub-cycling methods 1 and 2 are both reported using blue triangles and green crosses in Figure \ref{fig:relative_drift} and label-led, respectively, with \lq\texttt{wp1}\rq and \lq\texttt{wp2}\rq.
Notice that, while method 1 can be used for any $N_{\rm sub}\ge 1$ (we employ $N_{\rm sub}=5$), method 2 works only when $N_{\rm sub}$ is even (we set $N_{\rm sub}=4$). 

Note that computations have been carried out using both the CTU and RK2 time stepping methods and results are identical.
Indeed, in absence of spatial gradients, the two methods become coincident as it can be easily verified from Equation (\ref{eq:CTU_corr}) and (\ref{eq:RK2-v2}) with ${\cal L} = 0$.

\subsection{Non-Resonant Bell Instability}
\label{sec:bell_instability}
%
%

\begin{figure*}
  \centering
  \includegraphics[width=0.32\textwidth]{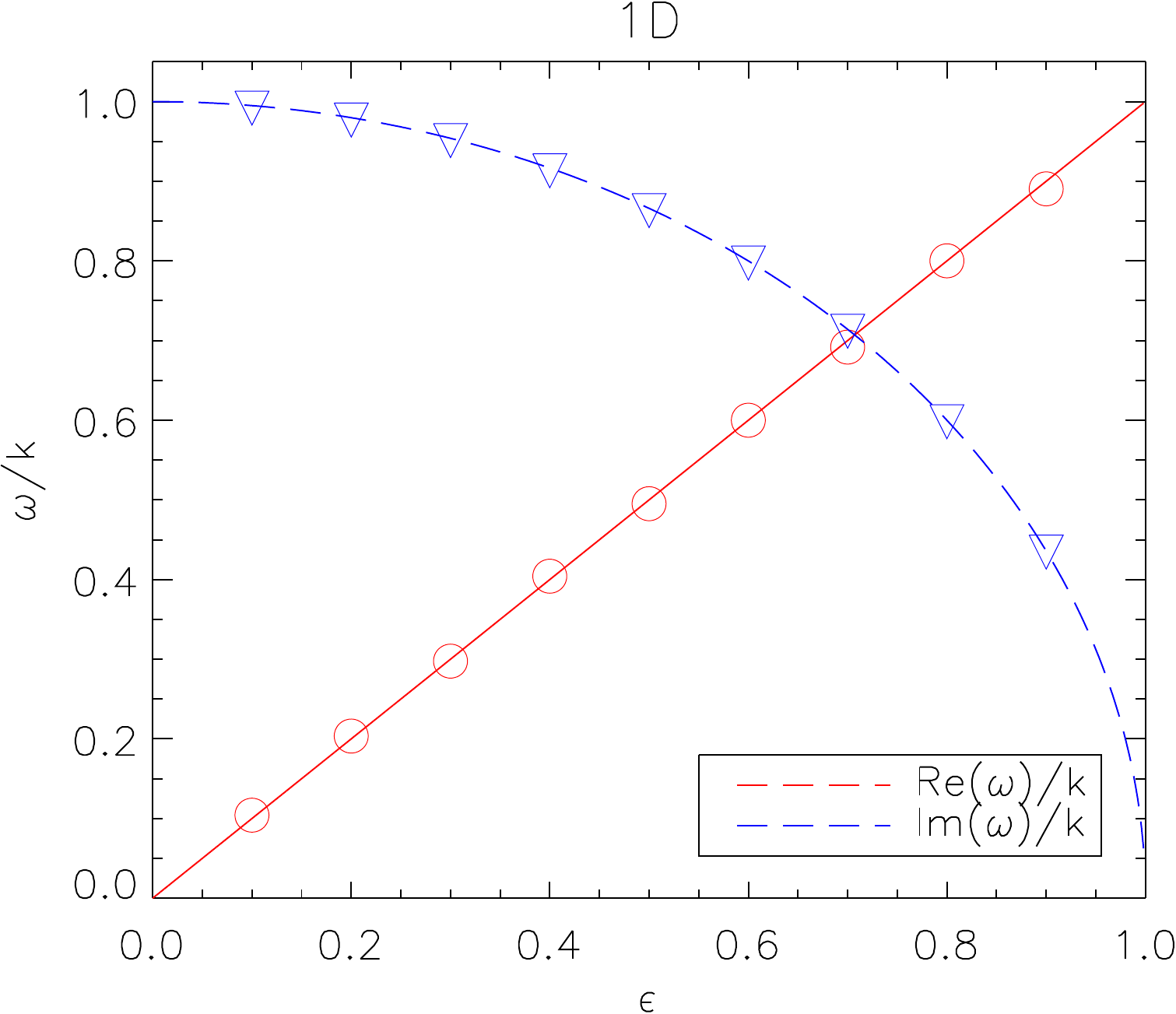}\hfill
  \includegraphics[width=0.32\textwidth]{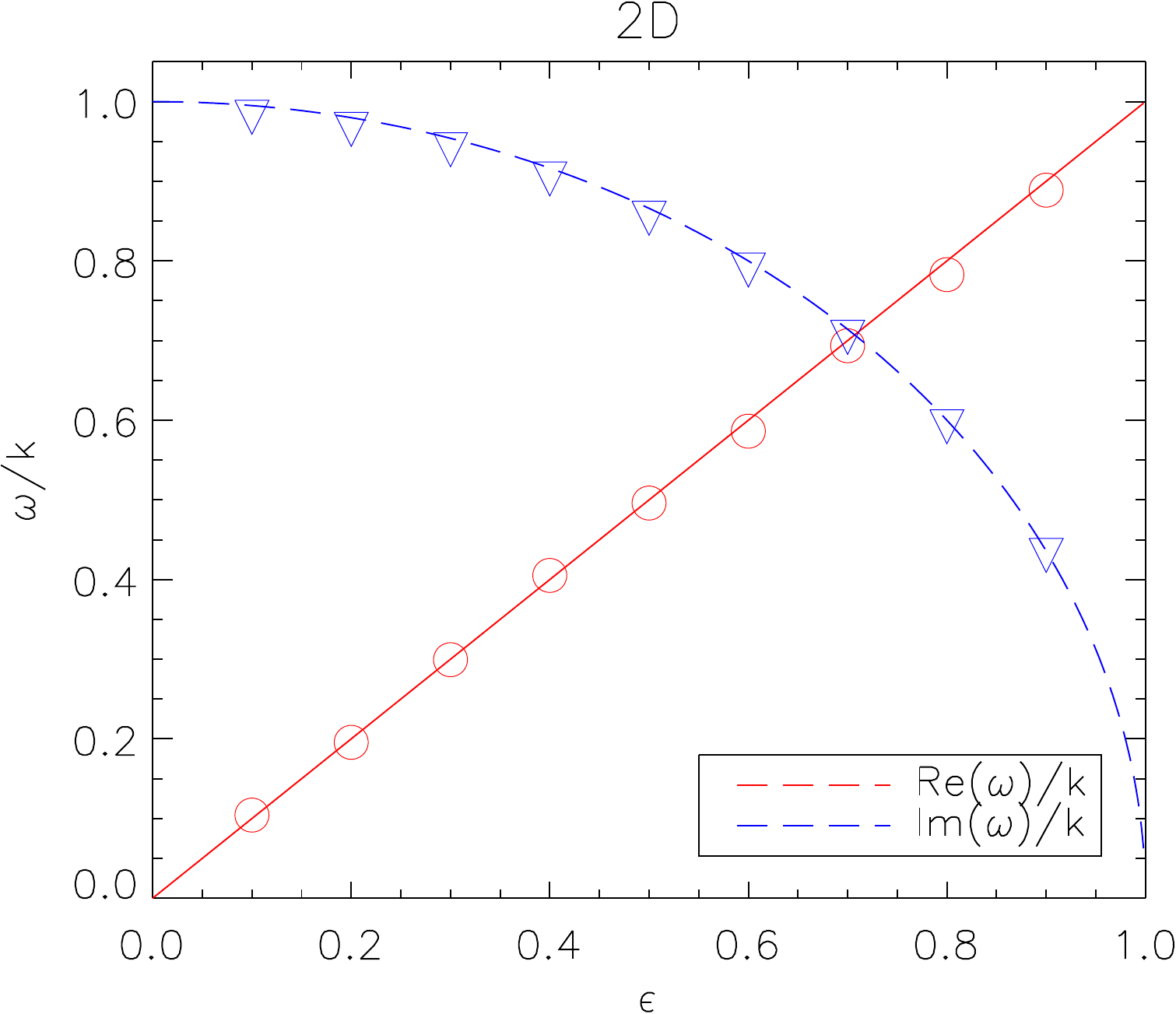}\hfill
  \includegraphics[width=0.32\textwidth]{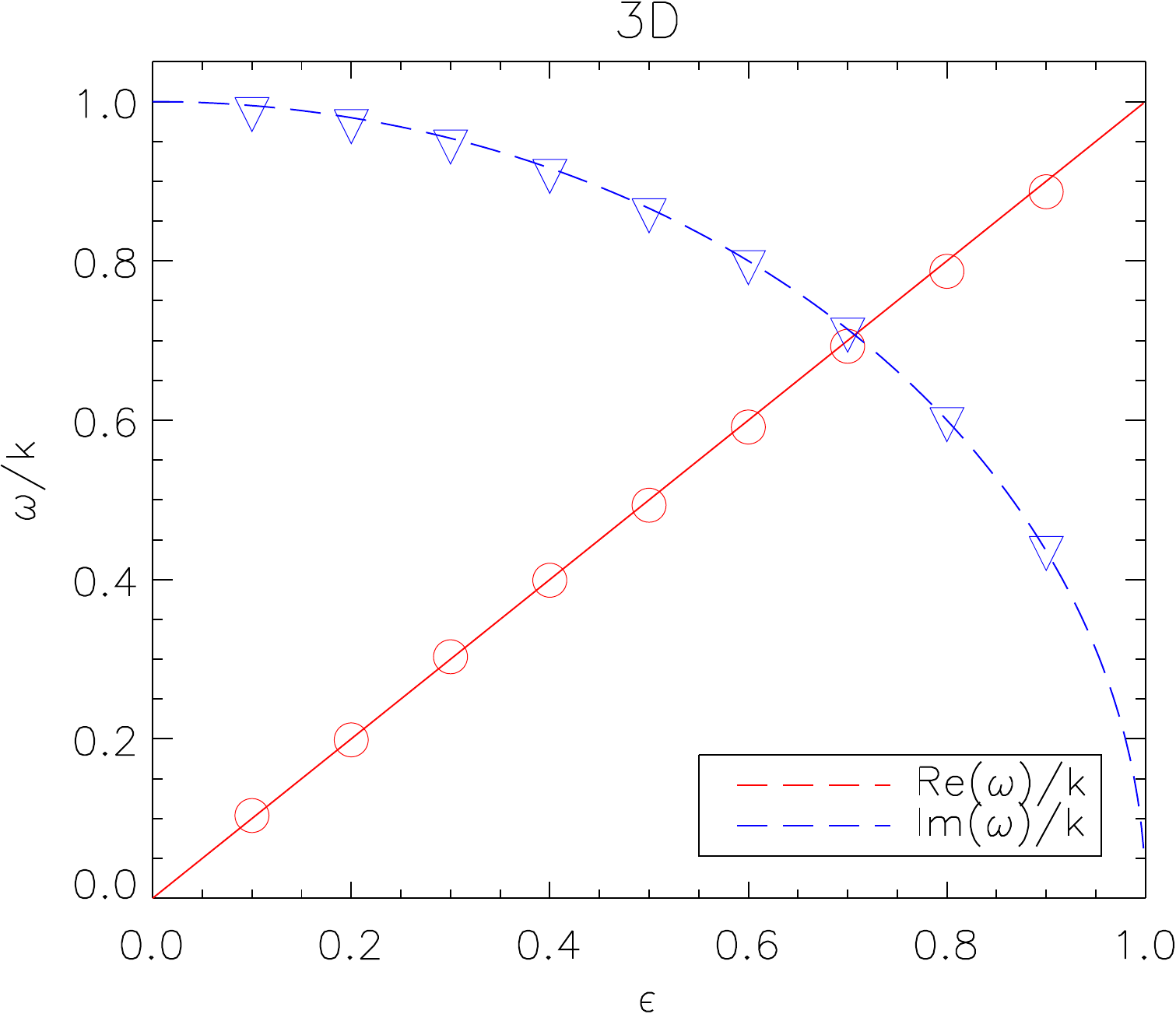}
  \caption{\footnotesize Real (red) and imaginary (blue) part of the growth rate 
           for the non-resonant Bell instability problem using different values
           of the $\epsilon$ parameter. 
           Solid and dashed lines give the theoretical expectation, 
           Equation (\ref{eq:Bell_DR_k0}), while symbols (triangles and circles) are the 
           results measured from the simulations using the CTU scheme.
           \label{fig:bell_instability.MH}}
\end{figure*}
\begin{figure*}
  \centering
  \includegraphics[width=0.32\textwidth]{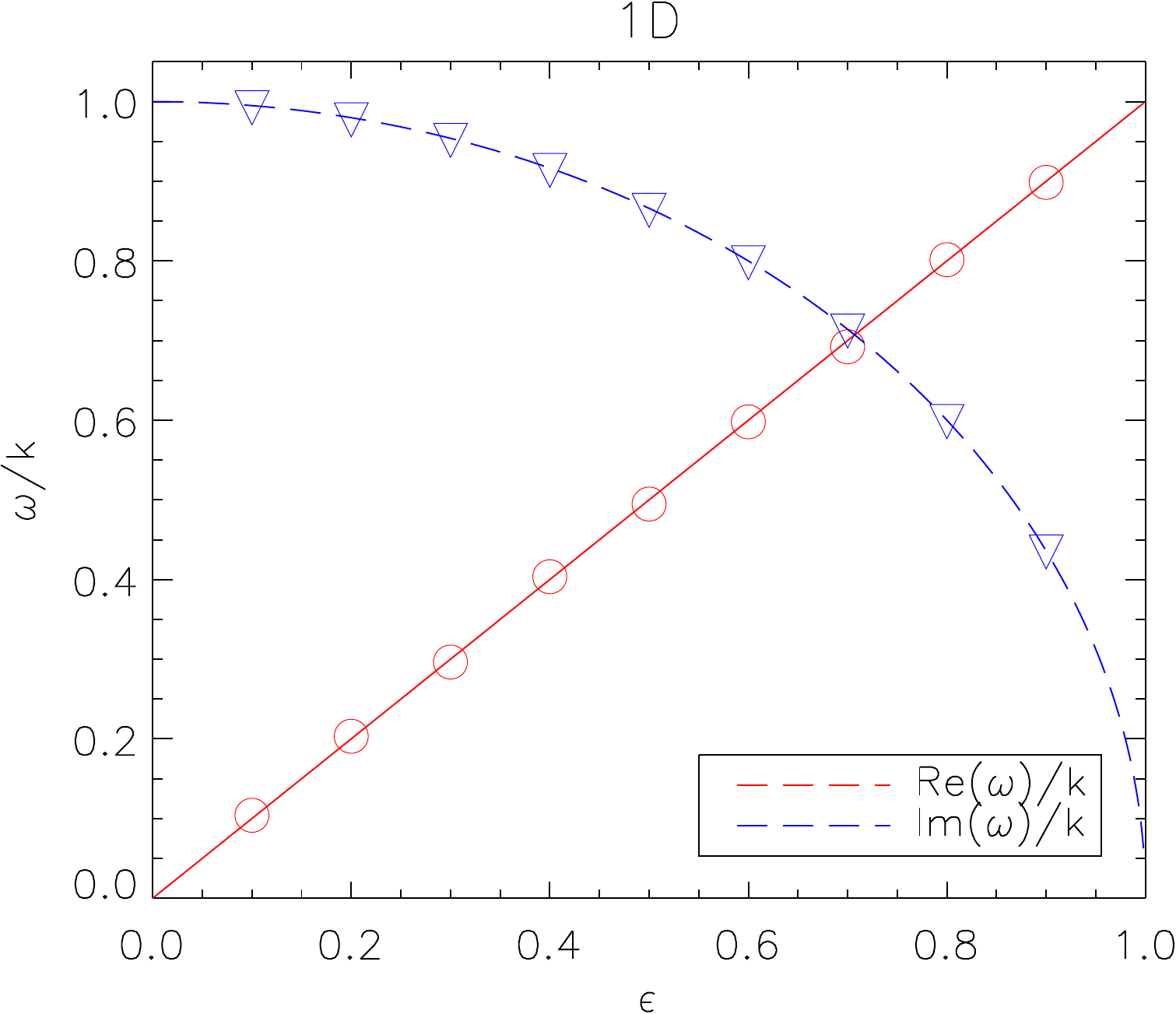}\hfill
  \includegraphics[width=0.32\textwidth]{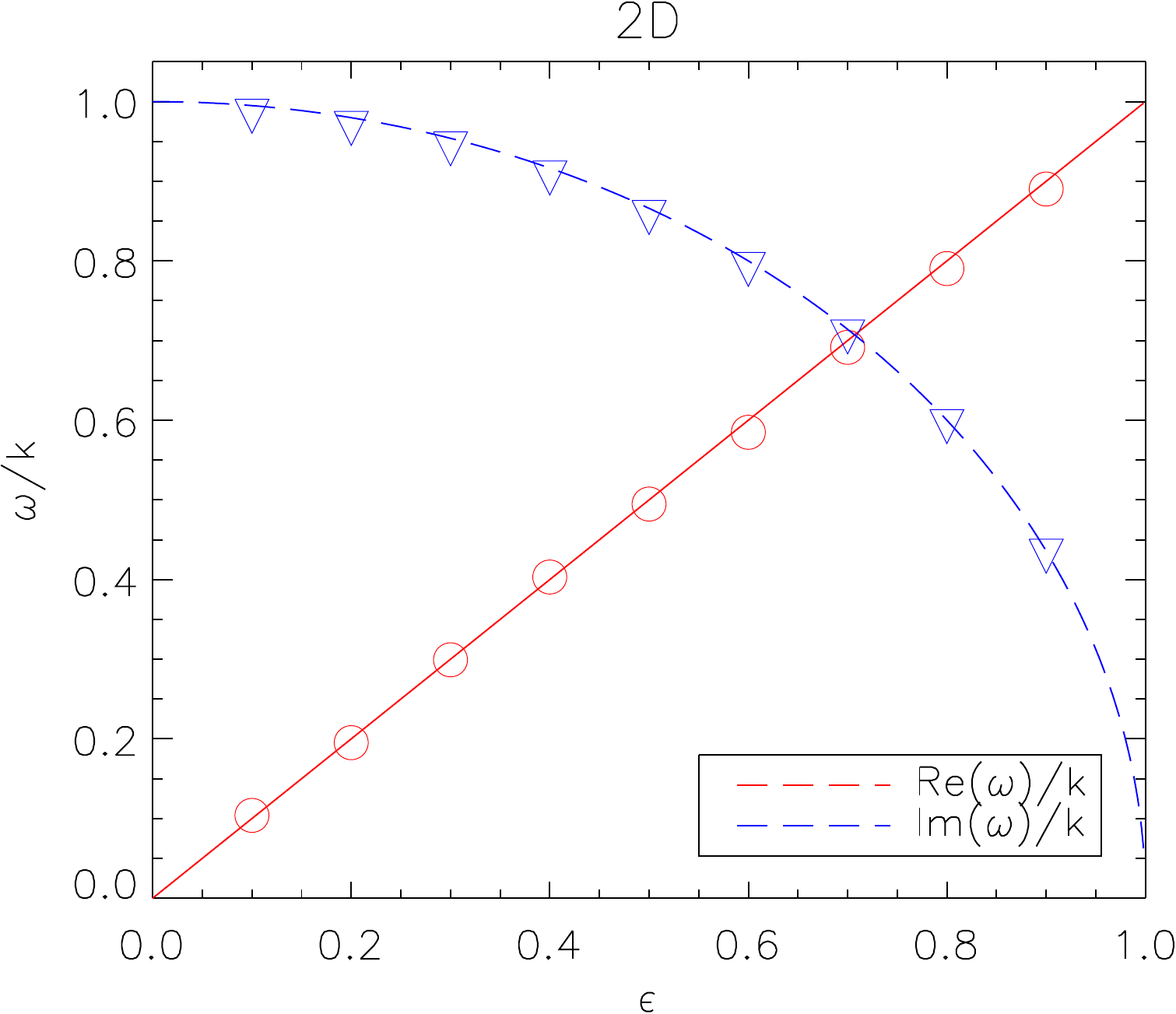}\hfill
  \includegraphics[width=0.32\textwidth]{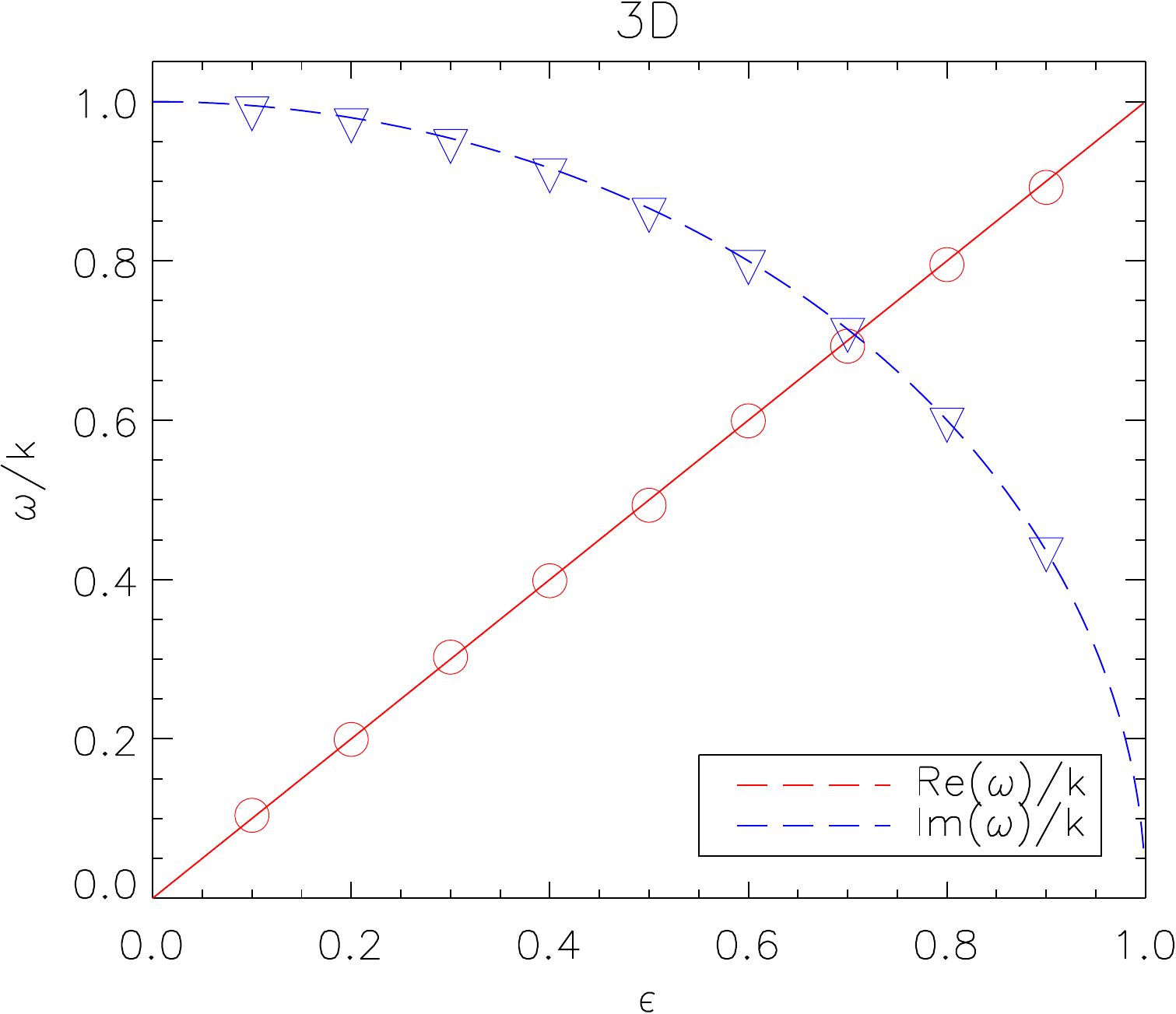}
  \caption{\footnotesize Same as Figure \ref{fig:bell_instability.MH}
           but for the RK2 time stepping. \label{fig:bell_instability.RK2}}
\end{figure*}

In the next test we verify the implementation of our MHD-PIC module by investigating the linear growth of the non-resonant Bell instability \citep{Bell.2004} in 1, 2 and 3 dimensions.
The instability is driven by the relative streaming between gas and CR particles along magnetic field lines and it takes place when the CR drift velocity exceeds the local Alfv{\'e}n speed.
The streaming of CR generates a return current in the thermal plasma (in the attempt to restore charge neutrality), so that small perturbations are amplified when the induced Lorentz force exceeds magnetic tension.
The instability excites nearly purely growing modes with wavelengths shorter than the Larmor radius and does not saturate when $\delta \vec{B}/\vec{B}\sim 1$, but it continues growing to produce amplified magnetic fields much larger than the initial field \citep{Bell.2013}.
This mechanism is believed to operate in the upstream regions of high Mach number SNR shocks, leading to efficient magnetic field amplification and the development of turbulence. 
Magnetic field fluctuations in the upstream magnetic field are then responsible for the scattering of CR and their confinement close to the shock front \citep{Bai_etal.2015, Bell.2004} thereby providing an efficient mechanism to trigger diffusive shock acceleration.

\cite{Bai_etal.2015} have carried a linear stability analysis by including the CR-Hall term that was previously neglected.
The quantity
\begin{equation}
  \Lambda = R\frac{|\vec{v}_\CR - \vec{v}_g|}{v_A}\,,
\end{equation}
where $v_A$ is the Alfv{\'e}n velocity, determines the importance of the CR-Hall term and leads essentially to a reduction of the growth rate which saturates at the ion cyclotron frequency of the background plasma when $J_\CR$ is increased, in the limit $\Lambda \gg 1$.

Here we consider the opposite limit ($\Lambda \ll 1$) which is also the same regime used in \cite{Bell.2004} (regime II).
Assuming incompressible perturbations proportional to $e^{i(kx-\omega t)}$ in the fluid rest frame, the dispersion relation becomes
\begin{equation}\label{eq:Bell_DR}
  \frac{\omega(k)}{k_0v_A} = \epsilon 
     + \sqrt{ \left(\frac{k}{k_0}\right)^2 - 2\frac{k}{k_0} + \epsilon^2}
\end{equation}
where $\epsilon = v_A/v_\CR$ and
\begin{equation}\label{eq:Bell_k0}
  k_0 = \frac{J_\CR}{2B_0c}
\end{equation}
is the most unstable wavenumber with $\vec{J}_\CR = en_\CR \vec{v}_\CR$ the CR current density.
The maximum growth rate is obtained when $k=k_0$ yielding 
\begin{equation}\label{eq:Bell_DR_k0}
  \omega_0 = k_0v_A\Big(\epsilon + {\rm i}\sqrt{1 - \epsilon^2}\Big) \,.
\end{equation}

To setup the problem, we consider a periodic box $x\in[0,\,L_x]$, $y\in[0,\,L_y]$ and $z\in[0,\,L_z]$ initially filled with a plasma with uniform density and pressure (we set $\rho =1,\,p=1$) and threaded by a constant background magnetic field $\vec{B}_0 = (1,0,0)$.
A monochromatic beam of CR particles is set to travel along the $x$ direction with velocity $v_\CR=v_A/\epsilon$, where $\epsilon \in(0,1)$ is now a free parameter.
In order to ensure that the CR current remains constant during the evolution, particles must have a large inertia and this is achieved by setting the charge to mass ratio of CR particles to be very small, i.e.,  $\alpha_p = 10^{-6} v_Ak_0/B_0$.
Using the definition of the CR current together with Equation (\ref{eq:Bell_k0}) for the most unstable wavenumber, one obtains that the CR density satisfies $\varrho_p = 2\times10^6\epsilon B_0^2/v_A^2$.
Finally, to ensure that $R \ll 1$ we set $\alpha_i = 10^3$ while the speed of light is fixed to $\C = 10^6$.

Perturbations in velocity and magnetic field at $t=0$ are introduced by using the exact eigenvectors obtained from the 1D linear dispersion relation in the limit $\Lambda = Rv_\CR/v_A \ll 1$ \citep[see Appendix of][]{Bai_etal.2015} according to which
\begin{equation}\label{eq:Bell_dv}
  \delta\vec{v}_g = v_A\frac{b_\perp}{B_0}\left[0,\,
                              \cos(\phi - \theta),\,
                              \sin(\phi - \theta)
                              \right]
\end{equation}
and 
\begin{equation}\label{eq:Bell_dB}
    \delta \vec{B} 
   = b_{\perp}\left[0,\, \cos(\phi),\, \sin(\phi) \right]
\end{equation}
where $\phi = k_0x$, $\theta = \sin^{-1}\epsilon$ and $b_\perp=10^{-5}$ is the initial perturbation amplitude.

In 1D, we set $k_0 = 2\pi$ so that, by choosing the box size $L_x=1$, we fit exactly one (most unstable) wavelength in the computational domain.
In 2D and 3D, the initial configuration is rotated so that the new wavevector is not grid-aligned but has orientation 
\begin{equation}
  \vec{k}'_0 = \frac{2\pi}{L_x}\left(1,\, \tan\alpha,\, \tan\beta\right)
\end{equation}
where $\tan\alpha = L_x/L_y$ and $\tan\beta = L_x/L_z$ still satisfy $|\vec{k}'_0| = 2\pi$.
Vectors are then rotated using $\vec{v}_g' = \tens{R}_{\gamma\alpha}\delta\vec{v}_g$ and $\vec{B}' = \tens{R}_{\gamma\alpha}(\vec{B}_0+\delta\vec{B})$, where the rotation matrix $\tens{R}_{\gamma\alpha}$ is defined as \citep[see also, e.g.,][]{MigTze.2010}:
\begin{equation}
  \tens{R}_{\gamma\alpha} = \left(\begin{array}{ccc}
  \cos\alpha\cos\gamma & -\sin\alpha & -\cos\alpha\sin\gamma  \\ \noalign{\medskip}
  \sin\alpha\cos\gamma &  \cos\alpha & -\sin\alpha\sin\gamma  \\ \noalign{\medskip}
  \sin\gamma           &     0       &  \cos\gamma
  \end{array} \right)
\end{equation}
where $\tan\gamma = \cos\alpha\tan\beta$.
In 2D, we employ $L_x = 2L_y = \sqrt{5}$ using $64\times 32$ zones while in 3D we set $L_x = 2L_y = 2L_z = 3$ using $96\times 48\times 48$ zones.
We run 9 simulations corresponding to $\epsilon = 0.1,..., 0.9$, for each case.
For the sake of comparison we perform computations using the CTU scheme with a MUSCL-Hancock predictor step and the RK2 scheme.
The CFL number is set to $0.45$ except for the 3D run using the Runge-Kutta scheme for which we lower it to $0.3$.

In order to measure the growth rate we first evaluate, at each time $t$, the transverse magnetic energy as $p_{m\perp} = (\tens{R}^{-1}_{\gamma\alpha}\vec{B}' - \vec{B}_0)^2/2$ and then find the value $t_{\rm max}$ at which a maximum is reached.
The imaginary part is then computed as the difference between $p_{m\perp}$ at $t_e=3t_{\max}/4$ and $t_b=t_{\max}/4$:
\begin{equation}
  {\rm Im}(\omega) = \frac{1}{t_e - t_b}\log\left[\frac{p_{m\perp}(t_e)}
                                                       {p_{m\perp}(t_b)}\right]
\end{equation}
Likewise, we compute the real part by measuring the distance traveled by a wave crest from $t_e$ to $t_b$:
\begin{equation}
  {\rm Re}(\omega) = k_x\frac{ x_{\max}(t_e) - x_{\max}(t_b) }{t_e - t_b}
\end{equation}
where $x_{\max}(t_b)$ denotes the horizontal position of the first maximum of the $z$ component of $\vec{B}'_\perp = \vec{B}' - (\vec{k}\cdot\vec{B})\vec{k}/k_0^2$.

Results obtained with the CTU and RK2 schemes are shown, respectively, in Figures (\ref{fig:bell_instability.MH}) and (\ref{fig:bell_instability.RK2}) where we plot, from left to right, the real and imaginary parts of the growth rate (red and blue symbols) together with their analytic values (red and blue lines) as given by Equation (\ref{eq:Bell_DR_k0}) in 1D, 2D and 3D, respectively.
Our results show a good agreement with the analytic predictions and a quantitative analysis show that the relative error computed as
\begin{equation}\label{eq:bell_instability_error}
 \Delta = \max_\epsilon\left(\left|\frac{\omega(\epsilon)}{\omega_0(\epsilon)}
                              -1\right|\right)
\end{equation}
never exceeds $\sim 4\%$ for the real part and $\sim 1.5\%$ for the imaginary part.
In Equation (\ref{eq:bell_instability_error}), $\omega(\epsilon)$ refer to the (real or imaginary part of the) measured value of the growth rate while $\omega_0(\epsilon)$ is given by Equation (\ref{eq:Bell_DR_k0}).
Error values are reported in Table \ref{tab:bell_instability} for the CTU and RK2 schemes in 1, 2 and 3 dimensions.

\begin{table}
\begin{center}
\caption{\footnotesize Relative errors for the Bell instability problem.
         \label{tab:bell_instability}}
\begin{tabular}{cccccc}
\tableline \noalign{\smallskip}
        &  \multicolumn{2}{c}{CTU}     &  &  \multicolumn{2}{c}{RK2}  \\ \noalign{\medskip}
        & $\Delta_{{\rm Re}(\omega)}$  & $\Delta_{{\rm Im}(\omega)}$
    &   & $\Delta_{{\rm Re}(\omega)}$  & $\Delta_{{\rm Im}(\omega)}$ \\
          \cline{2-3}    \cline{5-6}      \\
    1D  &  4.21E-02 & 3.60E-03 &  & 4.08E-02 & 3.22E-03  \\ \noalign{\smallskip}
    2D  &  4.29E-02 & 1.51E-02 &  & 3.97E-02 & 1.50E-02  \\ \noalign{\smallskip}
    3D  &  3.77E-02 & 1.15E-02 &  & 4.06E-02 & 1.09E-02  \\ \noalign{\smallskip}
\tableline
\end{tabular}
\end{center}

\end{table}

\subsection{Application to Collisionless Shocks}
%
%

In this section we apply our MHD-PIC module to investigate particle (ion) acceleration in parallel MHD collisionless shock.
Our configuration reproduces the setup described by \cite{Bai_etal.2015} in their R2 (classical) and R2-REL (relativistic) fiducial computations.
Note that, while the fluid is always described by the classical MHD equations, the two runs differs essentially for the reduced speed of light ($\C=10^4$ and $\C = 10\sqrt{2}v_0$, respectively).

The computational box is defined by the 2D rectangular domain with $0 \le x\le L_x$ and $0\le y\le L_y$ where $(L_x,L_y) = (120,\, 3)\times 10^3$ for the run R2 while a larger box $(L_x,L_y) = (384,\, 4.8)\times 10^3$ is used for the relativistic case (run R2-REL).
Lengths are conveniently expressed in units of the ion skin depth $c/\omega_{\rm pi}$.
The initial condition consists of a constant density and pressure ($\rho_0 = 1,\,p_0=1$) supersonic inflow propagating to the left with velocity $v_{0} = -M_A$ where $M_A=30$ is the Alfv{\'e}nic Mach number.
An ideal equation of state with specific heat ratio $\gamma = 5/3$ is employed.
The magnetic field is initially constant and parallel to the flow velocity $\vec{B} = (B_0,0,0)$.
We set $B_0=1$ so that velocities will be normalized to the initial upstream Alfv{\'e}n speed.
This also sets the time unit as the inverse of the cyclotron frequency, $\Omega_L^{-1} = c/(\omega_{\rm pi}v_A)$.

We employ a uniform grid resolution of $[N_x, N_y] = [11520,\, 288]$ and $[N_x, N_y] = [30720,\, 384]$ for the two cases, respectively.
This choice corresponds to a mesh resolution of $\approx 10.4 \, c/\omega_{pi}$ and $\approx12.5\, c/\omega_{pi}$ per cell, therefore giving a significant efficiency gain when compared to hybrid codes which typically requires finer grids ($\approx 2 \, c/\omega_{pi}$) to properly describe microphysics. 
At the leftmost boundary ($x=0$) we apply conducting conditions so that a right-going shock receding from the wall forms immediately.
Constant flow injection holds at the right boundary $x=L_x$ while vertical boundary conditions are periodic.
The MHD-PIC equations are evolved until $t=3000\,\Omega_L^{-1}$ in the non relativistic run (run R2) while computations are stopped at $t=11520\,\Omega_L^{-1}$ for the relativistic case.

\subsubsection{Injection Recipe.}
%
%
Since the MHD-PIC approach cannot consistently model the injection physics, a prescription that mimics the generation of supra-thermal particles in the downstream region of the shock is necessary.
As shown by \cite{Caprioli_Spitkovsky.2014a, CPS.2015}, the ion distribution immediately behind the shock shows an intermediate region of particles with mildly non-thermal energies.
Hybrid simulations indicate that the fraction of injected particles can be effectively parameterized by defining a threshold energy, $E_{\rm inj}$, which marks the boundary between the thermal and non-thermal distributions, and the injection fraction $\eta$.
While \cite{Bai_etal.2015} prescribe an injection recipe that is strictly one-dimensional, we present here a different approach that can also be used in the context of multidimensional calculations.

Particles are injected at the end of an \quotes{accumulation} cycle $\Delta T_{\rm acc}$ consisting of a finite number of hydro steps during which we track the amount of mass swept by the shock.
To this end, we add to the MHD-PIC equations the evolution of a passive tracer $\T$ which is updated in the following way.

\begin{figure*}[!ht]
  \centering
  \includegraphics[width=0.48\textwidth]{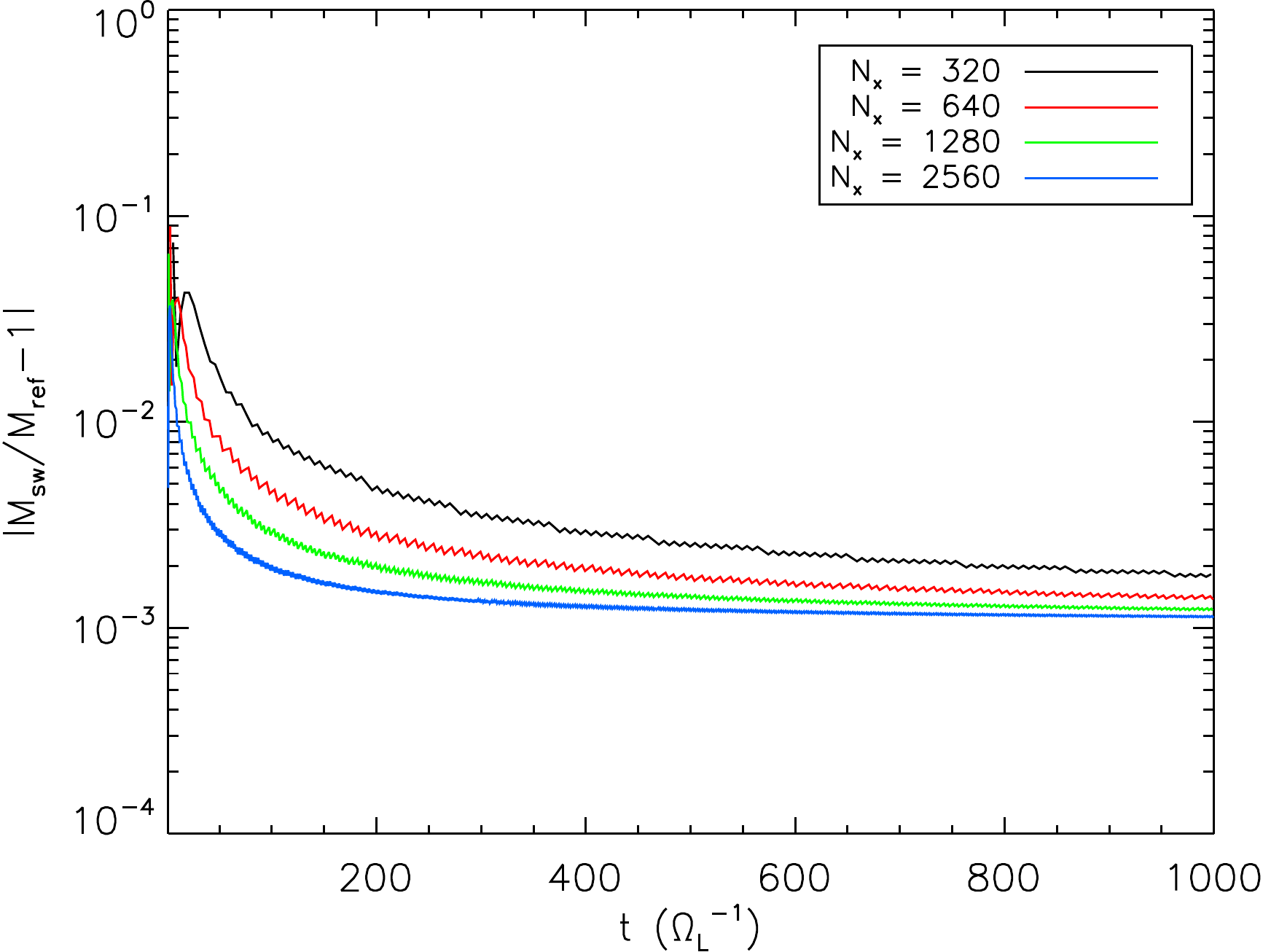}%
  \includegraphics[width=0.48\textwidth]{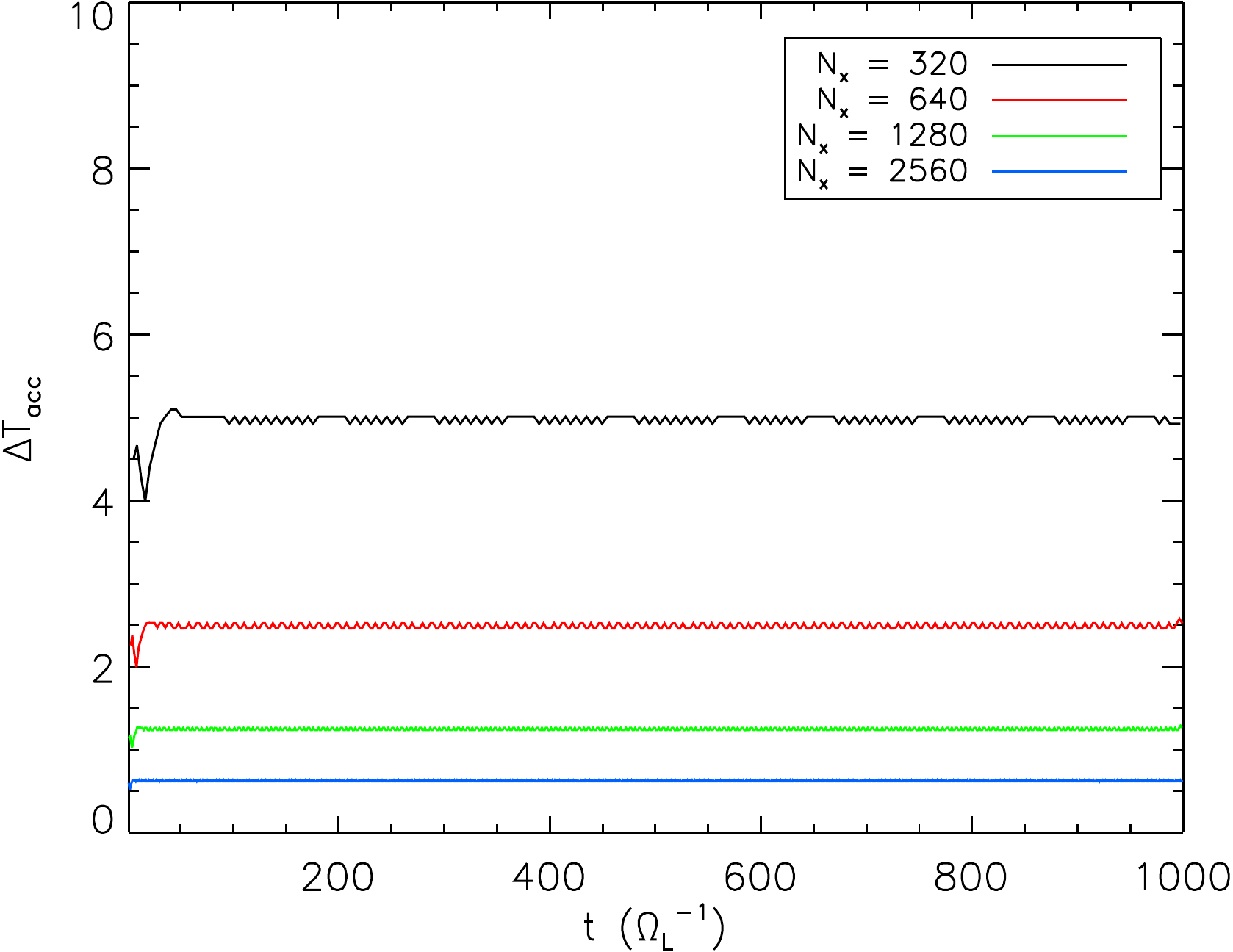}
  \caption{\footnotesize Left panel: relative error $|M_{\rm sw}/M_{\rm ref}-1|$
            of the cumulative swept mass as a function of time,
            where $M_{\rm sw} = \Delta x\Delta y\int_0^t
            \sum_{\vec{i}} (\rho_{\rm sh})_{\vec{i}}$ while
            $M_{\rm ref} = \rho_0v_{\rm sh}t L_y$ is the expected value for
            a 1D plane-parallel shock.
            Right panel: duration of an accumualation cycle as a function of time.
            Colors corresponds to different grid resolutions reported in the legend.
            \label{fig:check_Mswept}}
\end{figure*}

\begin{enumerate}

\item
At the beginning of an accumulation cycle we set $\T_{\vec{i}} = 0$ for all zones $\vec{i} = (i,j)$ in the computational domain.

\item
For each time step in the computation, we initialize the tracer to $1$ if a computational zone lies within a shock and then update $\T$ regularly using our conservative scheme.

The criterion for a zone $\vec{i}$ to be considered inside a shock demands: i) the divergence of fluid velocity to be negative and ii) the normalized second derivative of pressure to exceed a certain threshold:
\begin{equation}\label{eq:shock_detect1}
   \left\{\begin{array}{l}
   \DS  \sum_{d=x,y} \hvec{e}_d\cdot
      (\vec{v}_{g,\vec{i}+\hvec{e}_d} - \vec{v}_{g,\vec{i}-\hvec{e}_d})   < 0\,, 
      \\ \noalign{\medskip}
    \DS \sum_{d=x,y}
       \frac{|p_{\vec{i}+\hvec{e}_d} - 2p_{\vec{i}} + p_{\vec{i}-\hvec{e}_d}|}
            { p_{\vec{i}+\hvec{e}_d} + 2p_{\vec{i}} + p_{\vec{i}-\hvec{e}_d}}
      > \chi
  \end{array}\right.
\end{equation}
where $\vec{i} = (i,j)$, $\hvec{e}_x=(1,0)$, $\hvec{e}_y=(0,1)$ while we set the threshold $\chi = 0.2$.
An additional measure is necessary to avoid tracking the formation of secondary small discontinuities ahead or behind the shock during the turbulent regime.
We achieve this by selecting, among shocked zones, those with a large pressure jump:
\begin{equation}\label{eq:shock_detect2}
  \left\{\begin{array}{ll}
  \min\left(p_{\vec{i}+\vec{\delta}}\right)  &< \chi_{\min}\,,\quad
  \\ \noalign{\medskip}
  \max\left(p_{\vec{i}+\vec{\delta}}\right)  &> \chi_{\max}\,.
  \end{array}\right.
\end{equation}
where $\delta=[-1..1,-1..1]$ spans all of the 8 neighbor zones.
In order to detect the primary shock we use $\chi_{\min} = 15$ and $\chi_{\max} = 250$.
The criteria for choosing $\chi_{\min}$ and $\chi_{\max}$ depends on the shock that one wishes to track.
The details of the computation are not sensitive to their values, inasmuch as $\chi_{\min}$ ($\chi_{\max}$) is larger (smaller) than the upstream (downstream) pressure.

If conditions (\ref{eq:shock_detect1}) and (\ref{eq:shock_detect2}) are both satisfied, we consider the zone to lie within a shock and set a flag $f_{\vec{i}} = 1$ ($f_{\vec{i}} = 0$ otherwise).

\item
The passive scalar is evolved by repeating step 2 until the following condition is met:
\begin{equation}\label{eq:acc_cond}
       \sum_{\vec{i}} (\rho_{\rm sh})_{\vec{i}}
   > Q \sum_{\vec{i}} (\rho\T)_{\vec{i}} 
\end{equation}
where $\rho_{\rm sh} = \rho\T (1-f)$, the summation extends to all computational zones, $Q=0.8$ is a safety factor and $f$ is the current shock detector flag defined in step 2.
Equation (\ref{eq:acc_cond}) marks the end of the accumulation cycle $\Delta T_{\rm acc}$ and the summation on the left hand side represents the total mass (density) swept by the shock during this interval of time, by excluding zones that are currently flagged which would otherwise tend to overestimate the swept mass.

The reliability of our mass-tracking algorithm has been tested on the 1-D unperturbed propagation of the shock using different grid resolutions.
Figure \ref{fig:check_Mswept} shows (left panel) the relative error of the cumulative swept mass as a function of time: the uncertainity is larger at the beginning ($\approx 5-10 \%$ due to start-up error and wall-heating at the left boundary) while it progressively reduces to a few $10^{-3}$ at later times ($t\gtrsim 400 \Omega_L^{-1}$).
This error falls well within the uncertainty in estimating the mass fraction $\eta$ \citep[$\sim 10^{-3}-10^{-4}$, see][and the discussion below]{Caprioli_Spitkovsky.2014a} of particles crossing the shock and participating into the DSA process.
On the right hand panel, we plot the duration of an accumulation cycle as a function of time and point out that our injection recipe is not continuous in time but, rather, occurs periodically with a period $\Delta T_{\rm acc}$ that reduces as the mesh is refined.

For the grid resolution employed here the duration of a single accumulation cycle lasts  approximately $\Delta T_{\rm acc} \approx 2 \Omega_L^{-1}$.

\item
When condition (\ref{eq:acc_cond}) is fulfilled, particles injection takes place.
The amount of CRs injected in each zone $\vec{i}$ is proportional to the local swept mass distribution, that is, $(N_{\rm inj})_{\vec{i}} = N_{\rho_0}(\rho_{\rm sh})_{\vec{i}}$ where $N_{\rho_0}=4$ is the number of particles per cell at unit fluid density.
Particles will be mostly injected in the shock downstream (where $\rho_{\rm sh} \ne 0$) and their mass density is controlled by the parameter $\eta$ such that $N_{\rm inj}\varrho_p = \eta\rho_{\rm sh}$.
Following \cite{Bai_etal.2015}, we set the CR mass fraction $\eta = 2\times 10^{-3}$ \citep[see also section 3 of][for a thorough discussion]{Caprioli_Spitkovsky.2014a} and therefore $\varrho_p = \eta/N_{\rho_0}$.
In such a way, the mass of the injected particles is a fixed fraction $\eta$ of the shock swept mass.

Following \cite{Caprioli_Spitkovsky.2014a, CPS.2015}, we set the energy of injected particles to be $10E_{\rm sh}$ in the comoving shock frame, where $E_{\rm sh} = v_0^2/2$ is the shock specific kinetic energy.
In the lab frame the particle velocity is therefore initialized to 
\begin{equation}
  \vec{v}_p = 10\hvec{e}_x
              + \sqrt{20E_{\rm sh}}
              \left(\begin{array}{l}
                  \sin\theta\cos\varphi \\ \noalign{\medskip}
                  \sin\theta\sin\varphi \\ \noalign{\medskip}
                  \cos\theta  \end{array}\right)
\end{equation}
where $\theta$ and $\varphi$ are randomly distributed angles.

Finally, conservation of mass, momentum and energy is enforced by subtracting the corresponding injected amount from the gas.
This compensation procedure usually produce small variations and test runs without it show neglible variations.

After the injection process has completed, the tracer $\T$ is again reset to zero everywhere and a new accumulation cycle begins (step 2).
\end{enumerate}

Our injection prescription is independent of the shape and position of the shock front and, as such, it can easily adapts to curved and corrugated fronts.

During the first phase of injection ($t \lesssim t_i = 480\, \Omega_L^{-1}$), CR streaming is effective only in triggering the onset of the Bell instability as turbulent fluctuations are still small.
As pointed out by \cite{Bai_etal.2015}, particles giving rise to this transient flow do not participate in the shock acceleration process and are removed for $t > 2t_i$ in order to suppress spurious  effects once the Bell instability is fully developed.

\subsubsection{Non-Relativistic Regime}
%

\begin{figure*}[!ht]
  \centering
  \includegraphics[width=\textwidth]{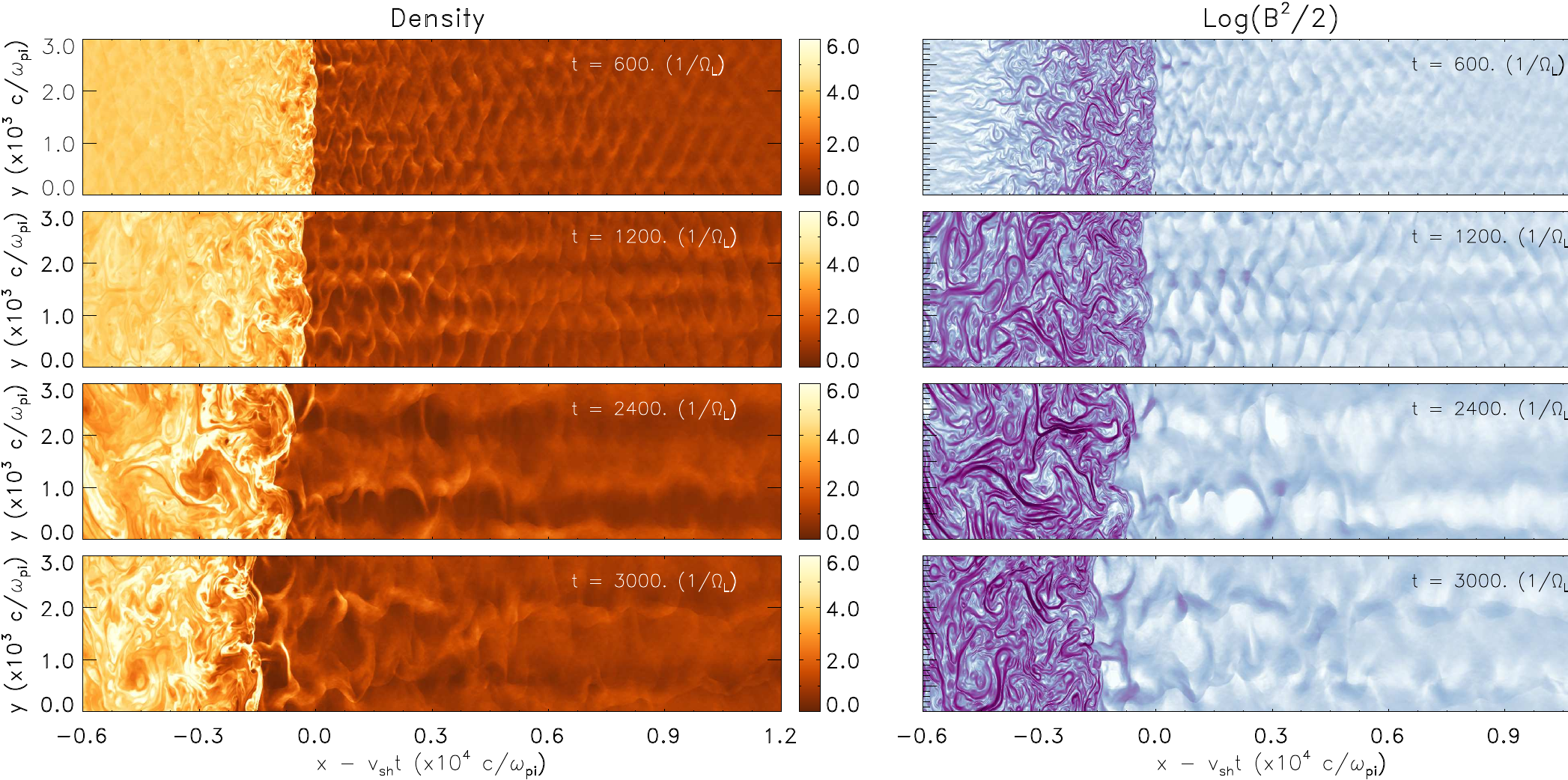}
  \caption{\footnotesize Density (left) and magnetic pressure (in log scale,
           right) snapshots for the collisionless shock problem (run R2) at
           four different times (reported in the panels).
           Only a reduced portion of the domain, in proximity of the unperturbed
           shock position $x_s = v_{\rm sh}t$, is shown. \label{fig:cs_evolution_R2}}
\end{figure*}
At the beginning, a shock is formed and reflected away from the wall at the left boundary.
CR particles injected at this early stages travel almost undisturbed along magnetic field lines without being efficiently scattered and propagate away from the shock.
The streaming of CRs in the upstream region triggers the Bell instability which grows linearly for a few hundreds Larmor periods.
As the instability enters the nonlinear stage, magnetic field fluctuations are amplified by a factor $\sim 4$ in the upstream region and a filamentary-like structure, alternating low and high density regions, becomes evident.
Snapshots of the evolution, showing both density and magnetic field, are given in Figure \ref{fig:cs_evolution_R2}.
Note that only a smaller portion of the computation domain is shown.
Magnetic field and density inhomogeneities are then further amplified once they cross the shock front enhancing strong turbulence in the downstream region for $t\gtrsim 1.2\times10^{3} \,\Omega_L^{-1}$.

\begin{figure}[!ht]
  \centering
  \includegraphics[width=0.5\textwidth]{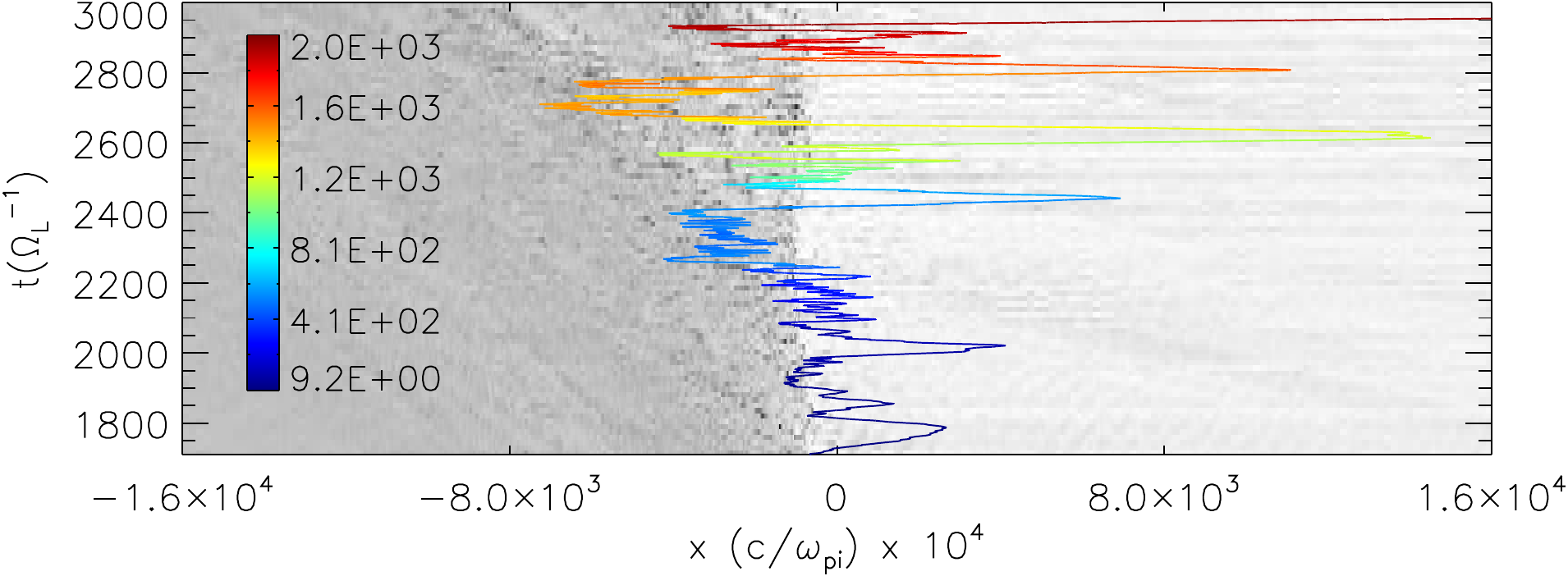}
  \caption{\footnotesize Space-time diagram in the $(x,t)$ plane showing the
           particle acceleration process.
           The coloured line gives the particle trajectory and the color
           indicate its specific kinetic energy.
           The background map in gray shows the $y-$averaged density structure
           of system at different times.
           \label{fig:cs_trajectory}}
\end{figure}

CR particles begin to be efficiently scattered at this stage and the diffusive shock-acceleration process commences.
Magnetic clumps provide the scattering centers and most of the particles suffer multiple head-on collisions across the shock resulting in a fractional energy gain. 
This process is best illustrated in Figure \ref{fig:cs_trajectory} where we show the space-time diagram of one among the most energetic particles in the reference frame in which the shock is stationary.
The color of the line indicates the particle energy as time advances while the background gray colormap is composed by superposing one-dimensional horizontal density profiles taken at the the particle $y$ coordinate.
\begin{figure}[!ht]
  \centering
  \includegraphics[width=0.5\textwidth]{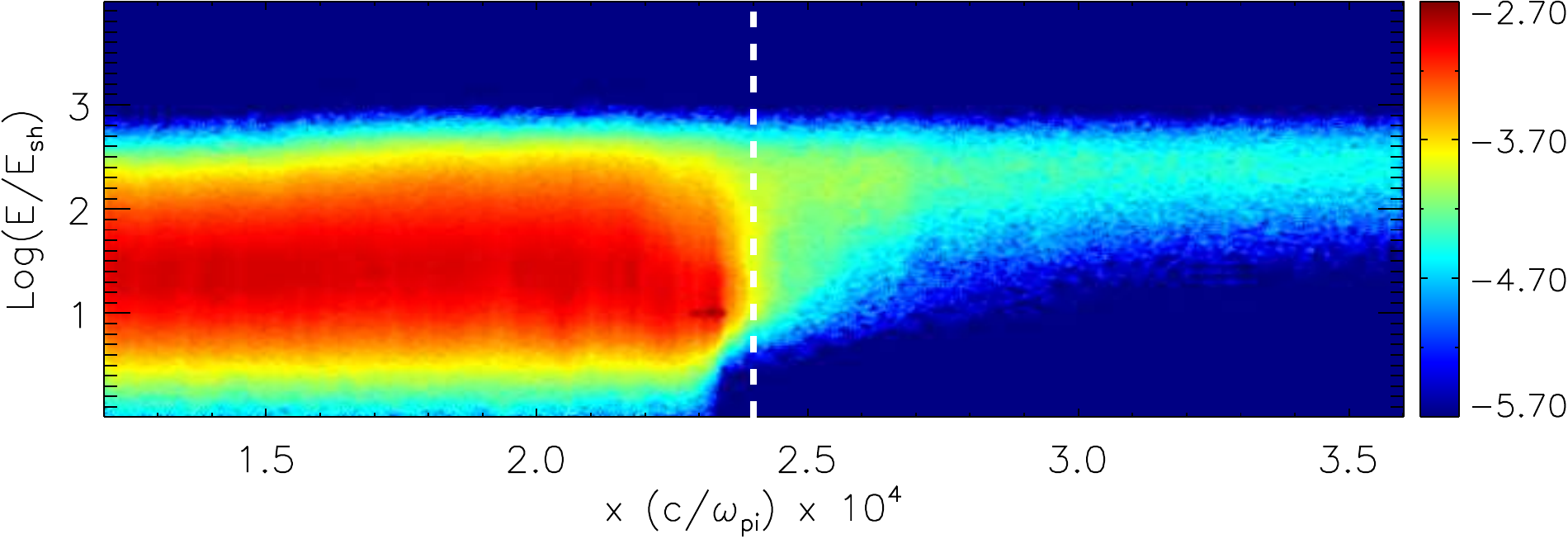}
  \includegraphics[width=0.5\textwidth]{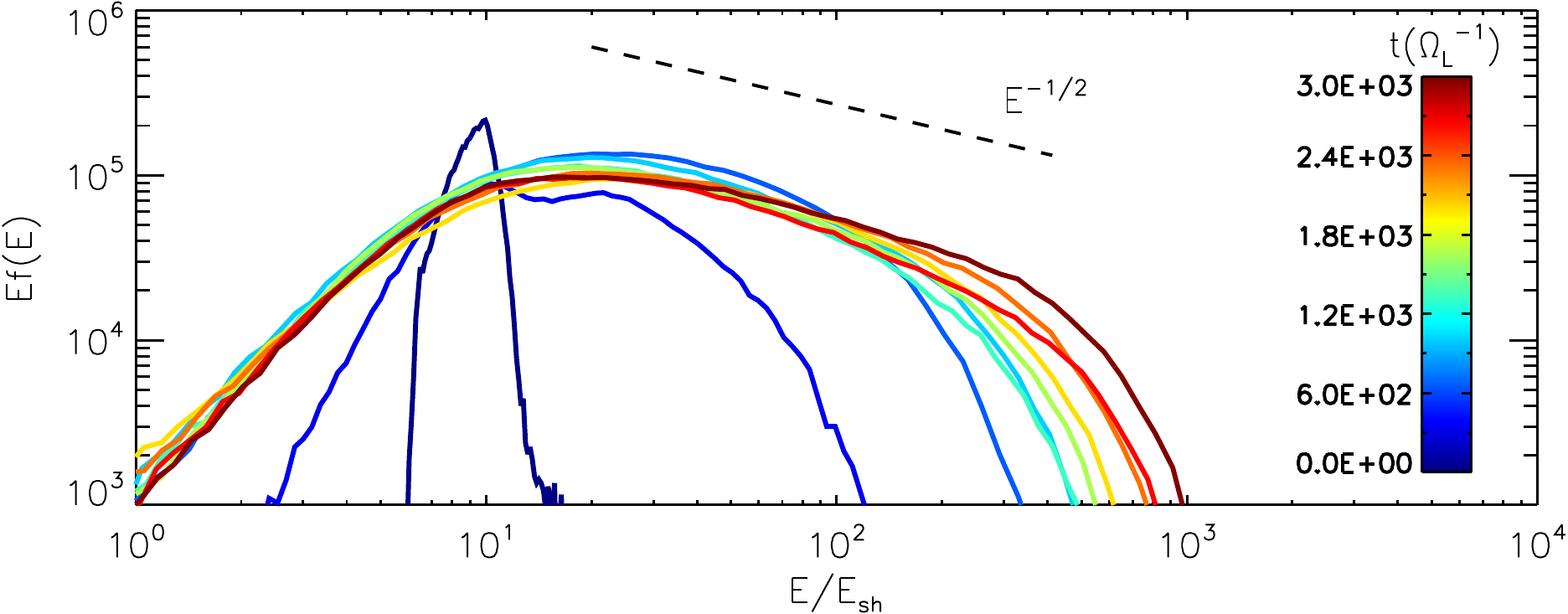}
  \caption{\footnotesize Top panel: particle energy distribution (in units of
           $E_{\rm sh} = v_0^2/2$) as a function of $x$ at $t=2400\Omega^{-1}_L$
           for the non-relativistic run R2.
           The white vertical dashed line gives the (unperturbed) position of
           the shock front.
           Only a smaller region around the shock is shown.
           Bottom panel: time evolution of the particle energy spectrum $Ef(E)$
           as a function of $E$.
           The different colors correspond to different simulation times and the
           spectrum is extracted by considering particles lying in a narrow strip
           of width $\approx 800, c/\omega_{\rm pi}$ in the downstream region.
           The black dahsed line shows the slope predicted by the Fermi
            acceleration model.
           \label{fig:cs_spectra_R2}}
\end{figure}

In the top panel of Figure \ref{fig:cs_spectra_R2} we show the energy spectrum $Ef(E)$ as a function of the horizontal coordinate $x$ and energy $E$ (in units of $E_{\rm sh}$) at $t = 2400\,\Omega_L^{-1}$.
The two-dimensional distribution is constructed by taking, for each $x$ coordinate, the spectra of all particles lying in a narrow vertical stripe which is $4$ zones wide. 
A tail of high energy particles penetrating into the shock upstream and driving the Bell instability is visible, in agreement with previous results\cite[see, e.g.,][and references therein]{Caprioli_Spitkovsky.2014a, Bai_etal.2015}.
Note that since our injection procedure tracks the shock front more accurately, no artificial protrusion appears for $E \approx 10 E_{\rm sh}$.

The particle spectrum is extracted from a narrow strip $[x_L,x_L+\delta]$ behind the shock, where $x_L = x_s - 2400\, c/\omega_{\rm pi}$ while $\delta = 800\, c/\omega_{\rm pi}$.
The distribution function is normalized to the number of particles, i.e., $\int f(E)\, dE = N_{[x_L,x_L+\delta]}$.
For isotropic scattering the expected particle distribution $f(E)$ should depend only on the compression ratio $r$ and take the form $f(E) \sim E^{(1-q)/2}$, where $q=3r/(r-1)$.
In the limit of strong shocks, one retrieves $Ef(E)\sim E^{-1/2}$ and this prediction is confirmed by the time evolution of the energy spectrum plotted at different times in the bottom panel of Figure \ref{fig:cs_spectra_R2}.
The plot indicates that the CR spectrum gradually broadens from the injected distribution (dark blue curve peaked around $\sim 10E_{\rm sh}$) towards a high energy power-law tail with slope consistent with $-3/2$.
A high energy cutoff at $\sim 10^3 E_{\rm sh}$ is reached towards the end of the simulation, in agreement with previous findings \citep[see, e.g.,][]{Caprioli_Spitkovsky.2013, Caprioli_Spitkovsky.2014a} and with the results of \cite{Bai_etal.2015}.

\subsubsection{Relativistic Regime}
%

We have further investigated the diffusive shock acceleration mechanism in the relativistic regime by repeating the R2-REL run discussed in \cite{Bai_etal.2015}.
A reduced value of the speed of light ($\C = 10\sqrt{2}v_0$) has been chosen in order to favour the transition from a non-relativistic injection condition to the final acceleration stage, where the most energetic particles become relativistic.
The typical particle velocity at injection is, indeed, $v_p = \sqrt{10}v_0 \approx 0.22\C$ corresponding to $\gamma_p \approx 1.026$.
The transition to the relativistic regime occurs at approximately $E_t\approx \C^2/2$ when $\gamma_t\approx 1.5$.

\begin{figure*}[!ht]
  \centering
  \includegraphics[width=\textwidth]{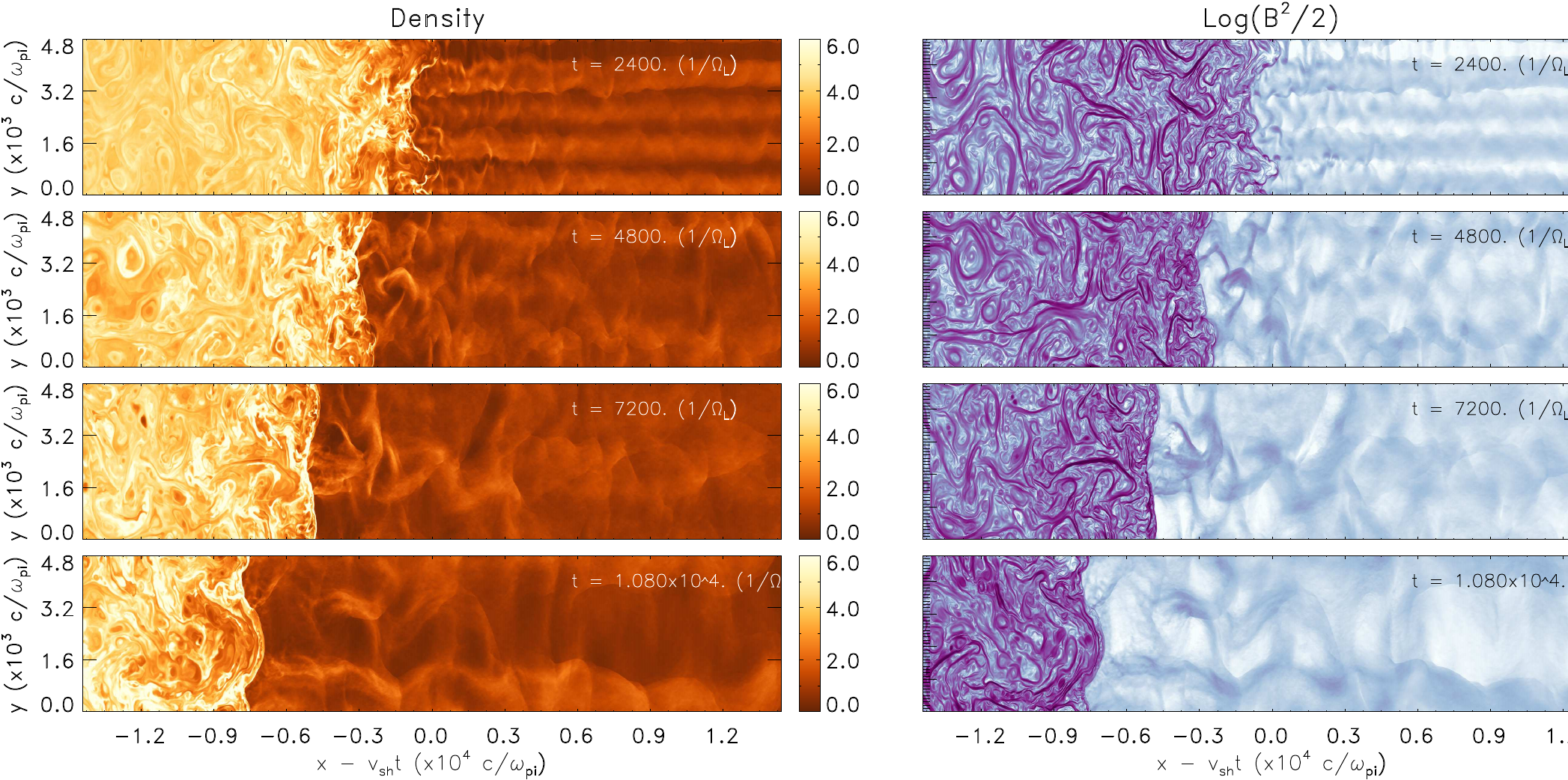}
  \caption{\footnotesize Density (left) and magnetic pressure (in log scale,
           right) snapshots for the collisionless shock problem in the
           relativistic run R2-REL, at four different
           times  (reported in the titles).
           Shown here is only a smaller portion of the domain centered around
           the unperturbed shock position $x_s = v_st$.
           \label{fig:cs_evolution_R2rel}}
\end{figure*}
Density and magnetic field strength are shown in the left and right panels of Figure \ref{fig:cs_evolution_R2rel} at different times.
Upstream of the shock we observe the formation of cavities and filamentary structures of larger size when compared to run R2, motivating the choice of a larger computational box.
This behavior can be attributed to the saturation of the CR current density which depends on the velocity of the particles and, for a reduced value of the speed of light, can not exceed $q_\CR \C$.
In other words, at relativistic velocities, an increase in the particles' energy does not correspond to an increase in the current density.
Indeed, from the linear analysis of the Bell instability (Equation \ref{eq:Bell_k0}), we expect the most unstable wavenumber to be smaller in the relativistic case. 
Similarly, the level of turbulence is somewhat reduced  and a sharper shock transition layer is formed, in agreement with the results of \cite{Bai_etal.2015}.

As a significant fraction of the fluid energy is transferred to CR during the acceleration process and the effective adiabatic index of the fluid decreases from its nominal value $5/3$ \citep[see Section 6.2 of][for a thorough discussion]{Caprioli_Spitkovsky.2014a} to a smaller value $\tilde{\gamma}$.
As a consequence, the shock compression ratio becomes slightly larger ($r\approx 4.2$) towards the end of the simulation.
In addition, since we expect $v_{\rm sh} = - (\tilde{\gamma}-1)/2v_0$ to hold for a strong shock, the front slows down and straggles with respect to its nominal position.
This can be clearly observed in the snapshots sequence in Figure \ref{fig:cs_evolution_R2rel}.  

%
%
%
%

\begin{figure}[!ht]
  \centering
  \includegraphics[width=0.5\textwidth]{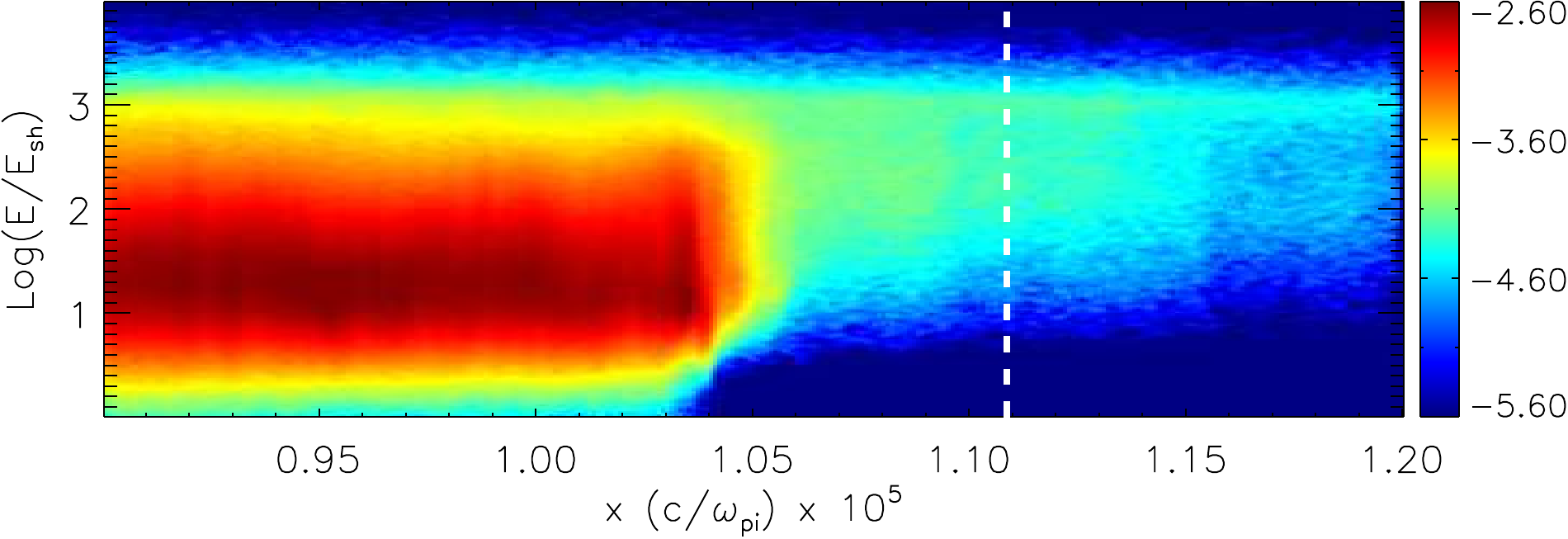}
  \includegraphics[width=0.5\textwidth]{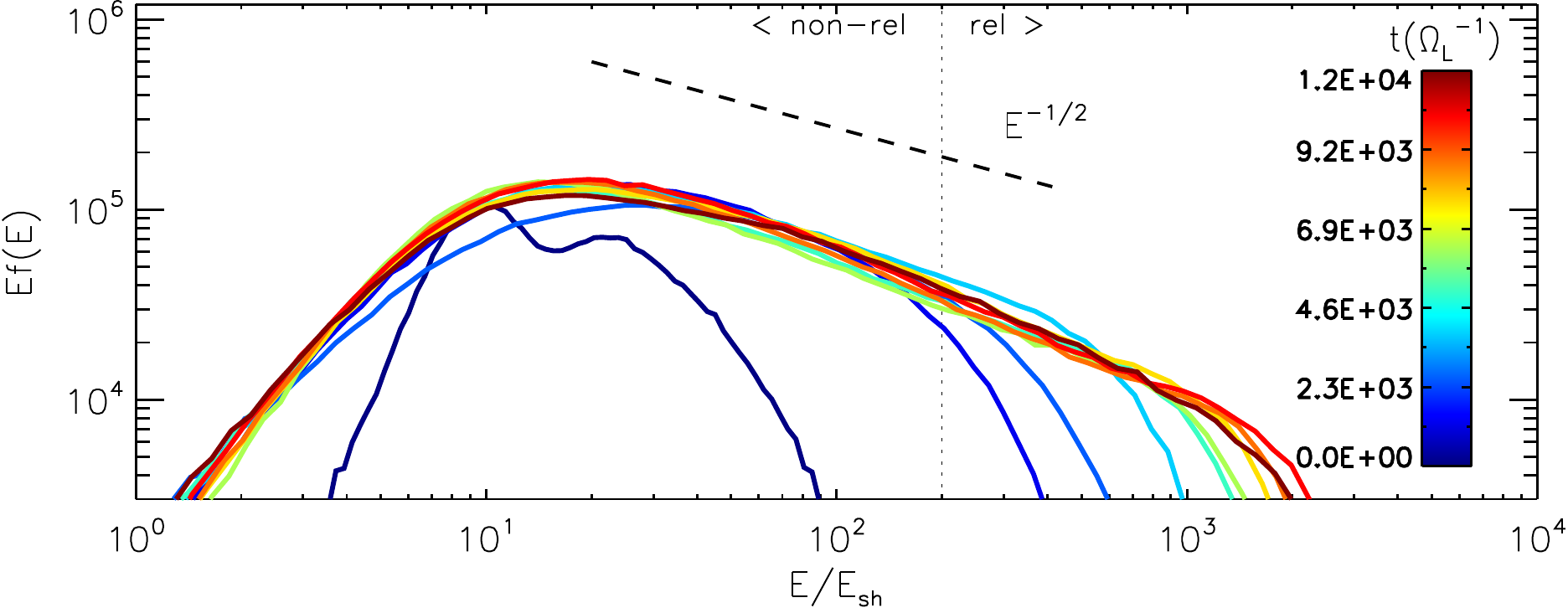}
  \includegraphics[width=0.5\textwidth]{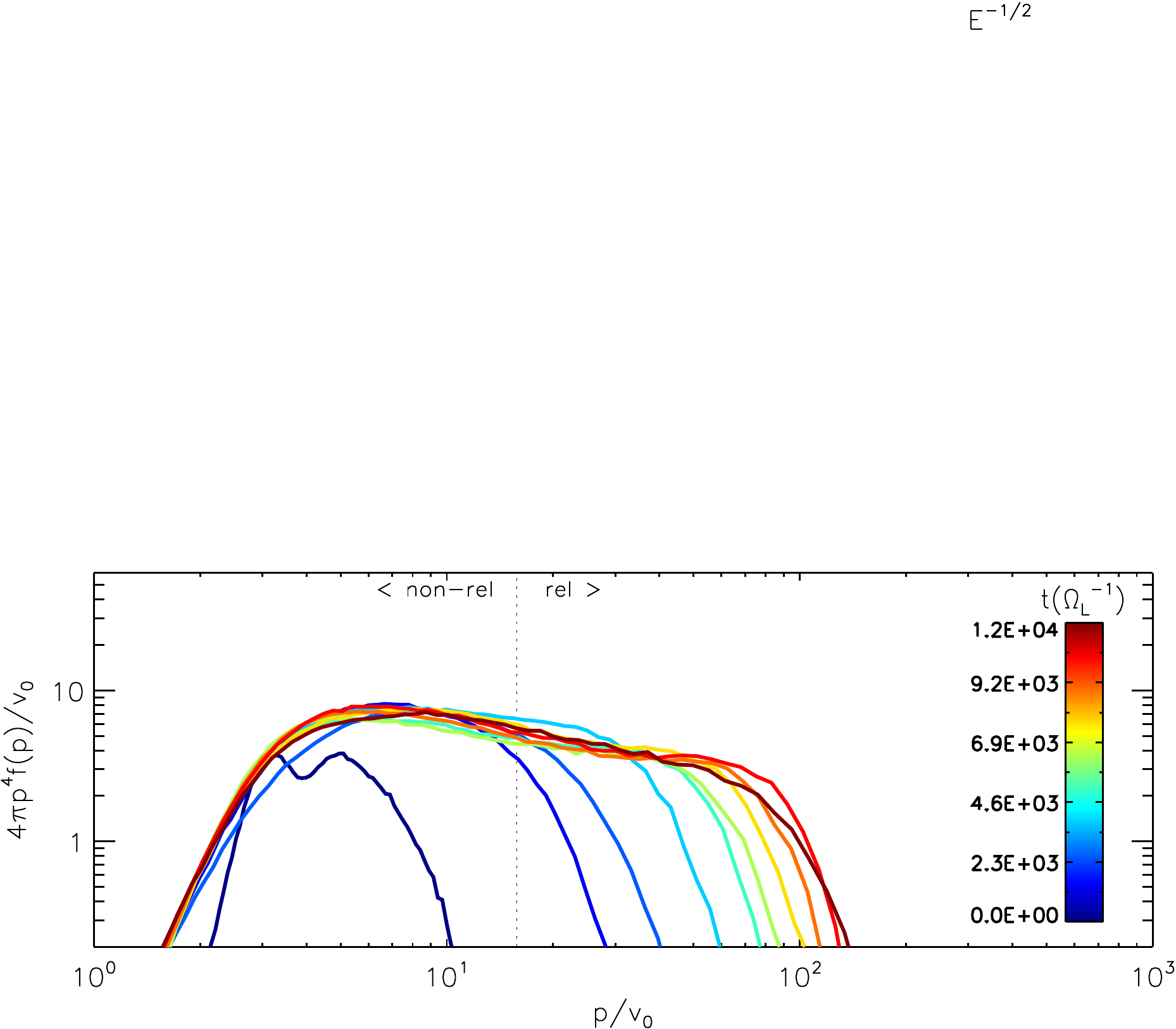}
  \caption{\footnotesize Top: spatial distribution of the particle energy
           (in units of $E_{\rm sh} = v_0^2/2$) as a function of $x$ and $E$
           at $\Omega_Lt=11088$.
           The white vertical dashed line gives the (unperturbed) position of
           the shock front.
           Only a smaller region around the shock is shown.
           Middle panel: time history of the energy spectrum $Ef(E)$ in
           dimensionless form.
           Curves with different colors correspond to the times indicated
           by the legend.
           Bottom panel: momentum spectrum $p^4f(p)$ as a function of $p/v_0$.
           The thin vertical dotted lines mark the transition from non-relativistic
           ($E \lesssim \C^2/2$) to relativistic energies ($E\gtrsim \C^2/2$) 
           while the black dashed line in the top panel represents the
           theoretical expected slope in the classical regime.
           \label{fig:cs_spectra_R2rel}}
\end{figure}

The energy and momentum distributions of CRs are shown in the three panels of Figure \ref{fig:cs_spectra_R2rel}. 
In the top one, we show a 2D color map of the spatial distribution of $Ef(E)$ at $t \approx 11088 \Omega_L^{-1}$ obtained by averaging, for each $x$, particles lying in a narrow vertical strip 8-zones wide.
From the figure we see that most particles escaping into the upstream region have energy in excess of $10^2 E_{\rm sh}$ ($\gamma_p \gtrsim 1.25$).
Again, since particle injection tracks more accurately the location of the shock front, we do not observe any low-energy protrusion in the upstream region.

In the middle panel of Figure \ref{fig:cs_spectra_R2rel} we plot the time history of the energy spectrum extracted by averaging, as before, all CR particles lying in a narrow strip located at a distance $\approx 2400 c/\omega_{pi}$ behind the actual shock position.
The spectrum is again consistent with a power-law with spectral index $-3/2$ and presents a cut-off at $E\approx 2\times 10^3$ ($\gamma\approx3.5$).
Note that a thin vertical line marks the transition from the non-relativistic to relativistic energies ($E_t\approx \C^2/2$).

We also compute the momentum spectrum $f(p)$ which is related to the energy distribution $f(E)$ through the transformation
\begin{equation}\label{eq:fE_fp}
  f(E_k) = 4\pi p^2 f(p) \frac{dp}{dE_k}\,,
\end{equation}
where $p\equiv \gamma_p v_p$ is the particle momentum per unit mass while
\begin{equation}\label{eq:energy_momentum}
  E_k = (\gamma-1)\C^2 = \C^2\sqrt{1 + \left(\frac{p}{\C}\right)^2} - \C^2
\end{equation}
is the specific kinetic energy.
Using $dE_k/dp = p\C^2/(E_k + \C^2)$ we invert Equation (\ref{eq:fE_fp}) to obtain $f(p)$ and plot the time history in in the bottom panel of Figure \ref{fig:cs_spectra_R2rel}.
The figure shows $p^4f(p)/v_0$ as a function of $p$ and reveals an approximately flat curve in the region $p/v_0 \lesssim 15.8$ (the non-relativistic region) and in $p/v_0 \gtrsim 25$ (the relativistic region).

We remind the reader that the Fermi acceleration theory predicts that the momentum spectrum should scale universally as $f(p) \propto p^{-4}$ at relativistic and non-relativistic energies.
Then, from Equation (\ref{eq:fE_fp}), one expects the energy distribution to smoothly change slope while shifting to higher energies:
\begin{equation}
  f(E_k) \propto  
  \left\{\begin{array}{lcl}
    \DS E_k^{-3/2} &\quad {\rm if}\quad & E_k \ll \C^2 \\ \noalign{\medskip}
    \DS E_k^{-2}   &\quad {\rm if}\quad & E_k \gg \C^2 
  \end{array}\right.
\end{equation}
in the non-relativistic and relativistic parts of the spectrum, respectively.
The transition to $f(E)\propto E^{-2}$ should take place at $\gamma \approx 10$ but  it cannot be captured by the present simulation since such high energies have not been reached yet.


\subsection{Particle Acceleration near an X-point}
%

Next we consider, as a proof of concept, test particle acceleration near an X-type magnetic reconnection region.
Relativistic magnetic reconnection in strongly magnetized environments has been pointed out \citep{Sironi_Spitkovsky.2014} as an efficient particle acceleration process which may be able to account for high-energy nonthermal emission from pulsar wind nebulae \citep[PWN, see e.g.][and references therein]{CUB.2013}, active galactic nuclei \citep[AGN, see e.g.][]{Giannios.2013} and gamma-ray burst \citep{McKUzd.2012}.

Our setup is similar to \cite{Mori_etal.1998} and consists of a 2-D computational square with $-2L\le x,y \le 2L$, threaded by magnetic and electric fields given by  
\begin{equation}
  \vec{B} = B_0\left(\frac{y}{L},\, \frac{x}{L},\, \frac{B_z}{B_0}\right) \,,\quad
  \vec{E} = (0,\, 0,\, E_z).
\end{equation}
where $B_0 = 1$.
We choose the Alfv{\'e}n speed as our reference velocity and set the speed of light to be $\C = 100 v_A$.
Lengths are normalized to the gyration radius $v_A/\Omega_L$ and we set $L = 2\times10^3$.
Computations are stopped at $t=100$ employing $512^2$ grid zones with $4$ particles per cell.
The particle velocity distribution is initialized to a Maxwellian distribution with thermal velocity $0.1v_A$.
Only particles are evolved in time while fluid and electromagnetic quantities are kept constant to their initial values.

\begin{figure}[!h]
  \centering
  \includegraphics*[width=0.5\textwidth]{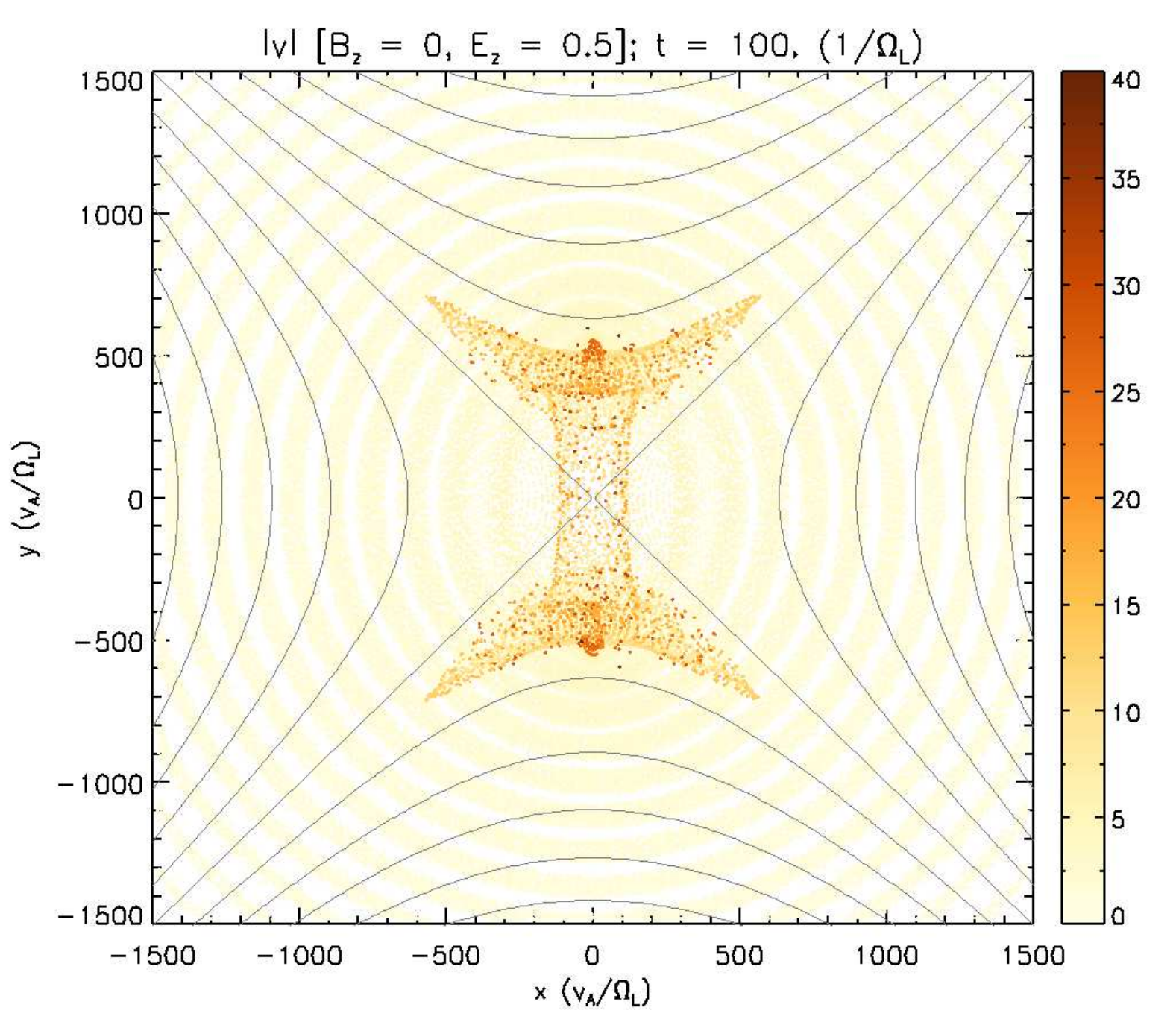}
  \includegraphics*[width=0.5\textwidth]{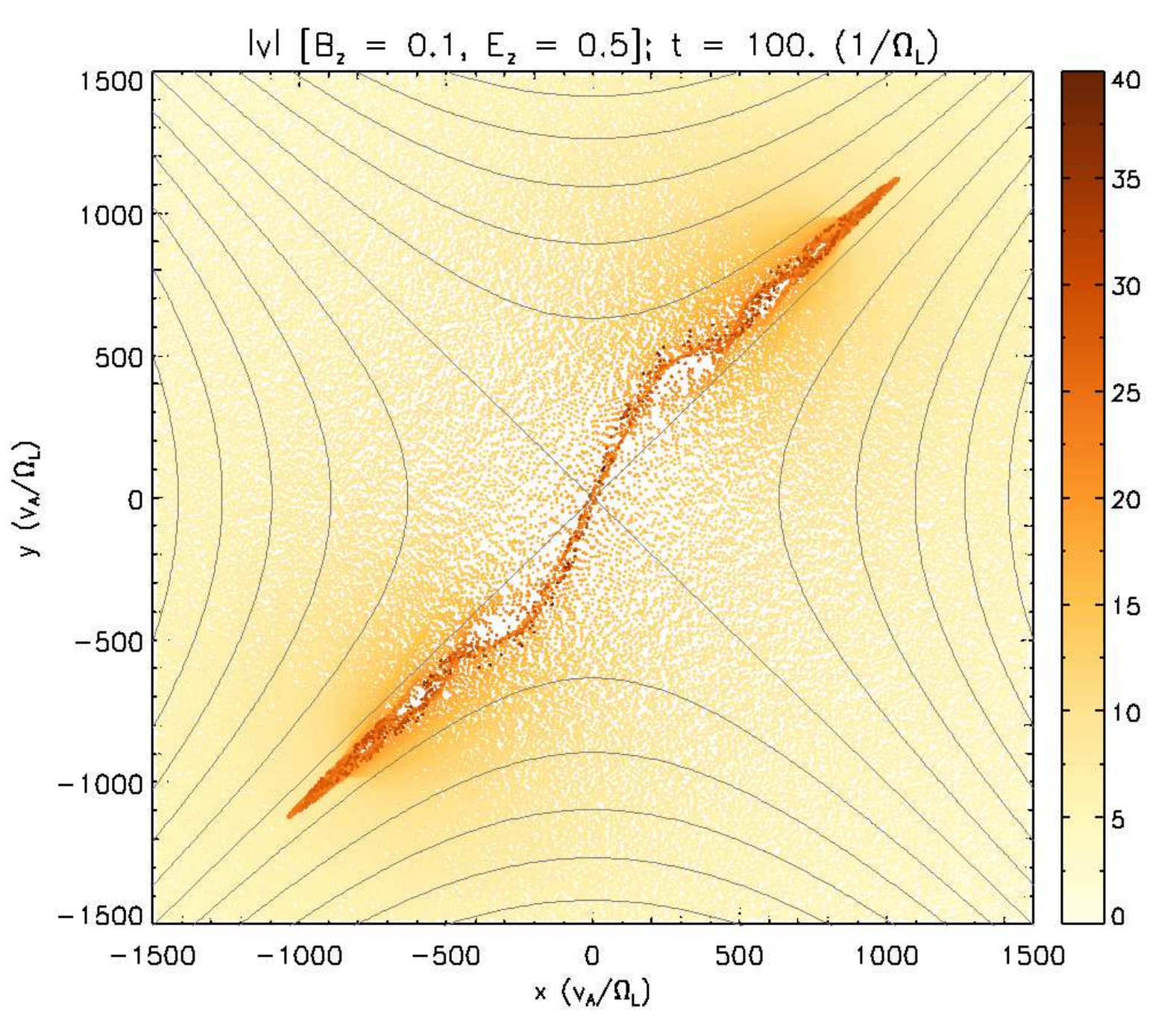}
  \caption{\footnotesize Test particle distribution for the X-point
  acceleration problem.
  Magnetic field lines in the $x-y$ plane are drawn using black lines while
  particles are coloured in orange by velocity magnitude.
  The top and bottom panels correspond, respectively, to the zero guide field
  case ($B_z = 0$) and to the guide field case with $B_z = 0.1$.
  In both cases the electric field $E_z=0.5$, directed out of the plane.
  \label{fig:xpoint}}
\end{figure}

\begin{figure}[!h]
  \centering
  \includegraphics[width=0.5\textwidth]{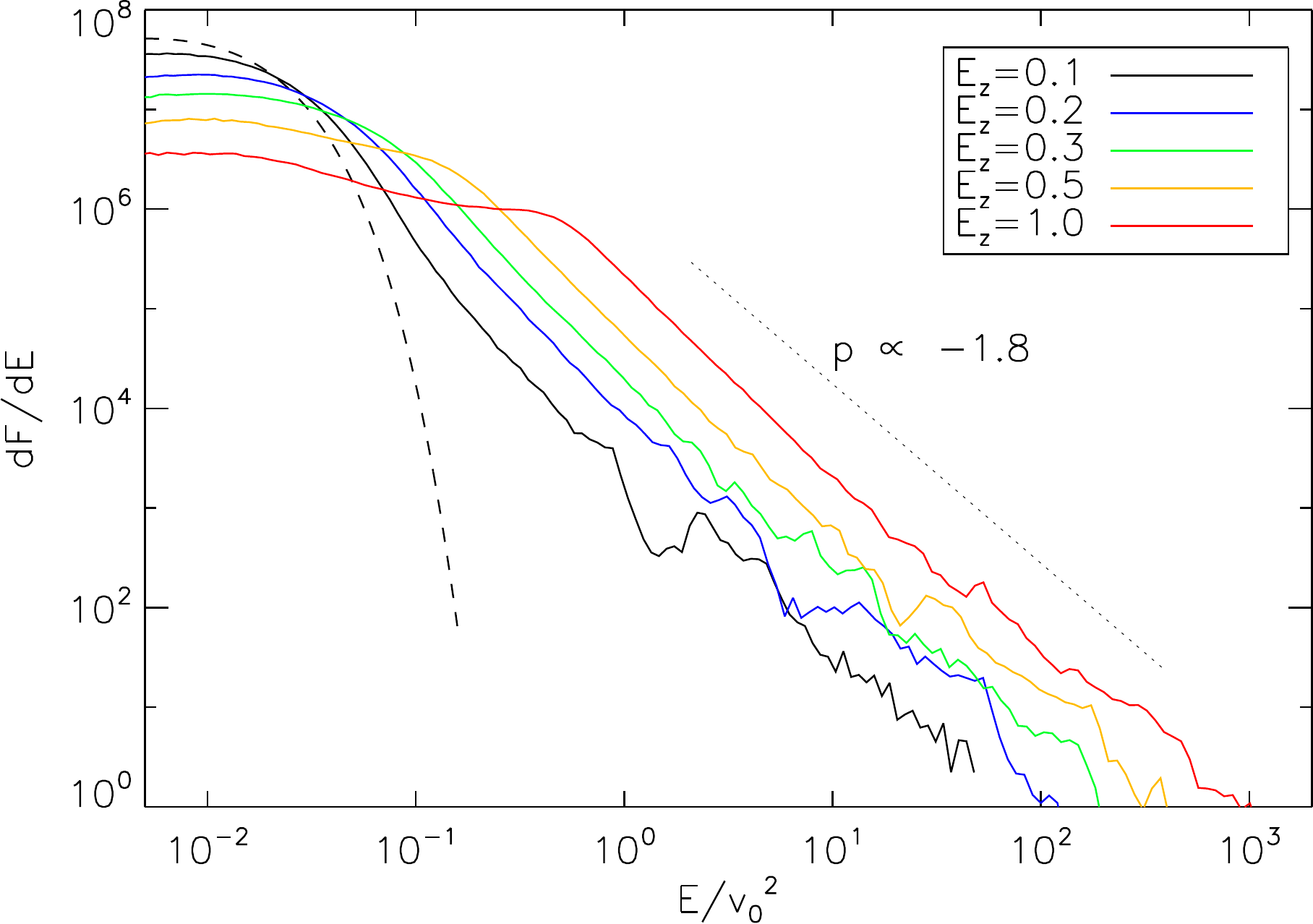}
  \includegraphics[width=0.5\textwidth]{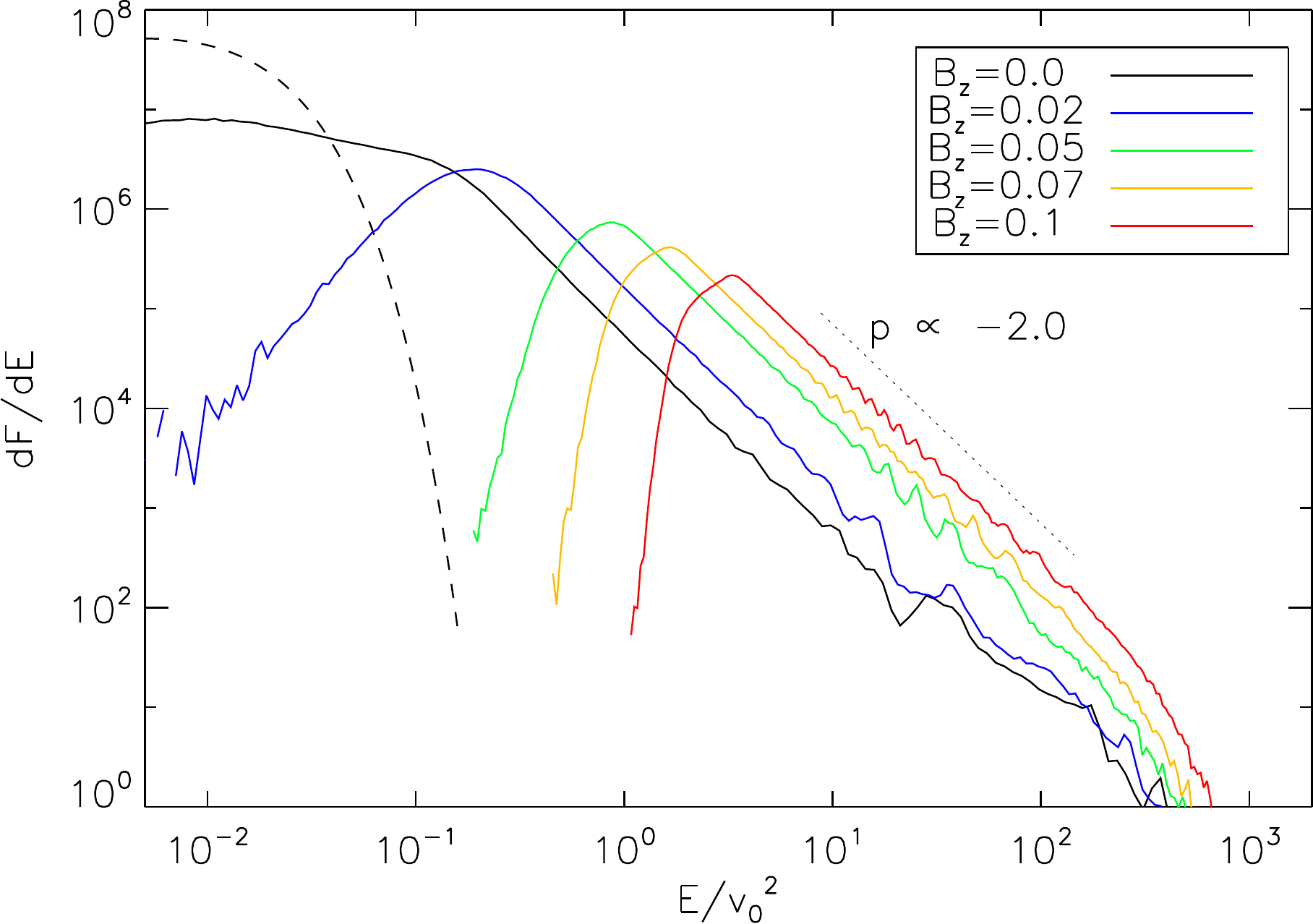}
  \caption{\footnotesize Particles energy spectra without (top panel) and
   with (bottom panel) the guide field.
   In the case with guide field, the electric field  $E_{z}$ is set to 0.5.
   The dashed line indicates the initial Maxwellian distribution whereas the
   the dotted line represents a power law with index -1.8 (top) and -2.0 
  (bottom).\label{fig:xpoint_spectra}}
\end{figure}

We first consider a configuration without a guide field ($B_z=0$) and vary the electric field strengths according to  $E_z = 0.1,\,0.2,\,0.3,\,0.5,\,1.0$.
Since $\vec{E}\cdot\vec{B}=0$ everywhere, particle acceleration takes place mostly in proximity of the null point where the electric field has a larger amplitude than the magnetic field. 
Outside of this region, no significant acceleration occurs.
Particle motion results from a combination of curvature, gradient and $\vec{E}\times\vec{B}$ drifts and produces a symmetric pattern with respect to the $y-$ axis, as shown in the top panel of Figure \ref{fig:xpoint}.
Owing to the perpendicular electric drift, particles with a large velocity in the $|y| < |x|$ regions have a bouncing oscillatory motion between the two separatrices while approaching the central X-point \cite[see][]{Vekstein_Browning.1997, Browning_Vekstein.2001}.
In these regions, the curvature and gradient drift have opposite direction with respect to the electric field and are thus unfavourable to acceleration.
Once the separatrix line is crossed, the situation is reversed and particles in the region $|y| > |x|$ move away from the null point because of the $\vec{E}\times\vec{B}$ drift.
Concurrently, the curvature and gradient drift take place in the positive $z$ direction and particles gain energy due to the strong electric field, thus producing the pattern of higher energy particle observed in the top panel of Figure \ref{fig:xpoint}.
A similar pattern is also shown by \cite{Mori_etal.1998}. 
The energy spectrum, plotted in the top panel of Figure \ref{fig:xpoint_spectra}, shows a high-energy tail that departs from the Maxwellian and extends to larger energies as the electric field is increased.
For strong electric fields ($E_z \gtrsim 0.3$) we observe that the energy distribution can be well approximated with a power law $\propto E^{-p}$ with a spectral index $p\approx 1.8$ (for $E_z = 1$).
This result is slightly smaller than the one found by \cite{Mori_etal.1998} who found $p\sim 2$.

In the second configuration we fix the value of the electric field to $E_z = 0.5$ and repeat the computations using different values of the guide field, $B_z = 0,\, 0.02,\, 0.05,\, 0.07,\, 0.1$.
The spatial distribution, shown in the bottom panel of Figure \ref{fig:xpoint}, indicates that the presence of a non-zero guide field breaks the symmetry with respect to the $y$-axis  and the most energetic particles distribute on an elongated stripe approximately laying along the separatrix line $y=x$ which is determined by the sign of the parallel components of the electric and magnetic fields.
The same behaviour has been reported by previous investigations, e.g. \cite{Browning_Vekstein.2001}.
As the guide field becomes stronger, parallel acceleration (as discussed in \S\ref{sec:parallel_EB}) increases and becomes significant.
In fact, since $\vec{E}\cdot\vec{B}\ne 0$ everywhere, acceleration takes place for all particles including those away from the null point.
This can be clearly seen in the energy spectra (bottom panel in Figure \ref{fig:xpoint_spectra}) showing a systematic shift to larger energies as the amplitude of the guide field grows.
The effects of the perpendicular drift on the parallel motion remain still relevant so that largest acceleration are observed close to the origin.
However, the strength of the guide field seems to affect more the low-energy part of the spectra rather than the high-energy cutoff.
Again we observe that the high-energy tail of the spectrum behaves as a power-law with spectral index, for $B_z = 0.1$) $p\sim 2$, in agreement with \cite{Mori_etal.1998}.

\subsection{Code Performance and Parallel Scaling}
%
%
%
%
%

Let $\Delta t_p$ and $\Delta t_h$ be, respectively, the computational time required to update a single particle and a single grid zone using the MHD solver.
The total CPU time for a single cell update may then be approximately expressed by
\begin{equation}
  \Delta t =\Big(m\Delta t_p + \Delta t_h\Big)\,,
\end{equation}
where $m$ is the number of particles per cell.
In order to measure $\Delta t_p$ we have repeated the same computation with different values of $m$ while leaving all other parameters unchanged so that
\begin{equation}\label{eq:delta_p}
  \Delta t_p \approx 
                     \frac{\Delta t_1 - \Delta t_2}{m_{1} - m_{2}} \,,
\end{equation}
where $m_{1}$ and $m_{2}$ are different numbers of particles per cell while $\Delta t_1$ and $\Delta t_2$ are the corresponding single cell integration times.
Code performance has been benchmarked on a 3 GHz Intel Xeon E5 processor using the relative drift test (Section \ref{sec:relative_drift}) without subcycling and grid resolution of $64$ zones in each direction.

\begin{figure*}
  \centering
  \includegraphics[width=0.48\textwidth]{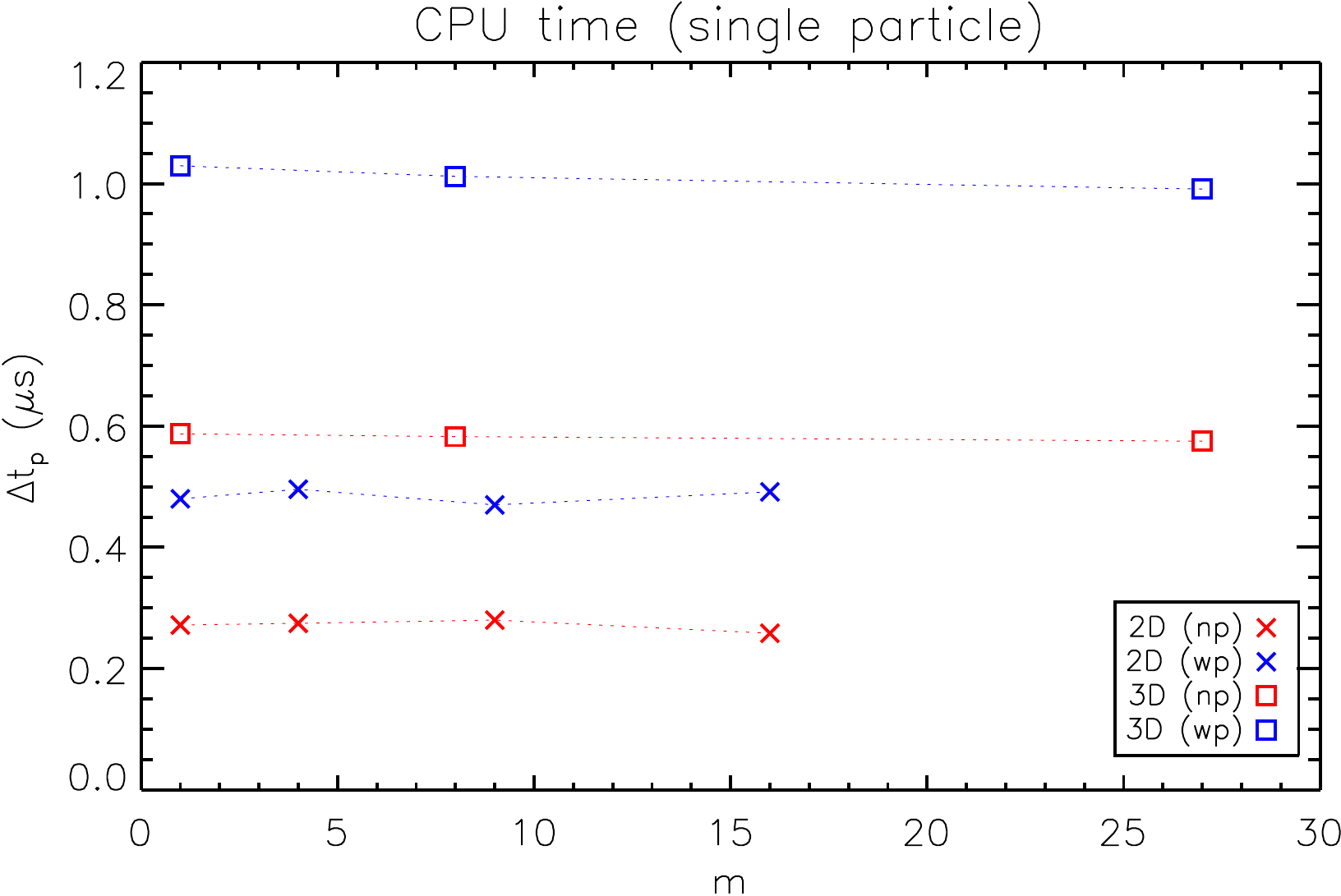}%
  \includegraphics[width=0.48\textwidth]{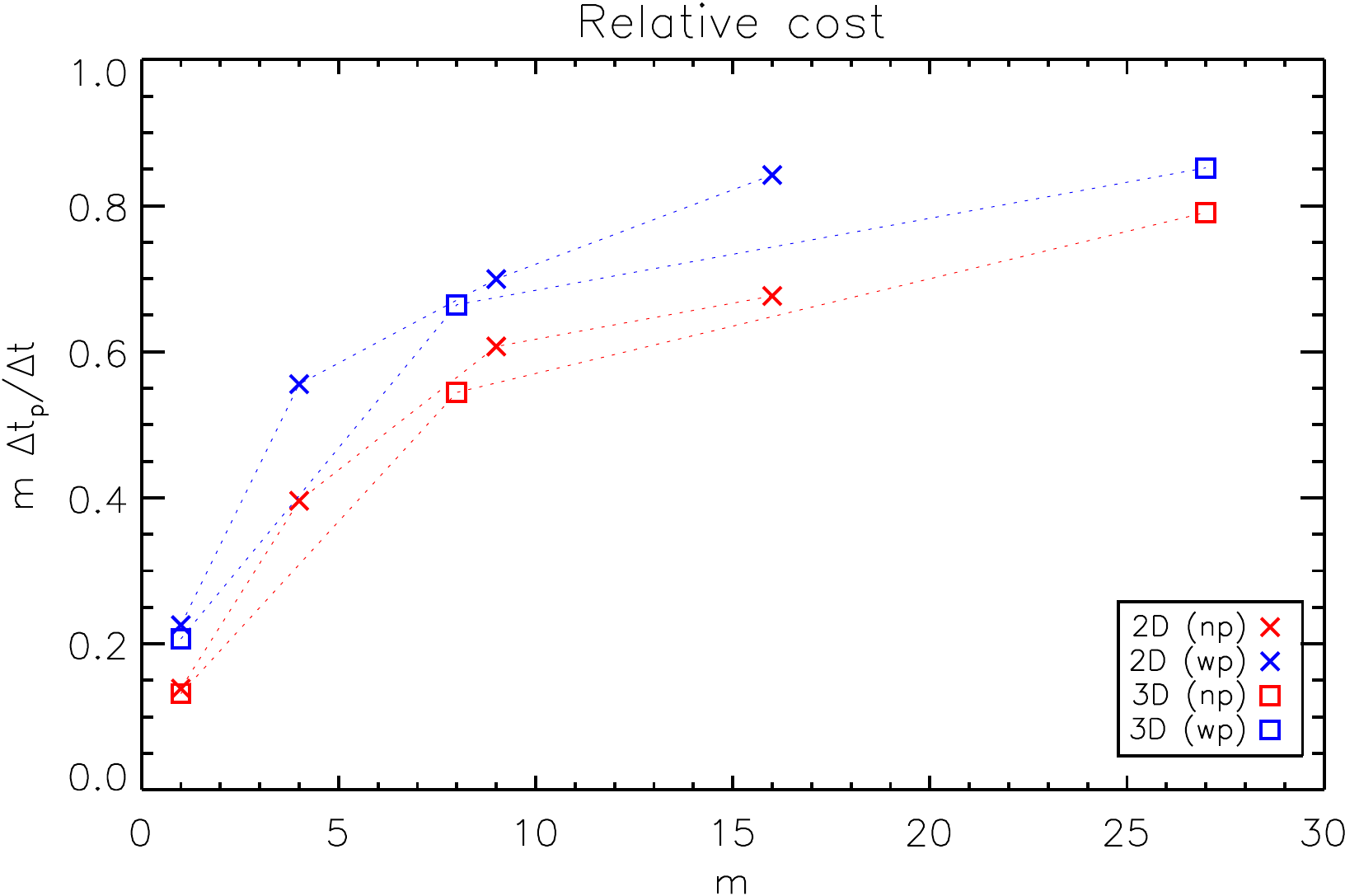}
  \caption{\footnotesize Left panel: single particle CPU integration time (per
   time step) without predictor (red symbols, \lq\texttt{np}\rq) and with the
   predictor step (blue symbols, \lq\texttt{wp}\rq).
   Crosses (squares) corresponds to 2D (3D) computations.
   Right: relative cost between particle integration time and overall
   CPU time for a single cell update.
   \label{fig:timing}}
\end{figure*}
The left panel in Figure (\ref{fig:timing}) shows the CPU time (in $\mu s$) computed using Equation (\ref{eq:delta_p}) with and without the predictor step (\lq\texttt{wp}\rq and \lq\texttt{np}\rq, respectively).
As expected, the computational time is essentially independent of $m$ and, on average, we find $\Delta t_p \approx 0.3\, \mu{\rm s}$ (in 2D) and $\Delta t_p \approx 0.6 \,\mu{\rm s}$ (in 3D).
We then include the predictor step (\S \ref{sec:CR_predictor}, blue symbols in the figure) and observe an average increase of $\sim 70\%$.
The right panel of Fig. (\ref{fig:timing}) shows $m\Delta t_p/\Delta t$ - the particle CPU time relative to a single cell update - as a function of $m$.
With $m \approx 6-7$ particle per cells, the code spends $\approx 50 \%$ of the total computational time in evolving the particles (without predictor step), in both 2D and 3D.
Inclusion of the predictor step leads again to an increase of the relative cost.

\begin{figure}
  \centering
  \includegraphics[width=0.5\textwidth]{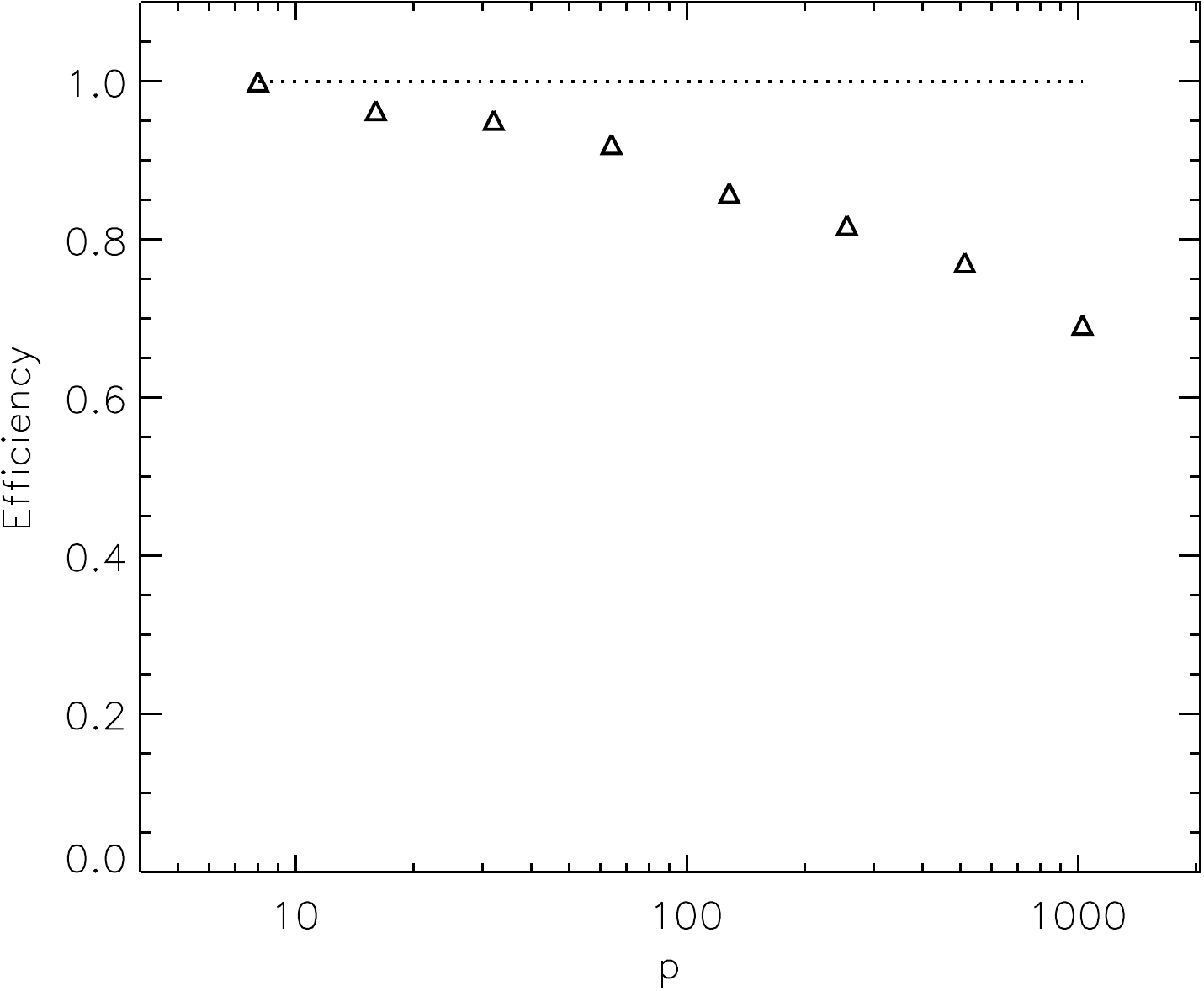}
  \caption{\footnotesize Parallel efficiency $E = \Delta T_1/(p\Delta T_{p})$, where
  $\Delta T_{p}$ is the computational time obtained with $p$ processors
  and the normalization has been chosen so that $\Delta T_1 = 8\Delta T_8$.
  The test under consideration is the 3D Bell instability test problem with
  final integration time $t=10$ and a resolution $256\times128^2$.
  The thin dotted line gives the ideal scaling ($E = 1$).
  \label{fig:parallel_scaling}}
\end{figure}

The MHD-PIC module has been parallelized using the Message Passing Interface (MPI) library.
In our implementation each processor updates only the particles lying on its physical domain \citep{Vaidya_etal.2016}.
Particles must be transferred between neighbors when they cross a processor boundary: in such a way, each processor communicates only with its neighbors.
Parallel performance (in strong scaling) has been tested on the Marconi cluster equipped with Xeon Phi 7250 CPU (Knights Landing) processors at 1.40 GHz, available at the CINECA supercomputing facility. 
For the present scaling test, we have chosen the 3D Bell instability test problem (section \ref{sec:bell_instability}) with grid resolution of $256\times128^2$, one particle per cell and RK2 time-stepping.
Figure \ref{fig:parallel_scaling} plots the parallel efficiency, measured as $\Delta T_1/(p\Delta T_{p})$ (where $\Delta T_{p}$ is the CPU time per step per zone using $p$ processors) obtained for $p = 8,16,... 1024$.
The efficiency remains above $\sim 0.8$ up to $p\sim 256$ processors and it decreases to $\sim 0.7$ at the largest number of processors.
Note that, given large inertia of CR, the number of particles per cell remains constant throughout the computation.
Numerical simulations with uneven particle distributions are likely to be less efficient.

\section{Summary}
\label{sec:summary}
%
%
%
%

A method paper describing a fluid-particle hybrid model for the dynamical interaction between a thermal plasma and a population of collisionless non-thermal particles (cosmic rays) has been presented as part of the PLUTO code.
The model equations can be formally derived starting from a three-component plasma in which thermal ions and (massless) electrons are combined together into a single-fluid whereas the non-thermal component is treated kinetically.
The single-fluid equations are those of MHD augmented by source terms accounting for momentum and energy feedback from the CR particles.
Ohm's law is derived from the electrons equation of motion and, neglecting electron-scale physics, it is modified by the presence of the CR-induced Hall term, which accounts for the relative drift between fluid and CRs.
The resulting system of conservation laws is equivalent to the MHD-PIC equations previously derived by \cite{Bai_etal.2015}.
In absence of momentum and energy feedback the particle module can also be employed to investigate the dynamics of test particles embedded in a MHD fluid.

The MHD-PIC approach can be employed on scales that are much larger than the ion skin depth thus offering a significant computational efficiency gain when compared to a Particle-In-Cell (PIC) numerical approach.
In this way, the MHD-PIC formlism paves the way for investigating kinetic effects at nearly macroscopic scales at a more affordable computational cost. 
At the same time, however, the formulation assumes that all electrons are thermal and the charge density ratio between CR particles and fluid is required to be small.
This limits the applicability of the model by compromising micro-physical effects arising at scales smaller than the ion skin depth.

The system of equations describing the composite system of plasma and CRs is solved numerically by combining finite volume Godunov methods for the MHD fluid with PIC techniques for the particle component.
In particular, we have presented a combined algorithm in which the fluid can be evolved using either Corner-Transport-Upwind method or Runge-Kutta time-marching schemes, both available in the PLUTO code.
Particles equations of motion are integrated using a second-order Boris pusher which is time-reversible and features good conservation properties for long-time simulations.
When particles feedback is included, we have presented a modification of the Boris algorithm that preserves second-order accuracy in time.
The correction consists of a predictor step where the electric field can be properly advanced at the half-time level and it does not affect the time-reversibility of the algorithm.
Furthermore, we have suggested two novel particle sub-cycling algorithms that can be applied when the CR dynamical time-scale becomes faster than the fluid evolution. 
By excluding particle feedback on the fluid, the same implementation can be used to study test particles in a dynamically evolving or a static fluid.
The MHD-PIC model have been implemented in the PLUTO code for astrophysical plasma \cite{PLUTO.2007, PLUTO.2012} and it is part of a more general fluid/particle hybrid framework allowing different types of physics to be incorporated.
A companion paper \citep{Vaidya_etal.2018} describes yet another implementation for solving the cosmic-ray transport equation of ultra-relativistic electrons with a time-dependent distribution.

We have verified our implementation through a number of numerical benchmarks including both test-particle dynamics in a fixed electromagnetic field or fully coupled evolution for the composite system (that is, with feedback).
When possible, results obtained from numerical computations have been compared to analytical or reference solutions.

Test-particles configurations have been proposed in order to investigate CR trajectories in both orthogonal and parallel field configuration, reproducing the expected solution within very good accuracy.
A simple benchmark configuration to inspect particle acceleration near a reconnecting X-point has been presented, confirming results from previously known studies \citep{Vekstein_Browning.1997, Browning_Vekstein.2001}.

The solution of the full MHD-PIC system of equations has been verified to be genuinely second-order accurate and a numerical investigation of the non-resonant Bell instability in multiple spatial dimensions \citep{Bell.2004} has shown excellent agreement with the results from linear analysis \citep{Bai_etal.2015}.
The MHD-PIC model has been applied to investigate diffusive shock acceleration in 2D parallel MHD-shocks.
Since a non-thermal population of CR cannot consistently originate from the thermal component within the proposed MHD-PIC framework, an \quotes{ad-hoc} recipe to inject particles in the shock downstream has been proposed.
The proposed injection method is more general than the one used by \cite{Bai_etal.2015} and can be extended to shocks with arbitrary shape provided its energy can be specified.
Being an imposed prescription, the injection process still depends on a free parameter ($\eta$) which controls the ratio between the mass of the generated CR particles and the mass swept by the shock.
Our result reproduce, within statistical fluctuations, the findings of \cite{Bai_etal.2015} confirming that efficient acceleration takes place through Fermi mechanism.
The system evolution is characterized by the development of strong turbulence, initially driven by the Bell instability in the shock precursors, accompanied by the formation of large cavities and filamentation and ensued by strong magnetic field amplification through the shock front.
Particles become accelerated on a few thousands Larmor scales and, in the case of non-relativistic particles, the resulting energy spectrum shows a power-law tail $f(E)\propto E^{-3/2}$.
We have also investigated particle transition to the relativistic regime by considering a second simulation with a larger computational box and used a reduced value of the speed of light.
Although the overall dynamical features are similar to the non-relativistic case, the particle momentum spectrum behaves as $f(p) \propto p^{-4}$ as predicted by Fermi I acceleration.
Our results are in agreement with the findings of previous authors, e.g., \cite{Caprioli_Spitkovsky.2014a} among others.

Our implementation will be made publicly available to the astrophysical community as a new particle module in the PLUTO code.
Future extension of this work will take into account relativistic extension, more accurate injection recipes enabling reconnection physics to be studied and extension to adaptive grids.

\acknowledgments
We acknowledge the CINECA award under the ISCRA initiative, for the availability of high performance computing resources and support.
Our work has been partially supported by the Prin MIUR grants 2015L5EE2Y and Prin Inaf 2014.
We also like to thank the anonymous referee who gave insightful comments during the development of this work.

\bibliography{numerics,astrophysics}
\bibliographystyle{aasjournal}

\clearpage
\appendix

\section{Derivation of the MHD-PIC Equations}
\label{app:MHD-PIC}
%
%

\paragraph{Standard Derivation of the Single Fluid Equations}
The fluid equations for ions and electrons can be obtained by taking moments of the distribution function directly from the Vlasov equations for the two species.
The derivation can be found on many plasma physics textbooks \citep[here we follow the book][]{Chiuderi_Velli.2015}.
We use the subscript $s$ to denote the two species  ($s=e,i$ for electrons and ions, respectively) with mass density $\rho^{(s)}$.
The continuity, momentum and energy equations for the two species take the form
\begin{align}
  \DS  \pd{\rho^{(s)}}{t} + \pd{}{x_k} (\rho v_k)^{(s)}
       & = 0
  \label{eq:app:2F_continuity} 
  \\ \noalign{\medskip}
  \DS  \pd{}{t}   (\rho v_{j})^{(s)}
       + \pd{}{x_k} (\rho v_jv_k + \Pt_{jk})^{(s)}
       -   q^{(s)} E_j
       -   q^{(s)} \left(\frac{\vec{v}^{(s)}}{c}\times \vec{B}\right)_j
       & =  0 
  \label{eq:app:2F_momentum} 
  \\ \noalign{\medskip}
  \DS   \pd{}{t}  \left(\HALF \rho v^2 + \frac{{\rm Tr}(\Pt)}{2}\right)^{(s)}
      + \pd{}{x_k}\left[ \left(
     \frac{1}{2} \rho v^2 + \frac{{\rm Tr}(\Pt)}{2}\right) v_k +
       v_j \Pt_{jk} + {\cal Q}_k\right]^{(s)}
     - q^{(s)}\vec{E}\cdot\vec{v}^{(s)}_i & = 0
  \label{eq:app:2F_energy} 
\end{align}
where $\vec{v}^{(s)}$ is the average velocity, $q^{(s)}$ is the charge density, $\vec{\cal Q}$ is the heat flux vector, $\vec{\E}$ is the electric field, $\vec{B}$ is the magnetic field.
Equations (\ref{eq:app:2F_continuity})-(\ref{eq:app:2F_energy}) are written in terms of average velocity defined as the first-order moment of the distribution function for the $s$ species:
\begin{equation}
  v^{(s)}_j = \av{V_j}^{(s)}  \,,
\end{equation}
where $V_j$ is the velocity coordinate in phase space and $\av{.}^{(s)}$ represents the average taken over the distribution function of the $s$ species.
The pressure tensors and heat flux vector are defined in terms of the peculiar velocities $w^{(s)}_j = V_j - v^{(s)}_j$:
\begin{equation}\label{eq:app:pressureTensor}
  \Pt_{jk}^{(s)}   = \rho^{(s)} \av{w_j w_k}^{(s)} \,,\qquad
  {\cal Q}^{(s)}_k = \rho^{(s)} \av{\frac{w^2}{2}w_k}^{(s)}\,.
\end{equation}
Note that, since $\vec{v}^{(e)}$ will in general be different from $\vec{v}^{(i)}$, the ion and electron pressure tensors as well as the heat flux vector are referred to different fluid velocities.

In order to obtain the single-fluid equations, one needs to add the two momentum equations and likewise the two energy equations.
In this process, however, the pressure tensors of the two species should be redefined so that the ions and electrons peculiar velocities refer to the same fluid speed, 
\begin{equation}\label{eq:app:sF_speed}
  \vec{v}\equiv \vec{v}_g = \frac{  \rho^{(e)} \vec{v}^{(e)}
                    + \rho^{(i)} \vec{v}^{(i)}}{\rho^{(e)} + \rho^{(i)}} \,.
\end{equation}
We are thus entitled to re-introduce the peculiar velocities as $\vec{w}' = \vec{V} - \vec{v}_g$ implying that $\vec{w}'$ now has non-zero mean:
\begin{equation}
  \av{\vec{w}'}^{(s)} =\vec{v}^{(s)} - \vec{v}_g \ne 0 \,.
\end{equation}
By adding the two momentum equations and the two energy equations one arrives, after some algebra \citep[for a detailed derivation see Section 4.3 in the book by][]{Chiuderi_Velli.2015}, at
\begin{align} 
   \DS  \pd{\rho}{t} + \pd{}{x_k}\left(\rho v_k\right)  &= 0
  \label{eq:app:sF_continuity}
  \\ \noalign{\medskip}
   \DS  \pd{}{t} (\rho v_j)
      + \pd{}{x_k} (\rho v_jv_k + \Pt'_{jk})
      - q_g E_j - \left(\frac{\vec{J}_g}{c} \times \vec{B}\right)_j &= 0
  \label{eq:app:sF_momentum}
  \\ \noalign{\medskip}
  \DS  \pd{}{t}   \left(\HALF\rho v^2 + \frac{3}{2} P' \right) 
     + \pd{}{x_k} \left[
       \left( \frac{1}{2} \rho v_g^2 + \frac{3}{2} P' \right)v_k
      + \Pt'_{j k} v_k + {\cal Q}'_k\right] - \vec{J}_g\cdot\vec{E} & =  0\,,
  \label{eq:app:sF_energy}
\end{align}
where
\begin{equation}\label{eq:app:sF_density}
  \rho = \rho^{(e)} + \rho^{(i)}
\end{equation}
is the fluid density, while
\begin{equation}\label{eq:app:sF_cc}
  q_g = q^{(e)} + q^{(i)}
  \,;\qquad
  \vec{J}_g = q^{(e)}\vec{v}^{(e)} + q^{(i)}\vec{v}^{(i)}
\end{equation}
are the total charge density and current density, respectively.
Note also that $q^{(e)} < 0$ while $q^{(i)} > 0$.

The total pressure tensor is now defined by the sum of the ion and electrons tensors, 
\begin{equation}\label{eq:Pt'}
  \Pt'_{jk} = \Pt'^{(e)}_{jk} + \Pt'^{(i)}_{jk}
            = P'\delta_{jk} + \Pi'_{jk} \,,
\end{equation}
where each of the pressure tensors now refers to the the same fluid velocity, that is, 
\begin{equation}\label{eq:app:Pt's}
  \Pt'^{(s)}_{jk} = \rho^{(s)} \av{w'_jw'_k}^{(s)} \,.
\end{equation}
A similar argument applies to the heat conduction flux which is now given by ${\cal Q}'_k = {\cal Q}'^{(e)}_k + {\cal Q}'^{(i)}_k$ with ${\cal Q}'^{(s)}_k = \rho^{(s)}\av{(w')^2w'_k}^{(s)}/2$.
In Equation (\ref{eq:Pt'}) the pressure tensor has been decomposed, assuming isotropy, into a diagonal term containing the scalar pressure $P'$ and in the shear-stress tensor $\Pi'_{jk}$ including only the off-diagonal terms which are different from zero in the presence of viscous forces.

\paragraph{Equivalence of the Pressure Tensors.}
We now prove that, in the limit of massless electrons, the two pressure tensors $\Pt'$ and $\Pt$ are actually equivalent.
This statement can be proven by writing the single-fluid peculiar velocity as
\begin{equation}
  \vec{w}' = \vec{V} - \vec{v}_g
           = \vec{w}^{(s)} + \delta \vec{v}^{(s)} \,,
\end{equation}
where $\delta\vec{v}^{(s)} = \vec{v}^{(s)} - \vec{v}_g$, or more specifically,
\begin{equation}
  \delta \vec{v}^{(i)} = \frac{\rho^{(e)}}{\rho} (\vec{v}^{(i)} - \vec{v}^{(e)})
  \,,\qquad
  \delta \vec{v}^{(e)} = \frac{\rho^{(i)}}{\rho} (\vec{v}^{(e)} - \vec{v}^{(i)}) \,.
\end{equation}
Equation (\ref{eq:app:Pt's}) may now be written as
\begin{equation}\label{eq:app:Pt's_2}
  \Pt'^{(s)}_{jk}
  = \rho^{(s)} \left(  \av{w_j w_k}^{(s)}
                + \av{\delta v_j \delta v_k}^{(s)}\right)
  = \Pt^{(s)}_{jk} + \rho^{(s)} \delta v^{(s)}_j\delta v^{(s)}_k \,.
\end{equation}
Adding the two pressure tensors defined by Equation (\ref{eq:app:Pt's_2}) gives
\begin{equation}\label{eq:app:PeqP'}
  \begin{array}{lcl}
  \Pt'_{jk} &=&\DS
  \Pt^{(e)}_{jk} + \Pt^{(i)}_{jk} 
   + \rho^{(e)}\delta v_j^{(e)} \delta v_k^{(e)}       
   + \rho^{(i)}\delta v_j^{(i)} \delta v_k^{(i)}
  \\ \noalign{\medskip}
  &=&\DS
    \Pt_{jk} + \frac{\rho^{(e)}\rho^{(i)}}{\rho}
             (v^{(i)}_j - v^{(e)}_j)(v^{(i)}_k - v^{(e)}_k) \,.
  \end{array}
\end{equation}
In the limit $\rho^{(e)} \to 0$ we thus obtain $\Pt_{jk} = \Pt'_{jk}$.

\section{CTU-CT Integrator}
\label{app:CTU-CT}
%
%

We describe the implementation details of the Corner Transport Upwind (CTU) scheme combined with the Constrained Transport (CT) method for the solution of the MHD-PIC equations in the PLUTO code.
In what follows, we denote with $V = (\rho,\,\vec{v}_g,\,\vec{B},\, p)$ and $U = (\rho,\, \rho\vec{v}_g,\, \vec{B},\, E_g)$, respectively, the array of primitive and conservative variables.
In the CTU-CT scheme \citep[see, e.g.][]{GarSto.2005, PLUTO.2007}, conservative variables such as density, momentum and energy are stored as zone averages centered at the cell center $\indx \equiv (i,j,k)$ while the magnetic field has a staggered representation so that the primary variables are defined at zone faces, i.e., $B_{x,i+\HALF}$, $B_{y,j+\HALF}$ and $B_{z,k+\HALF}$. 
Note that, for the sake of clarity, we omit the integer subscripts $i$,$j$, and $k$ when unnecessary and only keep the half-increment notation in denoting face values.
The standard CTU-CT scheme must be modified in order to account for particle feedback during interface states computation, Riemann solver and the final update stage.

\begin{enumerate}
  \item At $t=t^n$, compute $\vec{F}^n_{\CR}$ from the particles to the grid
  cell centers.
  This is done using Equation (\ref{eq:Fcr2}) with current and charges obtained
  with Equation (\ref{eq:q+J_deposit}).

  \item Compute normal predictors in primitive variables $\V^*_{i,\pm}$
  (at x-faces), $V^*_{j,\pm}$ (at y-faces) and $V^*_{k,\pm}$ (at z-faces).
  In our notations, $V_{i,\pm} = \lim_{x\to x_{i\pm\HALF}^\mp}V_i(x)$ denotes
  the rightmost ($+$) and leftmost ($-$) reconstructed value 
  from within the cell.
  The reconstruction step can be carried out using either linear or piecewise
  parabolic interpolants, see \cite{PLUTO.2012} for details.
  The reconstruction is then followed by a time extrapolation step that can be
  performed in characteristic variables or using a simple Hancock
  step, see (for instance)  Sections 3.2 - 3.3 of \cite{PLUTO.2012}.     
  For a simple 2$^{\rm nd}$-order reconstruction in the $x$ direction,
  for example, one has the formal corrispondence
  \begin{equation}
    \V^{n}_{i,\pm} = \V^{n} \pm \frac{\delta_x V^n}{2}
  \end{equation}
  where $\delta_xV^n$ are limited slopes in the $x$ direction.
  The normal predictor is then constructed (e.g. following a MUSCL-Hancock
  scheme) as
  \begin{equation}
    V^*_{i,\pm} = V^n_{i,\pm}
                - \frac{\Delta t}{2\Delta x}\tens{A}\delta_x V^n
  \end{equation}
  where $\tens{A}$ is the Jacobian matrix of the one-dimensional primitive
  form of the equations (without CR contributions).
  The construction of the normal predictors in the $y$ and $z$ direction
  is done in a similar way.

  \item Convert normal predictors in primitive variables to conservative ones
  $V^*_{i,\pm} \to U^*_{i,\pm}$ and add CR feedback terms to momentum,
  magnetic field and energy for half time step:
  \begin{equation}\label{eq:CTU:conservativeCorrection}
    \begin{array}{lcl}
    (\rho\vec{v})^{*}_{i,\pm} &\leftarrow&
     \DS (\rho\vec{v})^*_{i,\pm} - \frac{\Delta t}{2} \vec{F}^n_{\CR}
    \\ \noalign{\medskip}
    \vec{B}^*_{i,\pm} &\leftarrow&
      \DS  \vec{B}^*_{i,\pm} + \frac{\Delta t}{2}
         \nabla_x\times\left(\frac{c\vec{F}^n_\CR}{q_i}\right)
    \\ \noalign{\medskip}
    \E^*_{g,i,\pm} &\leftarrow& \DS \E^*_{g,i,\pm}
     - \frac{\Delta t}{2}\left[\nabla_x\cdot\left(
                              \frac{c\vec{F}^n_\CR\times\vec{B}}{4\pi q_i}\right)
                      + \vec{F}^n_{\CR}\cdot\vec{v}^n_g\right]
    \end{array}
  \end{equation}
  where $\nabla_x = (\partial_x, 0, 0)$ is the nabla operator in the $x$-direction.
  Similar expressions hold for the $y$- and $z$-directions.
  Spatial derivatives are discretized using finite differences between
  flux terms computed at the rightmost (+) and leftmost (-) interface values
  from within the cell, e.g.,
  \begin{equation}
    \pd{F_{\CR,z}^n}{x}
    \approx \frac{(F^n_{\CR,z})_{i,+} - (F^n_{\CR,z})_{i,-}}{\Delta x}
  \end{equation}

  \item Solve a Riemann problem between normal predictors by means of a
  standard solver,
  \begin{equation}\label{eq:CTU:predictorFlux}
     \F^*_{i+\HALF} =
     {\cal R}\left(U^{*}_{i,+},\,
                   U^{*}_{i+1,-}\right)
  \end{equation}
  and correct magnetic field and energy fluxes to include contributions from CR:
  \begin{equation}\label{eq:CTU:fluxCorrection}
    \begin{array}{lcl}
    \F^{*,(B_y)}_{i+\HALF} &\leftarrow&\DS \F^{*,(B_y)}_{i+\HALF}
                        + \left(\frac{cF^n_{\CR,z}}{q^n_g}\right)_{i+\HALF}
    \\ \noalign{\medskip}
    \F^{*,(B_z)}_{i+\HALF} &\leftarrow&\DS \F^{*,(B_z)}_{i+\HALF}
                        - \left(\frac{cF^n_{\CR,y}}{q^n_g}\right)_{i+\HALF}
    \\ \noalign{\medskip}
    \F^{*,(E_g)}_{i+\HALF}
    &\leftarrow&\DS \F^{*,(E_g)}_{i+\HALF}
     - \left[\frac{(c\vec{F}^n_\CR\times\vec{B}^n)_x}{4\pi q_i}
       \right]_{i+\HALF}
    \end{array}
  \end{equation}
  when computing fluxes in the $x$ direction.
  The corrections are added by taking the upwind state depending on the sign
  of the density flux.
  Flux corrections in the $y$ and $z$ direction are obtained by
  cyclic permutations of the indices.

  \item Evolve cell-centered values by half time step:
  \begin{equation}
     U^{n+\HALF} = U^n + \frac{\Delta t}{2}\sum_d {\cal L}^*_d
                       + \frac{\Delta t}{2}S^n_\CR  
  \end{equation}
  where $S^n_\CR = (0,-\vec{F}_\CR, \vec{0}, -\vec{F}_\CR\cdot\vec{v}_g)^n$
  is the CR source term.
  In the previous equation, 
    \begin{equation}\label{eq:CTU:Lop}
     {\cal L}^*_x =  -\frac{\F^*_{i+\HALF} - \F^*_{i-\HALF}}{\Delta x}
     \,,\qquad
     {\cal L}^*_y =  -\frac{\F^*_{j+\HALF} - \F^*_{j-\HALF}}{\Delta y}
     \,,\qquad
     {\cal L}^*_z =  -\frac{\F^*_{k+\HALF} - \F^*_{k-\HALF}}{\Delta z}
  \end{equation}
  are the flux-difference right hand side operators.

  \item Advance face-centered magnetic field by half a step:
  \begin{equation}
    \begin{array}{lcl}
    B^{n+\HALF}_{x,i+\HALF} &=&\DS B^{n}_{x,i+\HALF}
    - \frac{\Delta t}{2\Delta y}\left(
       cE^{*}_{z,i+\HALF,j+\HALF} - cE^{*}_{z,i+\HALF,j-\HALF}\right) 
    + \frac{\Delta t}{2\Delta z}\left(
       cE^{*}_{y,i+\HALF,k+\HALF} - cE^{*}_{y,i+\HALF,k-\HALF}\right) 
    \\ \noalign{\medskip}
    B^{n+\HALF}_{y,j+\HALF} &=&\DS B^{n}_{y,j+\HALF}
    - \frac{\Delta t}{2\Delta z}\left(
       cE^{*}_{x,j+\HALF,k+\HALF} - cE^{*}_{x,j+\HALF,k-\HALF}\right) 
    + \frac{\Delta t}{2\Delta x}\left(
       cE^{*}_{z,i+\HALF,j+\HALF} - cE^{*}_{z,i-\HALF,j+\HALF}\right) 
    \\ \noalign{\medskip}
    B^{n+\HALF}_{z,k+\HALF} &=&\DS B^{n}_{z,k+\HALF}
    - \frac{\Delta t}{2\Delta x}\left(
       cE^{*}_{y,i+\HALF,k+\HALF} - cE^{*}_{y,i-\HALF,k+\HALF}\right) 
    + \frac{\Delta t}{2\Delta y}\left(
       cE^{*}_{x,j+\HALF,k+\HALF} - cE^{*}_{x,j-\HALF,k+\HALF}\right) 
    \end{array}
  \end{equation}
  In the previous equations $c\vec{E}^*$ has been reconstructed from the
  face-centered fluxes computed at the predictor step
  (Equation \ref{eq:CTU:predictorFlux}) to cell edges by using a
  suitable reconstruction procedure.
  In the present work we employ the UCT-Contact method by \cite{GarSto.2005}.

  \item Advance particles by a full step using the algorithm described in
  Section \ref{sec:particle_mover}.
  Also, compute the particle momentum and energy change over the time step
  and deposit them on the grid to obtain $S_\CR^{n+\HALF}$ using
  Equation (\ref{eq:CTU_feedback}).

  \item Correct states with transverse flux gradients to form corner-coupled
  states:
  \begin{equation}\label{eq:CTU:CCstates}
    U^{n+\HALF}_{i,\pm} = U^{*}_{i,\pm}
                         + \frac{\Delta t}{2}\sum_{d\ne x}{\cal L}^*_d
  \end{equation}
  where the summation include only the right-hand side operators in the
  transverse directions.
  Note that Equation (\ref{eq:CTU:CCstates}) does not contain the source term
  since this has already been added in Equation
  (\ref{eq:CTU:conservativeCorrection}).
  As usual, corner-coupled states in the $y$ and $z$ direction are obtained
  by suitable permutations.

  \item Solve Riemann problem between corner-coupled states:
  \begin{equation}\label{eq:CTU:correctorFlux}
     \F^{n+\HALF}_{i+\HALF} =
      {\cal R}\left(U^{n+\HALF}_{i,+}, U^{n+\HALF}_{i+1,-}\right)
  \end{equation}
  and correct fluxes in analogy with the predictor step, i.e.,
  Equations (\ref{eq:CTU:fluxCorrection}).

  \item Advance the zone-averaged conservative variables to the next time level:
  \begin{equation}
    U^{n+1} = U^n
   + \Delta t\sum_d {\cal L}^{n+\HALF}_{d} + \Delta t S^{n+\HALF}_{\CR}
  \end{equation}
  where ${\cal L}_d$ is obtained as in Equation (\ref{eq:CTU:Lop}) using the fluxes
  (\ref{eq:CTU:correctorFlux}).

  \item Advance face-centered magnetic field to the next time level:
  \begin{equation}
    \begin{array}{lcl}
    B^{n+1}_{x,i+\HALF} &=&\DS B^{n}_{x,i+\HALF}
    - \frac{\Delta t}{\Delta y}\left(
       cE^{n+\HALF}_{z,i+\HALF,j+\HALF} - cE^{n+\HALF}_{z,i+\HALF,j-\HALF}\right) 
    + \frac{\Delta t}{\Delta z}\left(
       cE^{n+\HALF}_{y,i+\HALF,k+\HALF} - cE^{n+\HALF}_{y,i+\HALF,k-\HALF}\right)
    \\ \noalign{\medskip}
    B^{n+1}_{y,j+\HALF} &=&\DS B^{n}_{y,j+\HALF}
    - \frac{\Delta t}{\Delta z}\left(
       cE^{n+\HALF}_{x,j+\HALF,k+\HALF} - cE^{n+\HALF}_{x,j+\HALF,k-\HALF}\right) 
    + \frac{\Delta t}{\Delta x}\left(
       cE^{n+\HALF}_{z,i+\HALF,j+\HALF} - cE^{n+\HALF}_{z,i-\HALF,j+\HALF}\right)
    \\ \noalign{\medskip}
    B^{n+1}_{z,k+\HALF} &=&\DS B^{n}_{z,k+\HALF}
    - \frac{\Delta t}{\Delta x}\left(
       cE^{n+\HALF}_{y,i+\HALF,k+\HALF} - cE^{n+\HALF}_{y,i-\HALF,k+\HALF}\right) 
    + \frac{\Delta t}{\Delta y}\left(
       cE^{n+\HALF}_{x,j+\HALF,k+\HALF} - cE^{n+\HALF}_{x,j-\HALF,k+\HALF}\right)
    \end{array}
  \end{equation}
  where $\vec{E}^{n+\HALF}$ has been reconstructed from the face centered flux
  to cell edges by using a suitable reconstruction procedure, e.g.,
  \cite{GarSto.2005}.

\end{enumerate}



\end{document}